\documentclass[a4paper,11pt]{article}
\pdfoutput=1

\usepackage{jheppub}
\usepackage{amsmath,amssymb,amscd}
\usepackage{bbm}
\usepackage{graphicx}
\usepackage{url}
\usepackage{longtable}
\usepackage{accents}
\usepackage[export]{adjustbox}

\graphicspath{ {fig/} }

\makeatletter
\def\hlinewd#1{%
  \noalign{\ifnum0=`}\fi\hrule \@height #1 \futurelet
   \reserved@a\@xhline}
\makeatother

\makeatletter
\renewcommand\@fpheader{}
\renewcommand\@journal{}
\makeatother

\newcommand\thickbar[1]{\accentset{\rule{.4em}{.8pt}}{#1}}
\newcommand{\Graph}[2][0.3]{\vcenter{\hbox{\includegraphics[scale=#1]{#2}}}}

\newcommand{\figgraph}[3]{\overset{(#3-2\epsilon)}{\Graph[#1]{#2}}}
\definecolor{darkgreen}{rgb}{0.,.3,0}
\definecolor{darkblue}{rgb}{0.0,0.0,0.5}

\newcommand{\ud}{\mathrm{d}}
\DeclareMathOperator{\Li}{Li}
 
\title{
Numerical Multi-Loop Calculations via Finite Integrals and One-Mass EW-QCD Drell-Yan Master Integrals
}

\preprint{MSUHEP-17-002, MITP/16-101, TCDMATH-17-03}

\author[a,b]{Andreas von Manteuffel,}
\author[\,c]{and Robert M.~Schabinger}

\affiliation[a]{Department of Physics and Astronomy, Michigan State University,\\
648 N. Shaw Lane, East Lansing, MI 48824, USA}
\affiliation[b]{PRISMA Cluster of Excellence, Johannes Gutenberg University,\\
Staudinger Weg 7, 55099 Mainz, Germany}
\affiliation[c]{Hamilton Mathematics Institute, Trinity College,\\
College Green, Dublin 2, Ireland}

\emailAdd{manteuffel@pa.msu.edu}
\emailAdd{schabr@maths.tcd.ie}

\abstract{
We study a recently-proposed approach to the numerical evaluation of
multi-loop Feynman integrals using available sector decomposition programs.
As our main example, we consider the two-loop integrals for the $\alpha \alpha_s$ corrections
to Drell-Yan lepton production with up to one massive vector boson in physical kinematics.
As a reference, we evaluate these planar and non-planar integrals by the method
of differential equations through to weight five.
Choosing a basis of finite integrals for the numerical evaluation
with {\tt SecDec\;3} leads to tremendous performance improvements and
renders the otherwise problematic seven-line
topologies numerically accessible.
As another example, basis integrals for massless QCD three loop form factors are evaluated with {\tt FIESTA\;4}. 
Here, employing a basis of finite integrals results in an overall speedup of more than
an order of magnitude.
}

\begin{document}
\LTcapwidth=\textwidth
\unitlength1cm
\maketitle
\allowdisplaybreaks
\section{Introduction}
In the early days of hadron collider physics and Quantum Chromodynamics (QCD), the Drell-Yan process was recognized to be of fundamental importance \cite{Drell:1970wh} and, over the years, it has received much attention.
The electroweak corrections of relative order $\alpha$ were calculated in
\cite{Baur:2001ze,Dittmaier:2001ay,Baur:2004ig}.
Another milestone of particular interest was set long ago when the corrections of
relative order $\alpha_s^2$ were calculated in references \cite{Hamberg:1990np,Harlander:2002wh,Anastasiou:2003ds,Melnikov:2006di}.
The fact that these corrections were found to be relatively large in size motivates one to
consider the subleading terms in the two-loop perturbative expansion of the Standard 
Model Drell-Yan production cross section as well.
The most important class of subleading two-loop contributions are those of relative order
$\alpha \alpha_s$, what we shall hereafter refer to as mixed Electroweak-Quantum Chromodynamic (EW-QCD) corrections. 
The mixed Electroweak-Quantum Chromodynamic corrections of relative order $\alpha\alpha_s$
have already been studied in certain approximations. First, in reference \cite{Kilgore:2011pa}, the virtual corrections without propagating $W$ or $Z$ bosons were calculated.
More recently, the full set of two-loop corrections were considered in the single massive gauge boson resonance region \cite{Dittmaier:2014qza,Dittmaier:2015rxo}.
Progress on real emission contributions and the infrared structure has been
presented in~\cite{Kilgore:2013uta,deFlorian:2015ujt,Bonciani:2016wya}.
Recently, the planar two-loop master integrals with up to two same-mass vector bosons in the loops
were calculated in the Euclidean region \cite{Bonciani:2016ypc}.
The non-planar master integrals relevant to the problem are vertex integrals, which have been available in the literature for some time \cite{Gonsalves:1983nq,Aglietti:2003yc}.

In this paper, we consider the two-loop master integrals for mixed EW-QCD corrections
to Drell-Yan production with up to a single massive vector boson exchanged
and focus on physical kinematics.
The integrals may be conveniently computed using
the method of differential equations \cite{Kotikov:1990kg,Kotikov:1991hm,Kotikov:1991pm,Bern:1992em,Bern:1993kr,Remiddi:1997ny,Gehrmann:1999as}.
We employ a normal form basis \cite{Kotikov:2010gf,Henn:2013pwa,Henn:2013tua},
integrate the $\epsilon$-expanded differential equations in terms of multiple polylgarithms \cite{LDanilevsky}, and impose regularity conditions to fix the boundary constants.
In some cases this is supplemented by explicit solutions worked out directly from Feynman parameters to all orders in the parameter of dimensional regularization, $\epsilon$. 
Analytical solutions for all of our $\epsilon$-expanded integrals are included with our {\tt arXiv} submission through to weight four.\footnote{In this work, 
we make extensive use of the {\tt GiNaC}-based implementation of the multiple polylogarithms \cite{Bauer:2000cp,Vollinga:2004sn} and therefore adopt the notation of reference \cite{Vollinga:2004sn} for our function
definitions.}
Although the weight four solution is one of the primary goals of our study, we
found it interesting to calculate to one order higher in $\epsilon$ than necessary
to facilitate checks in the second part of this work.

To cross-check an analytical solution for a multi-loop integral or to even produce the primary
result for a phenomenological application, numerical evaluations using
sector decomposition programs~\cite{Binoth:2000ps,Bogner:2007cr,Borowka:2015mxa,Smirnov:2015mct}
are very valuable.
One might hope that one could just pick any integral basis,
run one of the available programs for a physical point in phase space,
and then obtain a reasonably accurate and precise result in, say, a few days' time.
However, for our mixed EW-QCD integrals we found this not to be the case in practice.
Instead, picking a basis which is free of infrared and ultraviolet divergences
allows us to obtain reliable and relatively fast results with {\tt SecDec\;3}
\cite{Borowka:2015mxa}.
This method of finite integrals introduced by Erik Panzer and the authors
\cite{vonManteuffel:2014qoa,Panzer:2014gra} is generic and straightforward to automate.
The approach has successfully been employed in the calculation
of two-loop integrals for double Higgs production with exact top quark mass dependence \cite{Borowka:2016ehy,Borowka:2016ypz}
already.
Here, we present a detailed numerical study of performance gains stemming from the use of this technique.
We quantify the effect also for a very different setup:
the evaluation of the massless three-loop form factor integrals
with {\tt FIESTA\;4}, which performs particularly well for such Euclidean integrals.
Again, we observe drastic improvements with regard to the numerical convergence.

The outline of this article is as follows. In Section \ref{sec:notandcon}, we comment on the Feynman diagrams relevant to the mixed EW-QCD corrections and define
the integral families which we use to describe our master integrals.
In Section \ref{sec:deq}, we define our normal form basis, present the differential equations they satisfy,
and solve the integrals in terms of multiple polylogarithms.
In Section \ref{sec:mixedDY}, we present the results of our numerical analysis of the two-loop mixed EW-QCD Drell-Yan integrals
and discuss the crucial performance enhancements obtained by employing a basis of finite integrals.
We present our conclusions in Section \ref{sec:end}.
Finally, in Appendix \ref{sec:3Lff}, we present a self-contained numerical study of the master integrals for massless three-loop form factors
up to contributions of weight eight.

\section{Integral Families}
\label{sec:notandcon}

\begin{figure}[t!]
\centering
\includegraphics[scale=0.6]{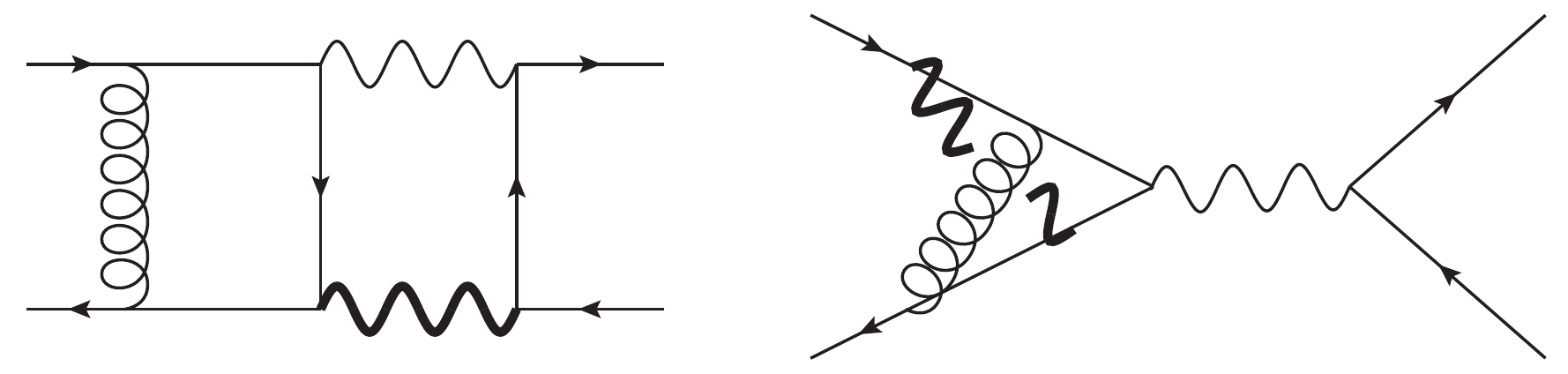}
\caption{Selected planar and non-planar two-loop Feynman diagrams for
Drell-Yan lepton production.
The heavy wavy lines denote a $Z$ boson, the light wavy lines denote a photon,
and the curly lines denote a gluon.
}
\label{fig:samplediagrams}
\end{figure}
We consider neutral-current and charged-current lepton pair production in quark-antiquark annihilation,
\begin{align}
 q(p_1) + \bar q(p_2) &\to \ell^-(p_3) + \ell^+(p_4)\\
\label{eq:ourprocess}
 q^\prime(p_1) + \bar q(p_2) &\to \ell^{-}(p_3) + \bar{\nu_\ell}(p_4)\\
  q^\prime(p_1) + \bar q(p_2) &\to \ell^{+}(p_3) + \nu_\ell(p_4)\,,
\end{align}
where $p_1^2 = p_2^2 = p_3^2 = p_4^2 = 0$.
As usual, the Mandelstam invariants are 
\begin{equation}
 s=(p_1+p_2)^2,\quad t = (p_1-p_3)^2, \quad u = (p_1-p_4)^2
 \quad\mbox{with}\quad s+t+u = 0\,.
\end{equation}
For the physical part of the phase space we have
\begin{equation}
 s>0, \quad -s<t<0,\quad -s<u<0
\end{equation}
and selected Feynman diagrams are depicted in Figure \ref{fig:samplediagrams} above.

We are interested in the two-loop master integrals required for the calculation
of the $\mathcal{O}(\alpha_s \alpha)$ corrections to these processes.
In this paper, we restrict ourselves to the subset of these integrals, which
involve at most one massive vector boson, $W$ or $Z$, propagating in the loops.
We will use the generic symbol $m$ for the vector boson mass. The relevant loop integrals can be indexed using the four integral families
shown in Table~\ref{tab:famsmixedDY}.
Integration by parts reduction~\cite{Tkachov:1981wb,Chetyrkin:1981qh,Laporta:2001dd}
allows us to rewrite any Feynman integral in these families as a linear combination
of master integrals.
The reductions for this paper were performed with the implementation
{\tt Reduze 2} \cite{vonManteuffel:2012np,Studerus:2009ye,Bauer:2000cp,fermat}
and we adapt its conventions for the labeling of sectors, etc.
An ideal choice for the master integrals will depend on the application
and we will discuss suitable options for different applications in the following
sections.

\begin{table}[t!]
\centering
\begin{tabular}{llll}
Family $\mathrm{A}$ & Family $\mathrm{B}$\hspace{5mm} & Family $\mathrm{C}$ & Family $\mathrm{D}$\hspace{5mm}\\[1mm]
\hlinewd{2pt}                                                                                                            
\rule{0pt}{2ex}~$k_1^2$                     & $k_2^2$               & $k_1^2$                     & $k_2^2$\\
\rule{0pt}{2ex}~$k_2^2$                     & $(k_2 - p_3)^2$       & $k_2^2$                     & $(k_2 - p_3)^2$\\
\rule{0pt}{2ex}~$(k_1-k_2)^2$               & $(k_2-p_1-p_2)^2$     & $(k_1-k_2)^2$               & $(k_2-p_1-p_2)^2$\\
\rule{0pt}{2ex}~$(k_1-p_1)^2$               & $(k_1-p_1)^2$         & $(k_1-p_1)^2$               & $(k_1-p_1)^2$\\
\rule{0pt}{2ex}~$(k_2-p_1)^2$               & $(k_1-k_2)^2$         & $(k_2-p_1)^2$               & $(k_1-k_2)^2$\\
\rule{0pt}{2ex}~$(k_1-p_1-p_2)^2$           & $(k_1-k_2+p_2)^2$     & $(k_1-p_1-p_2)^2$           & $(k_1-k_2+p_2)^2$\\
\rule{0pt}{2ex}~$(k_2-p_1-p_2)^2$           & $k_1^2$               & $(k_2-p_1-p_2)^2 - m^2$     & $k_1^2 - m^2$\\
\rule{0pt}{2ex}~$(k_1-p_3)^2$               & $(k_1-p_3)^2$         & $(k_1-p_3)^2$               & $(k_1-p_3)^2 - m^2$\\
\rule{0pt}{2ex}~$(k_2-p_3)^2$               & $(k_1-k_2+p_1+p_2)^2$ & $(k_2-p_3)^2$               & $(k_1-k_2+p_1+p_2)^2$    
\end{tabular}
\caption{The four integral families used for the two-loop mixed EW-QCD Drell-Yan master integrals with either no massive internal lines or exactly one massive internal line.}
\label{tab:famsmixedDY}
\end{table}

\section{Analytical Solution from Differential Equations}
\label{sec:deq}

We employ the method of differential equations to derive analytical solutions
for the master integrals.
Altogether, we find that we need to consider forty-nine master integrals
for the system of differential equations.
Some of these integrals are related by a permutation of external legs.
They are covered by crossed integral families, which we will denote by an overline ({\it e.g.} we write $\mathrm{\thickbar{F}}$:$x$ for the crossed version of sector $x$ from family $\mathrm{F}$). Note that, because our
four-particle integral topologies are naturally a function of the invariants $s$ and $t$ (and never $u = -s -t$), we may without loss of intelligibility suppress all explicit momentum labels; instead, we use directed external
lines and explicit function arguments to help clarify our basis integral definitions when necessary. As usual, we use dots to denote doubled propagators, heavy lines to denote massive propagators,
and explicitly write all numerator insertions in square brackets.
The starting point for the construction of our basis are the following integrals:

\begin{align*}
f_{1}^{\mathrm{A:38}} = \overset{(4-2\epsilon)}{\includegraphics[valign = m, raise = .3 cm, height = .125\linewidth, width = .125\linewidth,keepaspectratio]{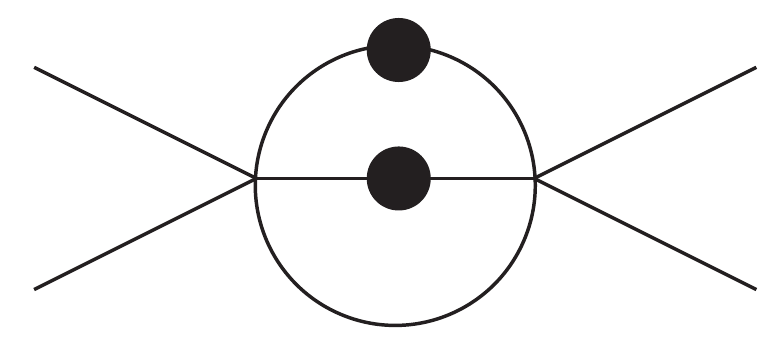}}(s)\quad
f_{2}^{\mathrm{\thickbar{A}:38}} = \overset{(4-2\epsilon)}{\includegraphics[valign = m, raise = .3 cm, height = .125\linewidth, width = .125\linewidth,keepaspectratio]{f1andf2}}(t)\quad
f_{3}^{\mathrm{A:99}} = \overset{(4-2\epsilon)}{\includegraphics[valign = m, raise = .3 cm, height = .175\linewidth, width = .175\linewidth,keepaspectratio]{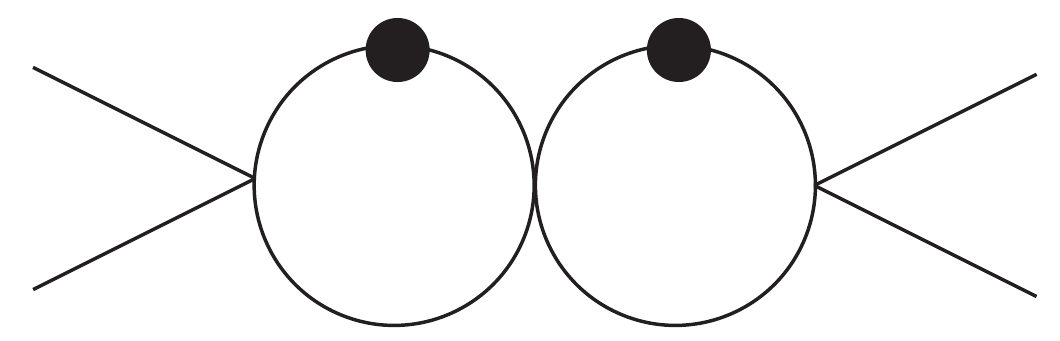}}(s)
\end{align*}

\begin{align*}
f_{4}^{\mathrm{\thickbar{A}:99}} = \overset{(4-2\epsilon)}{\includegraphics[valign = m, raise = .3 cm, height = .175\linewidth, width = .175\linewidth,keepaspectratio]{f3andf4}}(t)\quad
f_{5}^{\mathrm{A:53}} = \overset{(4-2\epsilon)}{\includegraphics[valign = m, raise = .2 cm, height = .135\linewidth, width = .135\linewidth,keepaspectratio]{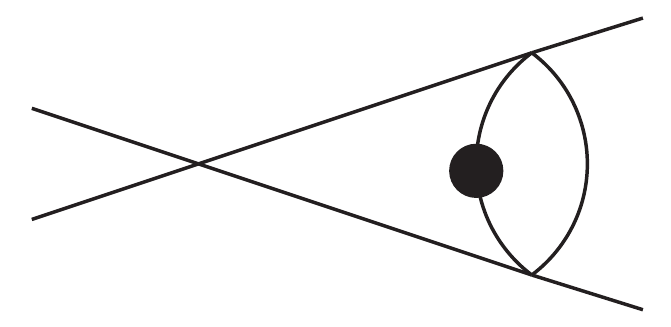}}(s)\quad
f_{6}^{\mathrm{\thickbar{A}:53}} = \overset{(4-2\epsilon)}{\includegraphics[valign = m, raise = .2 cm, height = .135\linewidth, width = .135\linewidth,keepaspectratio]{f5andf6}}(t)
\end{align*}

\begin{align*}
f_{7}^{\mathrm{A:174}} = \overset{(4-2\epsilon)}{\includegraphics[valign = m, raise = .2 cm, height = .135\linewidth, width = .135\linewidth,keepaspectratio]{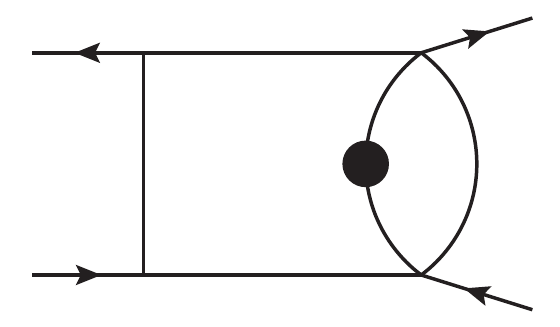}}(s,t)\quad
f_{8}^{\mathrm{\thickbar{A}:174}} = \overset{(4-2\epsilon)}{\includegraphics[valign = m, raise = .2 cm, height = .135\linewidth, width = .135\linewidth,keepaspectratio]{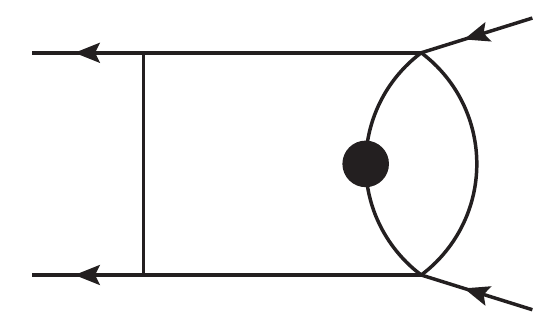}}(t,s)\quad
f_{9}^{\mathrm{A:182}} = \overset{(4-2\epsilon)}{\includegraphics[valign = m, raise = .2 cm, height = .135\linewidth, width = .135\linewidth,keepaspectratio]{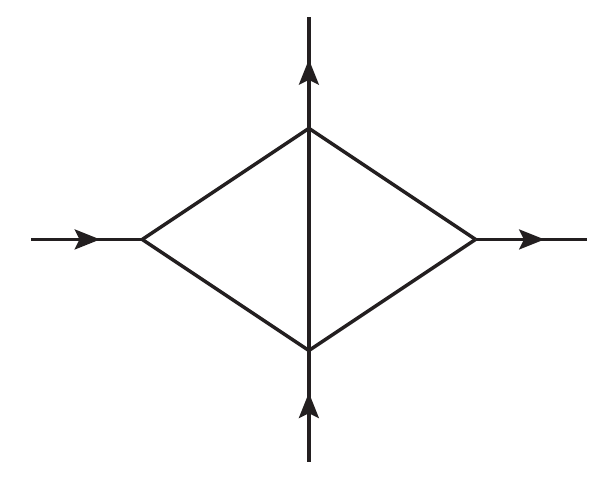}}(s,t)
\end{align*}

\begin{align*}
f_{10}^{\mathrm{A:247}} = \overset{(4-2\epsilon)}{\includegraphics[valign = m, raise = .22 cm, height = .15\linewidth, width = .15\linewidth,keepaspectratio]{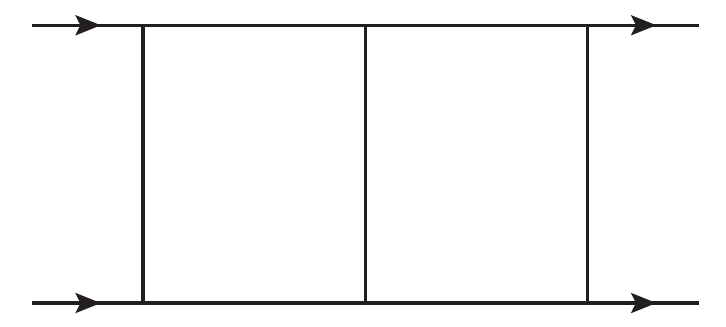}}(s,t)\quad
f_{11}^{\mathrm{A:247}} = \overset{(4-2\epsilon)}{\includegraphics[valign = m, raise = .22 cm, height = .15\linewidth, width = .15\linewidth,keepaspectratio]{f10andf11}}\left[(k_2 - p_3)^2\right](s,t)
\end{align*}

\begin{align*}
f_{12}^{\mathrm{B:125}} = \overset{(4-2\epsilon)}{\includegraphics[valign = m, raise = .3 cm, height = .175\linewidth, width = .175\linewidth,keepaspectratio]{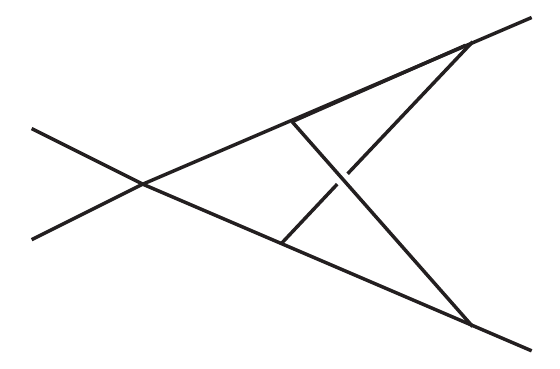}}(s)
\end{align*}

\begin{align*}
f_{13}^{\mathrm{C:97}} = \overset{(4-2\epsilon)}{\includegraphics[valign = m, raise = .3 cm, height = .2\linewidth, width = .2\linewidth,keepaspectratio]{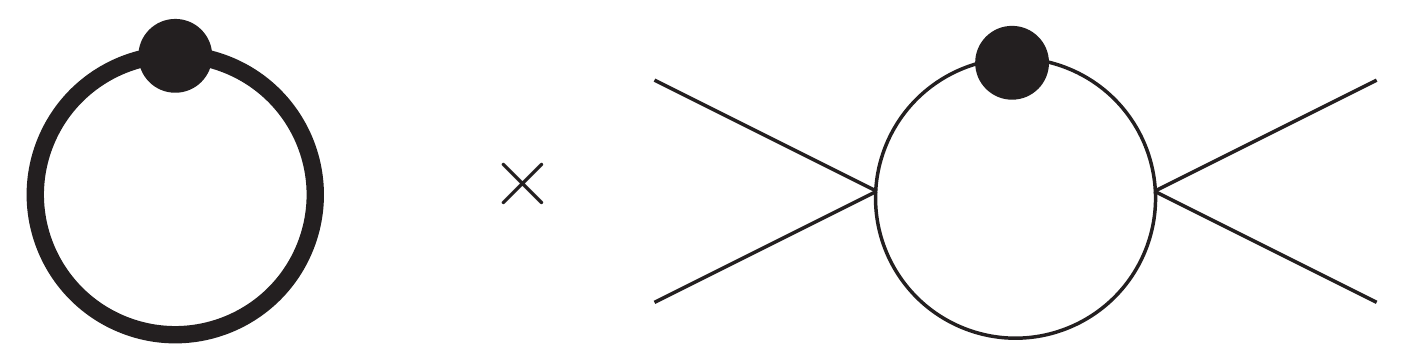}}(s)\quad
f_{14}^{\mathrm{\thickbar{C}:97}} = \overset{(4-2\epsilon)}{\includegraphics[valign = m, raise = .3 cm, height = .2\linewidth, width = .2\linewidth,keepaspectratio]{f13andf14}}(t)\quad
f_{15}^{\mathrm{C:76}} = \overset{(4-2\epsilon)}{\includegraphics[valign = m, raise = .3 cm, height = .05\linewidth, width = .05\linewidth,keepaspectratio]{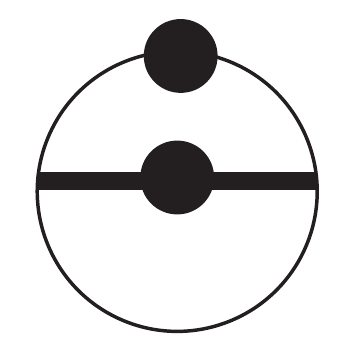}}\quad
\end{align*}

\begin{align*}
f_{16}^{\mathrm{C:69}} = \overset{(4-2\epsilon)}{\includegraphics[valign = m, raise = .3 cm, height = .125\linewidth, width = .125\linewidth,keepaspectratio]{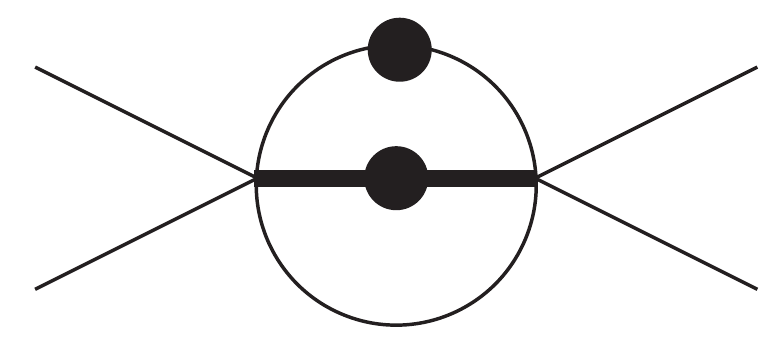}}(s)\quad
f_{17}^{\mathrm{C:69}} = \overset{(4-2\epsilon)}{\includegraphics[valign = m, raise = .3 cm, height = .125\linewidth, width = .125\linewidth,keepaspectratio]{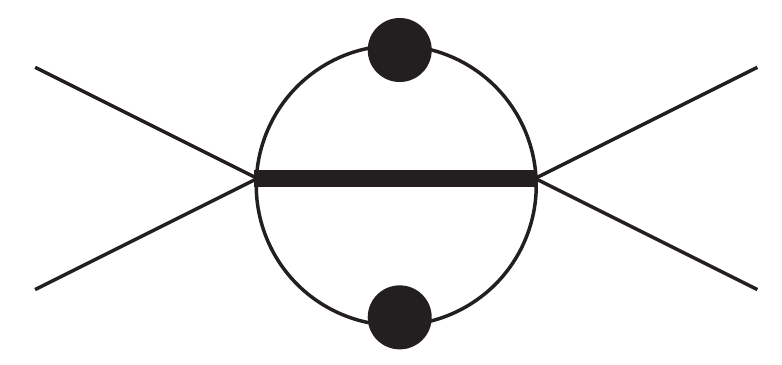}}(s)\quad
f_{18}^{\mathrm{C:99}} = \overset{(4-2\epsilon)}{\includegraphics[valign = m, raise = .3 cm, height = .175\linewidth, width = .175\linewidth,keepaspectratio]{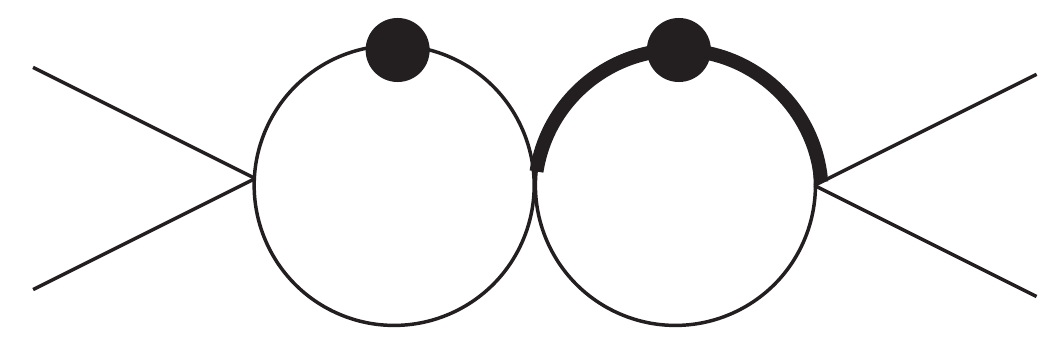}}(s)
\end{align*}

\begin{align*}
f_{19}^{\mathrm{C:102}} = \overset{(4-2\epsilon)}{\includegraphics[valign = m, raise = .6 cm, height = .175\linewidth, width = .175\linewidth,keepaspectratio]{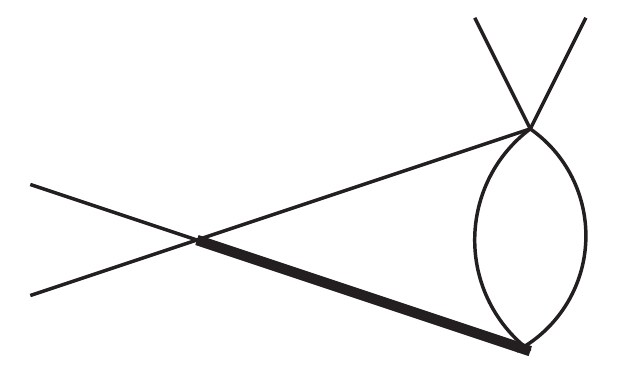}}(s)\quad
f_{20}^{\mathrm{C:78}} = \overset{(4-2\epsilon)}{\includegraphics[valign = m, raise = .1 cm, height = .175\linewidth, width = .175\linewidth,keepaspectratio]{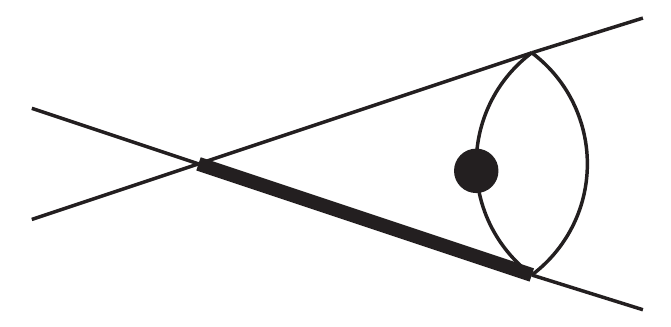}}(s)\quad
f_{21}^{\mathrm{C:212}} = \overset{(4-2\epsilon)}{\includegraphics[valign = m, raise = .1 cm, height = .15\linewidth, width = .15\linewidth,keepaspectratio]{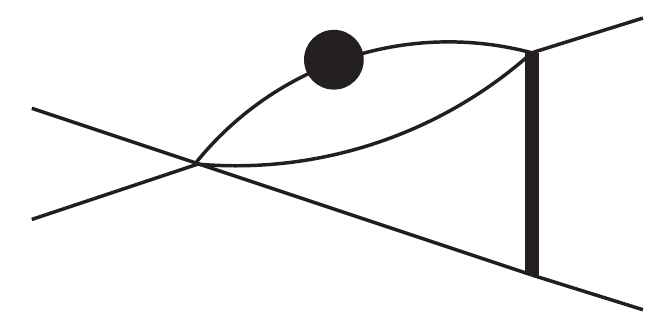}}(t)
\end{align*}

\begin{align*}
f_{22}^{\mathrm{\thickbar{C}:212}} = \overset{(4-2\epsilon)}{\includegraphics[valign = m, raise = .1 cm, height = .15\linewidth, width = .15\linewidth,keepaspectratio]{f21andf22}}(s)\quad
f_{23}^{\mathrm{C:204}} = \overset{(4-2\epsilon)}{\includegraphics[valign = m, raise = .1 cm, height = .175\linewidth, width = .175\linewidth,keepaspectratio]{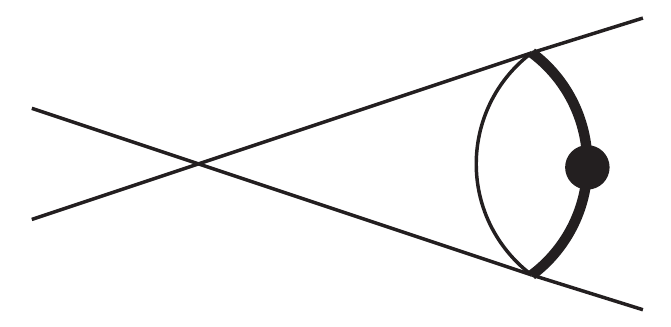}}(t)\quad
f_{24}^{\mathrm{C:204}} = \overset{(4-2\epsilon)}{\includegraphics[valign = m, raise = .1 cm, height = .175\linewidth, width = .175\linewidth,keepaspectratio]{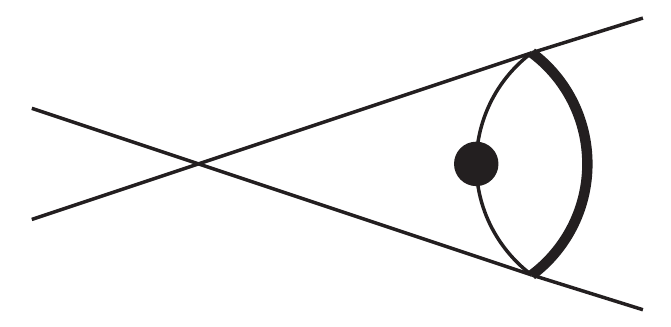}}(t)
\end{align*}

\begin{align*}
f_{25}^{\mathrm{\thickbar{C}:204}} = \overset{(4-2\epsilon)}{\includegraphics[valign = m, raise = .1 cm, height = .175\linewidth, width = .175\linewidth,keepaspectratio]{f23andf25}}(s)\quad
f_{26}^{\mathrm{\thickbar{C}:204}} = \overset{(4-2\epsilon)}{\includegraphics[valign = m, raise = .1 cm, height = .175\linewidth, width = .175\linewidth,keepaspectratio]{f24andf26}}(s)\quad
f_{27}^{\mathrm{C:472}} = \overset{(4-2\epsilon)}{\includegraphics[valign = m, raise = .2 cm, height = .175\linewidth, width = .175\linewidth,keepaspectratio]{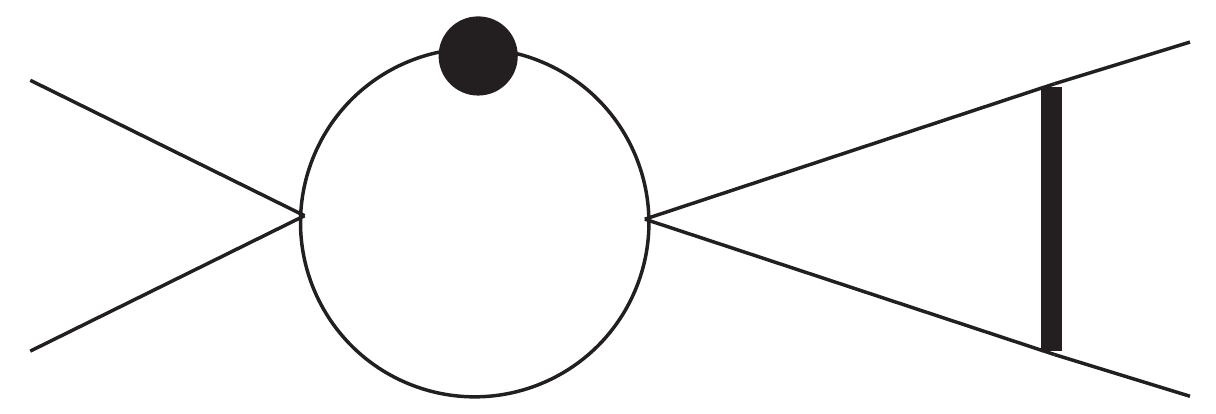}}(t)
\end{align*}

\begin{align*}
f_{28}^{\mathrm{C:372}} = \overset{(4-2\epsilon)}{\includegraphics[valign = m, raise = .2 cm, height = .15\linewidth, width = .15\linewidth,keepaspectratio]{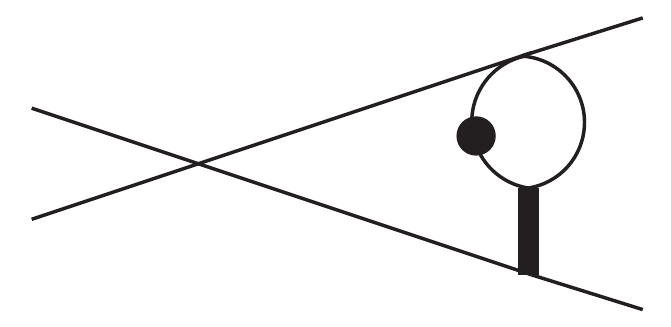}}(t)\quad
f_{29}^{\mathrm{C:244}} = \overset{(4-2\epsilon)}{\includegraphics[valign = m, raise = .7 cm, height = .175\linewidth, width = .175\linewidth,keepaspectratio]{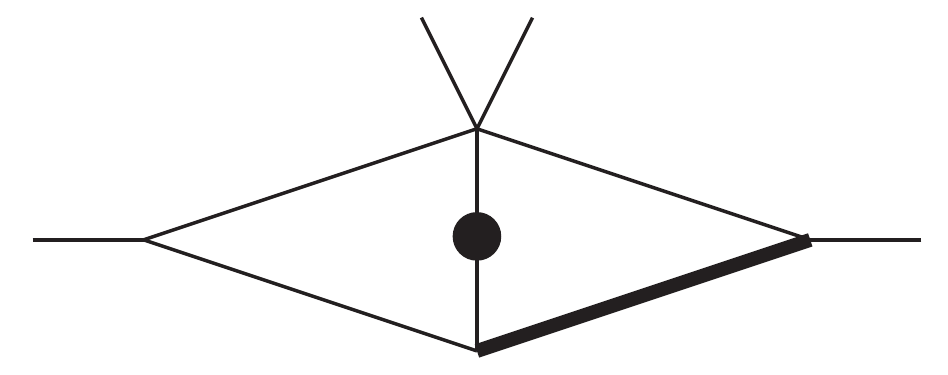}}(t)\quad
f_{30}^{\mathrm{\thickbar{C}:244}} = \overset{(4-2\epsilon)}{\includegraphics[valign = m, raise = .7 cm, height = .175\linewidth, width = .175\linewidth,keepaspectratio]{f29andf30}}(s)\quad
\end{align*}

\begin{align*}
f_{31}^{\mathrm{C:110}} = \overset{(4-2\epsilon)}{\includegraphics[valign = m, raise = .7 cm, height = .175\linewidth, width = .175\linewidth,keepaspectratio]{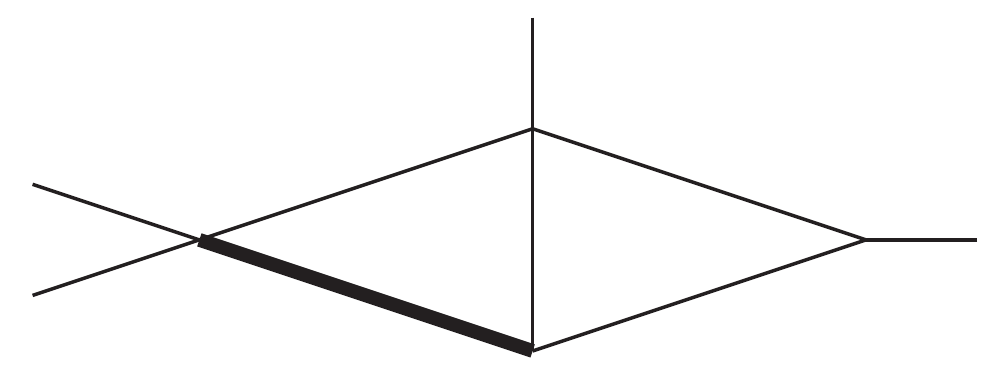}}(s)\quad
f_{32}^{\mathrm{C:220}} = \overset{(4-2\epsilon)}{\includegraphics[valign = m, raise = .7 cm, height = .175\linewidth, width = .175\linewidth,keepaspectratio]{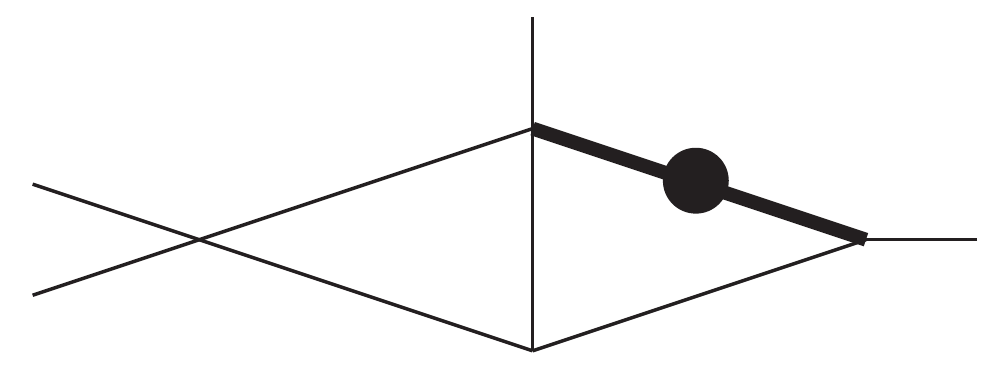}}(t)\quad
f_{33}^{\mathrm{\thickbar{C}:220}} = \overset{(4-2\epsilon)}{\includegraphics[valign = m, raise = .7 cm, height = .175\linewidth, width = .175\linewidth,keepaspectratio]{f32andf33}}(s)
\end{align*}

\begin{align*}
f_{34}^{\mathrm{C:117}} = \overset{(4-2\epsilon)}{\includegraphics[valign = m, raise = .7 cm, height = .175\linewidth, width = .175\linewidth,keepaspectratio]{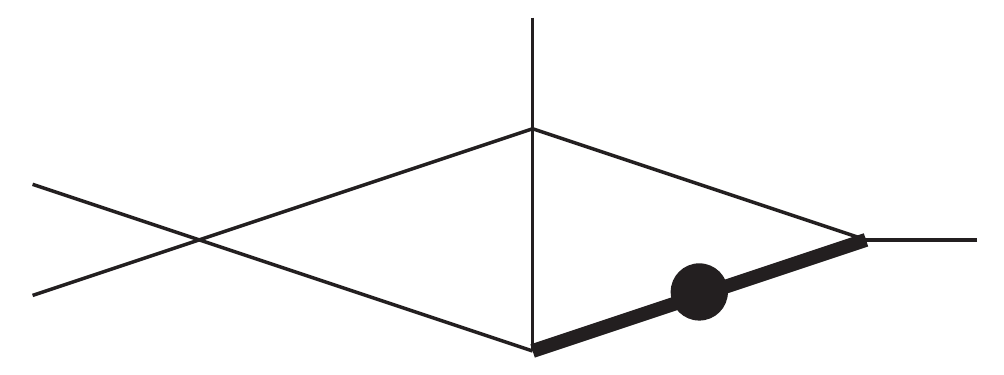}}(s)\quad
f_{35}^{\mathrm{C:117}} = \overset{(4-2\epsilon)}{\includegraphics[valign = m, raise = .7 cm, height = .175\linewidth, width = .175\linewidth,keepaspectratio]{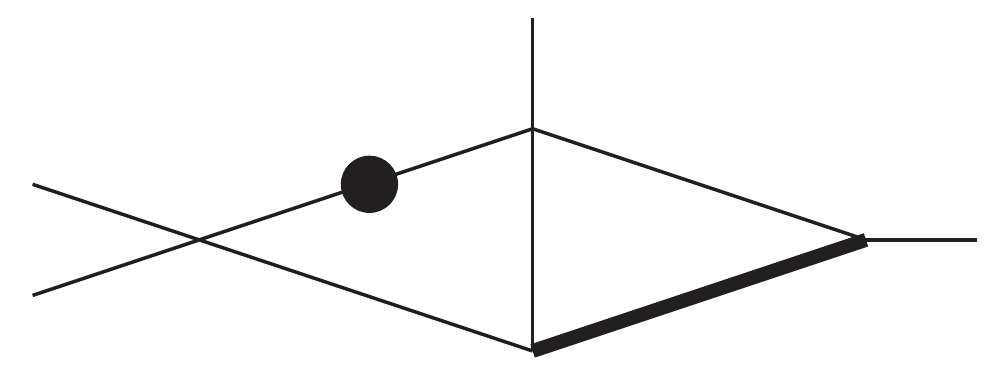}}(s)\quad
f_{36}^{\mathrm{C:214}} = \overset{(4-2\epsilon)}{\includegraphics[valign = m, raise = .2 cm, height = .135\linewidth, width = .135\linewidth,keepaspectratio]{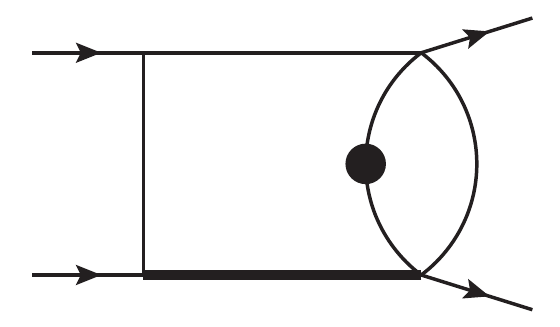}}(s,t)
\end{align*}

\begin{align*}
f_{37}^{\mathrm{C:341}} = \overset{(4-2\epsilon)}{\includegraphics[valign = m, raise = .2 cm, height = .135\linewidth, width = .135\linewidth,keepaspectratio]{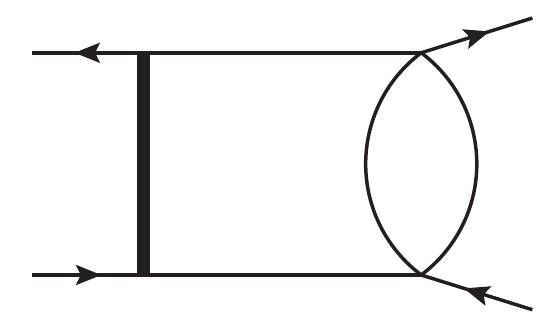}}(s,t)\quad
f_{38}^{\mathrm{C:341}} = \overset{(4-2\epsilon)}{\includegraphics[valign = m, raise = .2 cm, height = .135\linewidth, width = .135\linewidth,keepaspectratio]{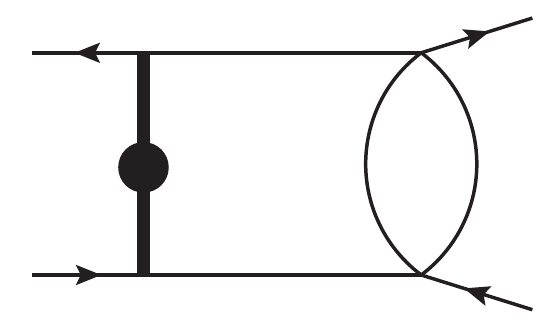}}(s,t)\quad
f_{39}^{\mathrm{C:213}} = \overset{(4-2\epsilon)}{\includegraphics[valign = m, raise = .2 cm, height = .135\linewidth, width = .135\linewidth,keepaspectratio]{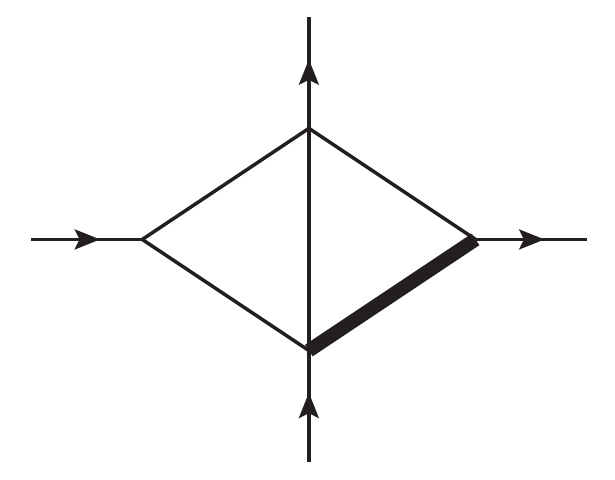}}(s,t)
\end{align*}

\begin{align*}
f_{40}^{\mathrm{C:213}} = \overset{(4-2\epsilon)}{\includegraphics[valign = m, raise = .2 cm, height = .135\linewidth, width = .135\linewidth,keepaspectratio]{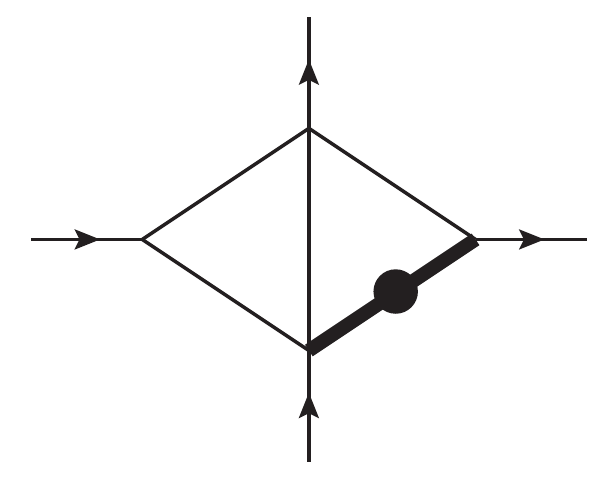}}(s,t)\quad
f_{41}^{\mathrm{C:374}} = \overset{(4-2\epsilon)}{\includegraphics[valign = m, raise = .45 cm, height = .15\linewidth, width = .15\linewidth,keepaspectratio]{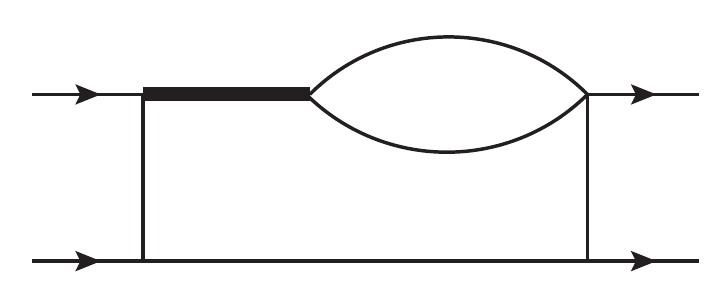}}(s,t)\quad
f_{42}^{\mathrm{C:246}} = \overset{(4-2\epsilon)}{\includegraphics[valign = m, raise = .45 cm, height = .15\linewidth, width = .15\linewidth,keepaspectratio]{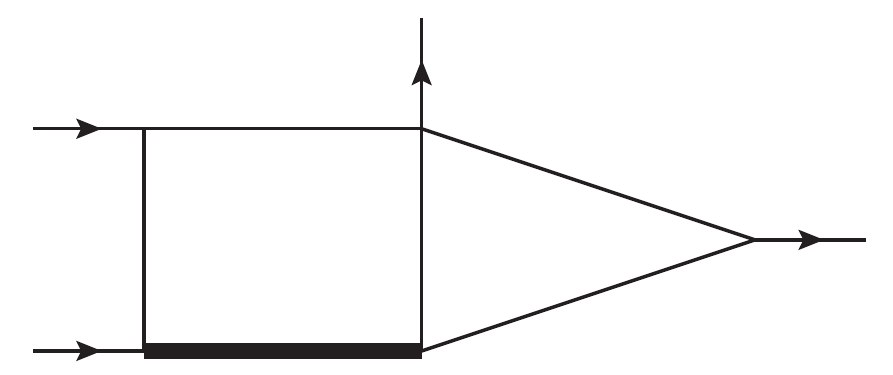}}(s,t)
\end{align*}

\begin{align*}
f_{43}^{\mathrm{C:245}} = \overset{(4-2\epsilon)}{\includegraphics[valign = m, raise = .45 cm, height = .15\linewidth, width = .15\linewidth,keepaspectratio]{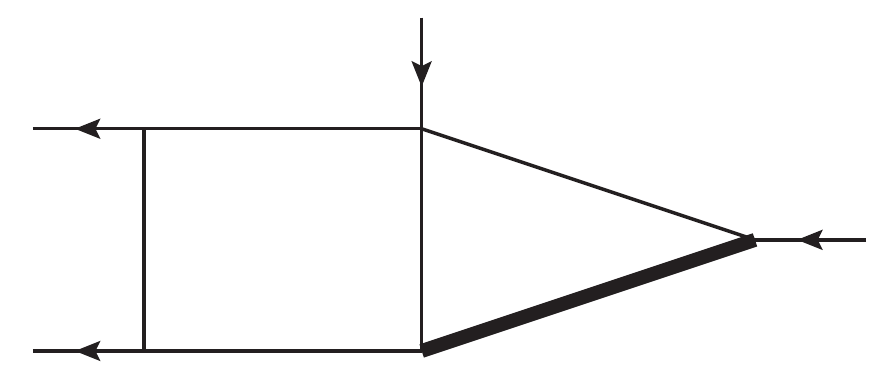}}(s,t)\quad\quad
f_{44}^{\mathrm{C:245}} = \overset{(4-2\epsilon)}{\includegraphics[valign = m, raise = .45 cm, height = .15\linewidth, width = .15\linewidth,keepaspectratio]{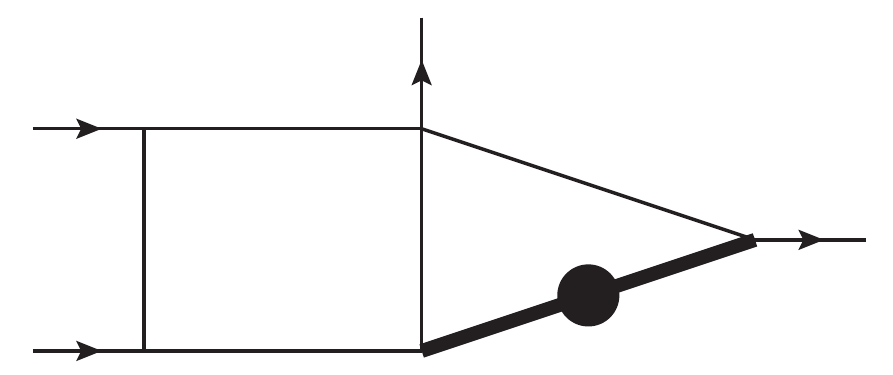}}(s,t)
\end{align*}

\begin{align*}
f_{45}^{\mathrm{C:247}} = \overset{(4-2\epsilon)}{\includegraphics[valign = m, raise = .22 cm, height = .15\linewidth, width = .15\linewidth,keepaspectratio]{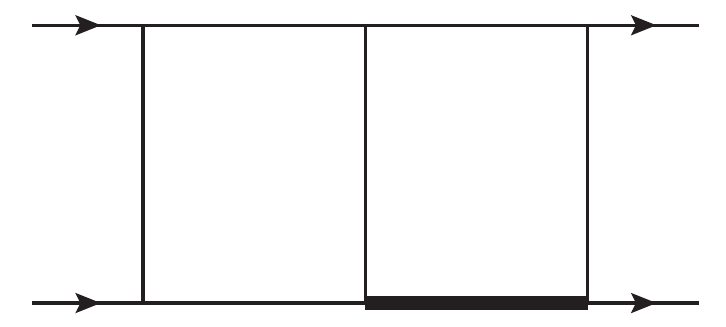}}(s,t)\quad
f_{46}^{\mathrm{C:247}} = \overset{(4-2\epsilon)}{\includegraphics[valign = m, raise = .22 cm, height = .15\linewidth, width = .15\linewidth,keepaspectratio]{f45andf46}}\left[(k_2 - p_3)^2\right](s,t)
\end{align*}

\begin{align*}
f_{47}^{\mathrm{D:341}} = \overset{(4-2\epsilon)}{\includegraphics[valign = m, raise = .3 cm, height = .175\linewidth, width = .175\linewidth,keepaspectratio]{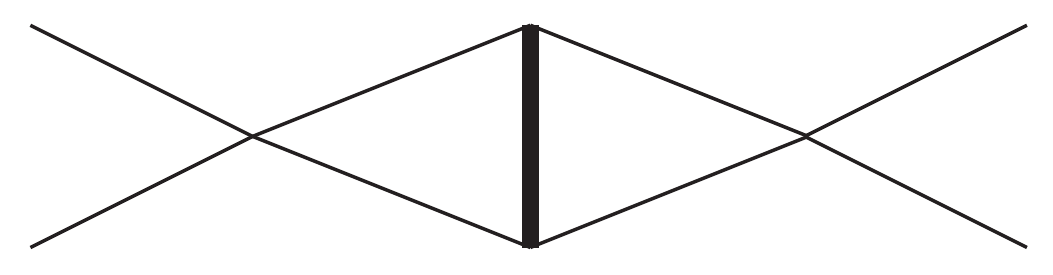}}(s)\quad\quad
f_{48}^{\mathrm{D:117}} = \overset{(4-2\epsilon)}{\includegraphics[valign = m, raise = .7 cm, height = .175\linewidth, width = .175\linewidth,keepaspectratio]{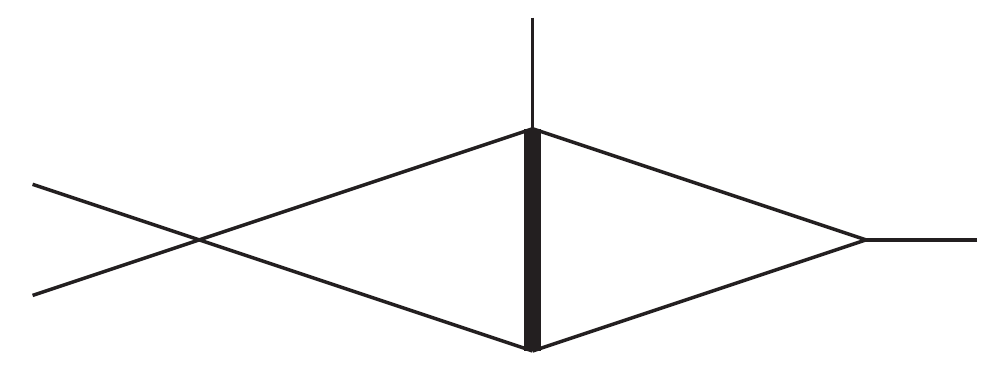}}(s)
\end{align*}

\begin{align}
\label{eq:reduzedef}
f_{49}^{\mathrm{D:125}} = \overset{(4-2\epsilon)}{\includegraphics[valign = m, raise = .22 cm, height = .15\linewidth, width = .15\linewidth,keepaspectratio]{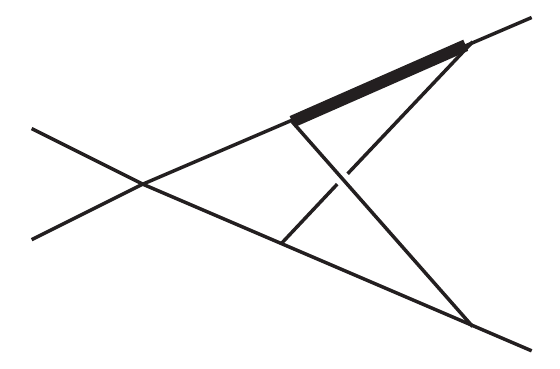}}\left[(k_1 - k_2 + p_1 + p_2)^2\right](s).
\end{align}

Building upon these definitions, we construct a convenient basis of integrals
for the method of differential equations
along the lines discussed in reference~\cite{Gehrmann:2014bfa}:\footnote{
We would like to point out further recent work on the analysis of master
integrals and the construction of a normal form basis
\cite{Lee:2013hzt,Argeri:2014qva,Lee:2014ioa,Tancredi:2015pta,Ablinger:2015tua,Ita:2015tya,Larsen:2015ped,
Primo:2016ebd,Meyer:2016slj,Georgoudis:2016wff,Prausa:2017ltv,Gituliar:2017vzm}.
}

\begin{align*}
m_{1}=\epsilon^2 s f_{1}^{\mathrm{A:38}}\quad
m_{2}=\epsilon^2 t f_{2}^{\mathrm{\thickbar{A}:38}}\quad
m_{3}=\epsilon^2 s^2 f_{3}^{\mathrm{A:99}}\quad
m_{4}=\epsilon^2 t^2 f_{4}^{\mathrm{\thickbar{A}:99}}\quad
m_{5}=\epsilon^3 s f_{5}^{\mathrm{A:53}}
\end{align*}
\begin{align*}
m_{6}=\epsilon^3 t f_{6}^{\mathrm{\thickbar{A}:53}}\quad
m_{7}=\epsilon^3 s t f_{7}^{\mathrm{A:174}}\quad
m_{8}=\epsilon^3 s t f_{8}^{\mathrm{\thickbar{A}:174}}\quad
m_{9}=\epsilon^4 (s + t) f_{9}^{\mathrm{A:182}}
\end{align*}
\begin{align*}
m_{10}=\epsilon^4 s^2 t f_{10}^{\mathrm{A:247}}\quad
m_{11}=\epsilon^4 s^2 f_{11}^{\mathrm{A:247}}\quad
m_{12}=\epsilon^4 s^2 f_{12}^{\mathrm{B:125}}\quad
m_{13}=\epsilon^2 s f_{1}^{\mathrm{A:38}}
\end{align*}
\begin{align*}
m_{14}=\epsilon^2 t f_{2}^{\mathrm{\thickbar{A}:38}}\quad
m_{15}=\epsilon^2 s f_{13}^{\mathrm{C:97}}\quad
m_{16}=\epsilon^2 t f_{14}^{\mathrm{\thickbar{C}:97}}\quad
m_{17}=\epsilon (1-\epsilon) m^2 f_{15}^{\mathrm{C:76}}
\end{align*}
\begin{align*}
m_{18}=\epsilon^2 s f_{16}^{\mathrm{C:69}}\quad
m_{19}=2 \epsilon^2 (s-m^2) f_{16}^{\mathrm{C:69}} + \epsilon^2 (s-m^2) f_{17}^{\mathrm{C:69}}\quad
m_{20}=\epsilon^2 s^2 f_{3}^{\mathrm{A:99}}
\end{align*}
\begin{align*}
m_{21}=\epsilon^2 t^2 f_{4}^{\mathrm{\thickbar{A}:99}}\quad
m_{22}=\epsilon^3 s f_{5}^{\mathrm{A:53}}\quad
m_{23}=\epsilon^3 t f_{6}^{\mathrm{\thickbar{A}:53}}\quad
m_{24}=\epsilon^2 s^2 f_{18}^{\mathrm{C:99}}
\end{align*}
\begin{align*}
m_{25}=\frac{\epsilon^2 s \Big(s(1-5\epsilon)-m^2(1-3\epsilon)\Big)}{4(1-3\epsilon)(s-m^2)}f_{1}^{\mathrm{A:38}}-\frac{\epsilon(1-\epsilon)m^4}{2(s-m^2)}f_{15}^{\mathrm{C:76}}+\frac{\epsilon^2 (1-2\epsilon)^2}{s-m^2}f_{19}^{\mathrm{C:102}}
\end{align*}
\begin{align*}
m_{26}=\epsilon^3 s f_{20}^{\mathrm{C:78}}\quad
m_{27}=\epsilon^3 t f_{21}^{\mathrm{C:212}}\quad
m_{28}=\epsilon^3 s f_{22}^{\mathrm{\thickbar{C}:212}}\quad
m_{29}=\epsilon^3 t f_{23}^{\mathrm{C:204}}\quad
m_{30}=\epsilon^3 t f_{24}^{\mathrm{C:204}}
\end{align*}
\begin{align*}
m_{31}=\epsilon^3 t f_{25}^{\mathrm{\thickbar{C}:204}}\quad
m_{32}=\epsilon^3 t f_{26}^{\mathrm{\thickbar{C}:204}}\quad
m_{33}=\epsilon^3 s t f_{7}^{\mathrm{A:174}}\quad
m_{34}=\epsilon^3 s t f_{8}^{\mathrm{\thickbar{A}:174}}
\end{align*}
\begin{align*}
m_{35}=\epsilon^4 (s+t) f_{9}^{\mathrm{A:182}}\quad
m_{36}=\epsilon^3 t^2 f_{27}^{\mathrm{C:472}}\quad
m_{37}=\epsilon^3 m^2 t f_{28}^{\mathrm{C:372}}\quad
m_{38}=\epsilon^3 m^2 t f_{29}^{\mathrm{C:244}}
\end{align*}
\begin{align*}
m_{39}=\epsilon^3 m^2 s f_{30}^{\mathrm{\thickbar{C}:244}}\quad
m_{40}=\epsilon^4 s f_{31}^{\mathrm{C:110}}\quad
m_{41}=\epsilon^3 m^2 t f_{32}^{\mathrm{C:220}}\quad
m_{42}=\epsilon^3 m^2 s f_{33}^{\mathrm{\thickbar{C}:220}}
\end{align*}
\begin{align*}
m_{43}=\epsilon^3 m^2 s f_{34}^{\mathrm{C:117}}\quad
m_{44}=\epsilon^3 s^2 f_{35}^{\mathrm{C:117}}\quad
m_{45}=\epsilon^3 (s-m^2) t f_{36}^{\mathrm{C:214}}\quad
m_{46}=\epsilon^3 (1 - 2\epsilon) t f_{37}^{\mathrm{C:341}}
\end{align*}
\begin{align*}
m_{47}=3\epsilon^3 (1 - 2\epsilon) m^2 f_{37}^{\mathrm{C:341}} + \epsilon^2 (1 - 2\epsilon) m^2 (m^2 + t) f_{38}^{\mathrm{C:341}}\quad
m_{48}=\epsilon^4 (s + t) f_{39}^{\mathrm{C:213}}\quad
\end{align*}
\begin{align*}
m_{49}=\epsilon^3 m^2 (s + t) f_{40}^{\mathrm{C:213}}\quad
m_{50}=\epsilon^3(1-2\epsilon)(s-m^2) t f_{41}^{\mathrm{C:374}}\quad
m_{51}=\epsilon^4 (s t -m^2 (s + t)) f_{42}^{\mathrm{C:246}}
\end{align*}
\begin{align*}
m_{52}=\epsilon^4 s t f_{43}^{\mathrm{C:245}}\quad
m_{53}=2\epsilon^4 m^2 s f_{43}^{\mathrm{C:245}} + \epsilon^3 m^2 s (m^2 + t) f_{44}^{\mathrm{C:245}}\quad
m_{54}=\epsilon^4 s (s - m^2) t f_{45}^{\mathrm{C:247}}
\end{align*}
\begin{align}
\label{eq:normformdef}
m_{55}=\epsilon^4 s^2 f_{46}^{\mathrm{C:247}}\quad
m_{56}=\epsilon^3(1-2\epsilon) s f_{47}^{\mathrm{D:341}}\quad
m_{57}=\epsilon^4 s f_{48}^{\mathrm{D:117}}\quad
m_{58}=\epsilon^4 s f_{49}^{\mathrm{D:125}}.
\end{align}
Note that these definitions repeat nine massless integrals to allow for an independent treatment of the
differential equations for the massless integrals, $m_1,\ldots,m_{12}$,
and the one-mass integrals, $m_{13},\ldots,m_{58}$.
We employ the integration measure
\begin{equation}
\label{eq:norm2}
\left(\frac{\Gamma(1-\epsilon) s^{\epsilon}}{i \pi^{2-\epsilon}}\right)^2 \int {\rm d}^{4-2\epsilon} k_1 \int {\rm d}^{4-2\epsilon} k_2
\end{equation}
with $\epsilon=(4-d)/2$, which renders our integrals $m_i$ functions
of the dimensionless variables
\begin{equation}
x = \frac{-t}{-s}\qquad {\rm and} \qquad y = \frac{m^2}{-s}
\end{equation}
only.

In this basis, we obtain a system of differential equations in normal form~\cite{Kotikov:2010gf,Henn:2013pwa,Henn:2013tua}
for the vector $\vec{m}=(m_i)$,
\begin{equation}\label{deq}
\ud \vec{m}(\epsilon,x,y) = \epsilon \sum_k \ud\ln\big( l_k(x,y)\big)\,A^{(k)} \vec{m}(\epsilon,x,y)
\end{equation}
with the letters $l_k(x,y)$ of the symbol alphabet
\begin{equation}\label{letters}
\{ l_1, \ldots, l_7 \} = \left\{ x, 1 + x, y, 1 + y, 1 - y, x - y, x + y + x y\right\}
\end{equation}
and matrices $A^{(k)}$ of rational numbers.

Expanding the masters integrals $m_i$ about $\epsilon=0$, the differential equations
fully decouple and can be integrated order-by-order in $\epsilon$ in terms of multiple polylogarithms.
The alphabet is linear in both variables, which implies that the integration path can be
chosen along the $x$ and $y$ axis, resulting in an expression written in terms of Goncharov $G$
functions with either $x$ or $y$ in the final argument.
This representation can introduce spurious singularities and therefore may not be optimal for
numerical evaluations~\cite{Gehrmann:2015ora}.

Instead, one can employ $\Li$ functions of fewer but possibly more involved arguments.
Recently, it has been demonstrated explicitly that $\ln$, $\Li_2$, $\Li_3$, $\Li_4$, and $\Li_{2,2}$ functions are sufficient to reduce a general $G$ function
of at most weight four~\cite{Frellesvig:2016ske}. In order to arrive at real-valued functions with a convergent power series representation in a specific region of phase space,
it can be useful to allow for $\Li_{2,1}$ functions as well. A representation written in terms of such functions can be obtained either by recasting
a solution written in terms of $G$ functions or by directly integrating the differential equations via an ansatz built out of the required functions.
We perform these calculations using in-house routines based on the symbol and coproduct calculus~\cite{Goncharov:2005sla,Brown:2011ik,Duhr:2011zq,Duhr:2012fh}.

Note that, for phenomenological applications, we are interested in the solutions of our integrals through to weight four.
In the next section, however, we consider alternative bases of master integrals for the purpose of numerical evaluation which partially involve weight five functions at the $\epsilon$ order required
for physics applications. In order to validate our numerical analysis of the solution, we therefore find it useful to obtain results through to weight five. 

We construct an ansatz for our solutions in terms of the functions
\begin{equation}
\ln,\quad \Li_2,\quad \Li_3,\quad \Li_{2,1},\quad \Li_4,\quad \Li_{2,2},\quad \Li_5,\quad \Li_{3,2},\quad \Li_{2,2,1}
\end{equation}
using the Duhr-Gangl-Rhodes algorithm \cite{Duhr:2011zq}.
We require the symbol of each function to not introduce letters beyond
those already present in our differential equation~\eqref{deq}
and we construct suitable power products of our letters~\eqref{letters}
accordingly.
For the arguments of the logarithms $\ln$ we choose
\[
\{ -l_1, l_2, -l_3, l_4, l_5, -l_6, -l_7 \}.
\]
As admissible arguments for the classical polylogarithms $\Li_n$, we obtain the following sixty-six power products of letters
\begin{align}
\bigg\{&
-l_1,l_2,-l_3,l_3,l_4,l_5,-l_7,-\frac{1}{l_7},\frac{1}{l_5},\frac{1}{l_4},-\frac{1}{l_3},\frac{1}{
   l_3},\frac{1}{l_2},-\frac{1}{l_1},\frac{l_1}{l_7},\frac{l_1}{l_6},\frac{l_1}{l_3},\frac{l_1}{l_2},l_2
   l_4,\frac{l_2}{l_6},\frac{l_2}{l_4},l_3^2,\frac{l_3}{l_7},
\nonumber\\ &
-\frac{l_3}{l_6},-\frac{l_3}{l_5},\frac{l_3}
   {l_4},l_4 l_5,-\frac{l_4}{l_6},-\frac{l_6}{l_4},\frac{1}{l_4
   l_5},\frac{l_4}{l_3},-\frac{l_5}{l_3},-\frac{l_6}{l_3},\frac{l_7}{l_3},\frac{1}{l_3^2},\frac{l_4}{l_2}
   ,\frac{l_6}{l_2},\frac{1}{l_2
   l_4},\frac{l_2}{l_1},\frac{l_3}{l_1},\frac{l_6}{l_1},\frac{l_7}{l_1},\frac{l_1 l_4}{l_7},
\nonumber\\ &
   \frac{l_1
   l_4}{l_6},-\frac{l_1 l_4}{l_3},-\frac{l_1}{l_2 l_3},\frac{l_2 l_3}{l_7},-\frac{l_2 l_3}{l_6},\frac{l_2
   l_4}{l_7},\frac{l_7}{l_2 l_4},-\frac{l_6}{l_2 l_3},\frac{l_7}{l_2 l_3},-\frac{l_2
   l_3}{l_1},-\frac{l_3}{l_1 l_4},\frac{l_6}{l_1 l_4},\frac{l_7}{l_1 l_4},\frac{l_1 l_4}{l_2
   l_3},-\frac{l_3^2}{l_4 l_5},
\nonumber\\ &
   -\frac{l_4 l_5}{l_3^2},\frac{l_2 l_3}{l_1 l_4},\frac{l_1^2 l_4}{l_6
   l_7},-\frac{l_2 l_3^2}{l_6 l_7},-\frac{l_6 l_7}{l_2 l_3^2},\frac{l_6 l_7}{l_1^2 l_4},\frac{l_1^2
   l_4}{l_2 l_3^2},\frac{l_2 l_3^2}{l_1^2 l_4}
   \bigg\}
\end{align}
which have the property that not only the argument $z$ but also $1-z$ factorize
over the original symbol alphabet.
Extending this set by the argument $1$, forming pairs and selecting those pairs
$(z_1,z_2)$, for which $1-z_1 z_2$ factorizes over the alphabet, gives us
618 admissible arguments for $\Li_{2,1}$, $\Li_{2,2}$, and $\Li_{3,2}$ functions which we
do not list here for the sake of brevity.
Similarly, we find 3342 admissible triples $(z_1,z_2,z_3)$ as arguments
for our $\Li_{2,2,1}$ functions.
While we have no proof that $\Li_{2,2,1}$ functions are strictly required, we
were successful in constructing the genuine weight five part of the solution
only when including them in addition to $\Li_{3,2}$ and $\Li_5$ functions.
This observation is independent of further constraints to be described below,
which motivate the inclusion of $\Li_{2,1}$ functions.

Not all of these functions are needed and we formulate further objectives
for our functional basis.
We prefer functions which are real-valued for physical kinematics
\begin{equation}
-1 < x < 0, \quad -\infty < y < 0\,.
\end{equation}
This requirement is the reason why we include $\Li_{2,1}$ functions in our
basis.
In addition to requiring real-valuedness, we prefer functions
which possess a convergent
power series representation. 
Insisting upon such a representation automatically avoids additional manipulations
which might otherwise be necessary to carry out numerical evaluations.

Indeed, we find that it is possible to use functions which are real-valued
over the entire physical region of the phase space with a single important exception.
The letter $l_4=1+y$ changes sign at $s=m^2$ and the functional form of our results will change in this region if we insist upon separating real and imaginary parts explicitly. 
This, however, is to be expected because there is a physical singularity at the point $s = m^2$, which would be regulated by the width of the massive vector boson inside the loop.
Our solution, in fact, contains just a single function which is sensitive to
the cut starting at $y=-1$:
\begin{alignat}{3}
&\ln(l_4) &\qquad&\text{for~}y>-1\\
&\ln(-l_4) + i \pi &&\text{for~}y<-1.
\end{alignat}
In contrast, no logarithms of the letter $l_6 = x - y$ appear in our results and all other elements of our functional basis which depend of $l_6$ are real-valued for both $x > y$ and $x < y$.

Employing our functional basis, we construct an ansatz for the solution
and match it against the differential equations~\eqref{deq}.
We employ regularity conditions and require real-valuedness in the Euclidean
domain to fix the integration constants.
In addition, we calculated $f_{1}^{\mathrm{A:38}}$, $f_{3}^{\mathrm{A:99}}$, $f_{5}^{\mathrm{A:53}}$,
$f_{13}^{\mathrm{C:97}}$, $f_{15}^{\mathrm{C:76}}$, $f_{16}^{\mathrm{C:69}}$, $f_{18}^{\mathrm{C:99}}$, $f_{27}^{\mathrm{C:472}}$, and $f_{29}^{\mathrm{C:244}}$ explicitly from Feynman parameters to all orders in $\epsilon$.
Most of these integrals are completely straightforward to evaluate. Some, such as
\begin{align}
f_{29}^{\mathrm{C:244}} &= \frac{s^{2 \epsilon }\left(m^2\right)^{-2-2 \epsilon} \Gamma(1-2 \epsilon) \Gamma^4(1-\epsilon) \Gamma(2+\epsilon) \Gamma(1+2 \epsilon) }{2\, \epsilon^3 \Gamma(2-\epsilon) \Gamma(-2 \epsilon)} 
{}_3 F_2\left(1,1,2+\epsilon;2,2-\epsilon;\frac{-t}{m^2}\right) \nonumber \\
& - \frac{s^{2 \epsilon } (-t)^{-1-2 \epsilon } \Gamma^5(1-\epsilon) \Gamma(1+2\epsilon)}{2\, m^2 \epsilon^3 \Gamma(1-3\epsilon)}
{}_3 F_2\left(1,-2\epsilon,1-\epsilon;1-2\epsilon,1-3\epsilon;\frac{-t}{m^2}\right)
\end{align}
turn out to be a bit more non-trivial.

As an additional sanity check, we found it convenient in some cases to look at asymptotic expansions of integrals in particular limits in order to explicitly confirm their scaling behavior. 
For this purpose, we found the {\tt Mathematica} package {\tt asy.m} \cite{Jantzen:2012mw} to be extremely useful. For the convenience of the reader, we provide our analytical solutions through to weight four
in the ancillary files \verb$sol-phys-AB.m$ and \verb$sol-phys-CD.m$ on {\tt arXiv.org}.

\section{Numerical Analysis}
\label{sec:mixedDY}

In an effort to check an analytical calculation such as the one presented
in the previous section, it is desirable to numerically evaluate the loop
integrals with an independent method.
In other cases, a numerical evaluation might even be the method of
choice, {\it e.g.}\ if it is not clear how to evaluate the integrals analytically.
A standard method for numerical evaluation is sector decomposition
and one might think that it is straightforward to numerically
check the integrals $f_i$ in \eqref{eq:reduzedef} using this approach.
Unfortunately, this is not the case.
Specifically, for double box integrals with additional numerators or dotted
propagators we often encountered cases where none of the available software packages
is able to provide reliable results.
In some cases, issues related to the presence of problematic singularity structures in the integrand can be tracked down and,
with dedicated effort, cured~\cite{Borowka:2013cma}.
For what concerns the two-loop mixed EW-QCD topologies considered in this paper,
the most challenging cases for the code are seven-line integrals which have $\epsilon^{-4}$
poles and doubled propagator denominators.

As discussed in earlier work by the authors and Erik Panzer
\cite{vonManteuffel:2014qoa,vonManteuffel:2015gxa}, it is possible to employ a
basis of finite Feynman integrals defined by allowing for higher-dimensional integrals with potentially
higher powers of the propagator denominators to extract {\it all} poles in $\epsilon$ analytically before performing any numerical integrations.\footnote{Let us remind the reader
that, due to the principle of analytical continuation, the existence of an on-shell
Euclidean region for all integrals guarantees that our procedure applies equally well to all integrals when physical kinematics is considered.}
The change of basis is performed with integration by parts reductions, whose
computational complexity is non-negligible but still lower than that of the reductions
for the amplitude. This is true both for the processes discussed in this work and for all other phenomenologically relevant problems which we have studied.

The finite integrals could, in principle, be evaluated by direct
numerical integration.
However, their integrands may still contain structures which
spoil the convergence of the numerical integrations, such as integrable but
large variations near the boundaries.
In our experiments with a limited number of Euclidean integrals,
we find that an additional sector decomposition step improves the
performance of the evaluation.
Furthermore, handling branch cuts requires a dedicated treatment for physical
kinematics.
We therefore find it convenient to employ existing sector decomposition
programs for the numerical integration of our finite integrals.

We employ the sector decomposition program {\tt SecDec\;3} to evaluate our
two-loop topologies using either conventional or finite master integrals.
As is apparent from Table \ref{tab:mixedDYnums}, working with a basis of
finite integrals improves the numerical performance tremendously.
Let us now explain in more detail how we used the code to produce Table \ref{tab:mixedDYnums}.
Although we made a good effort to use the program in an appropriate way, it is certainly 
possible that other researchers might manage to produce better results in less time. 
However, we at least took care to use identical program settings and hardware while carrying out our numerical studies of the conventional and finite integral bases.\footnote{We followed the
recommendations given in \cite{Borowka:2015mxa} as much as possible, using, for example, their new sector decomposition strategy {\tt G2} and {\tt Mathematica 9.0.1} for all of our calculations.
Using this version of {\tt Mathematica} avoided some parallelization issues
encountered with versions 8 or 10 in our experiments.}

Altogether, we ran for roughly sixty days on the four cores of a i7-4940MX processor. For our finite integral evaluations, just over an hour was required; the excessive run time is due to the fact that, in spite of allowing
the {\tt VEGAS} routine \cite{Lepage:1977sw} provided by {\tt CUBA} \cite{Hahn:2004fe} to sample up to $5 \times 10^8$ points, 
numerous conventional basis integrals failed to achieve the very modest goal
of $\mathtt{epsrel} = 1 \times 10^{-2}$ and $\mathtt{epsabs} = 1 \times 10^{-4}$ at the physical phase space point $p = \{s \rightarrow 17, t \rightarrow -7, m^2 \rightarrow 6241/1681\}$.
Analytical expressions for the finite basis integrals were obtained by using the explicit connection between the finite integral basis and the normal form integral basis. We found
good agreement in all cases. At worst, we recorded a fractional difference of
$6 \times 10^{-3}$
at weight four between our exact results evaluated at $p$ and the numerical results output by the program. The average relative accuracy 
at weight four was found to be
$9 \times 10^{-4}$,
quite acceptable given our modest accuracy goal.

Due to the fact that there is a substantial loss of precision in rotating from the finite integral basis to the normal form integral basis, 
we found that our initial $\mathtt{epsrel} = 1 \times 10^{-2}$ and $\mathtt{epsabs} = 1 \times 10^{-4}$
run of {\tt SecDec\;3} did not allow us to directly check the normal form basis to similar part per mille precision.
To achieve this, we had to carry out a second, higher precision run of the program with $\mathtt{epsrel} = 1 \times 10^{-4}$ and $\mathtt{epsabs} = 1 \times 10^{-8}$, which took roughly eleven days of run time.
Let us stress again that it would be hopeless to attempt this exercise using the conventional integral basis.

\begin{center}
\begin{longtable}[h!]{|c|c|c||c|c|c|}
\hline
$\figgraph{.2}{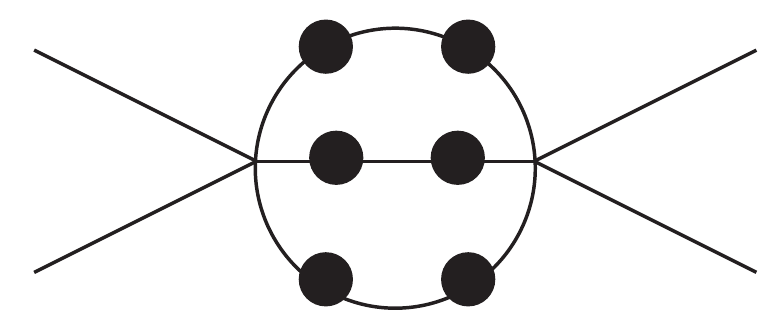}{8}(s)$     & $4~\mathrm{s}$     & $3.17 \times 10^{-4}$     &$\figgraph{.2}{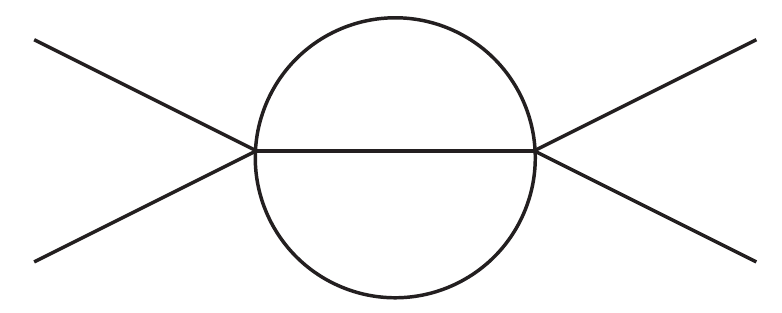}{4}(s)$      & $4~\mathrm{s}$     & $3.42 \times 10^{-3}$\\
\hline
$\figgraph{.2}{finf1andf2}{8}(t)$      & $4~\mathrm{s}$     & $8.39 \times 10^{-7}$     &$\figgraph{.2}{cornerf1andf2}{4}(t)$      & $4~\mathrm{s}$     & $1.40 \times 10^{-6}$\\
\hline
$\figgraph{.2}{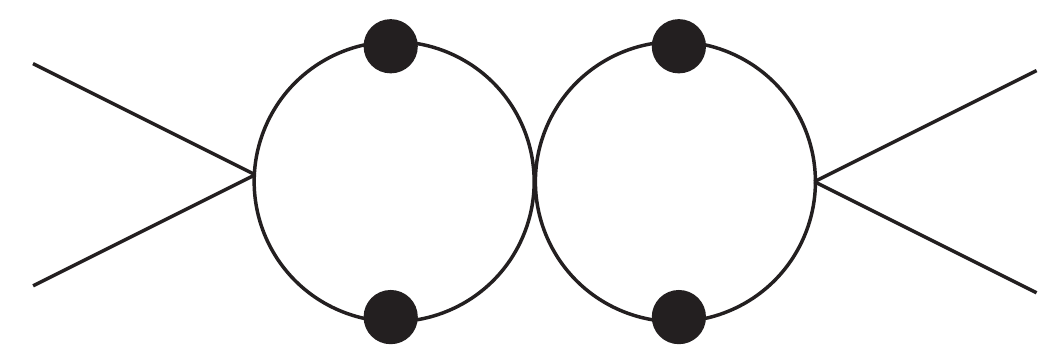}{6}(s)$      & $11~\mathrm{s}$     & $3.77 \times 10^{-4}$     &$\figgraph{.2}{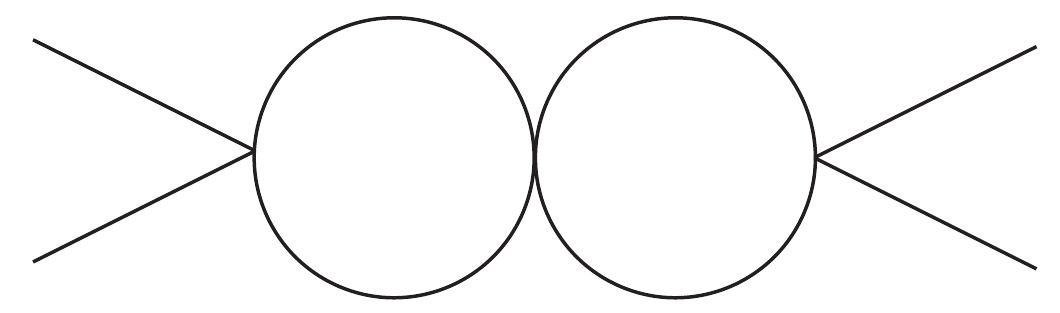}{4}(s)$      & $48~\mathrm{s}$     & $2.11\times 10^{-3}$\\
\hline
$\figgraph{.2}{finf3andf4}{6}(t)$     & $4~\mathrm{s}$     & $1.33 \times 10^{-5}$     &$\figgraph{.2}{cornerf3andf4}{4}(t)$      & $12~\mathrm{s}$     & $4.98 \times 10^{-5}$\\
\hline
$\figgraph{.2}{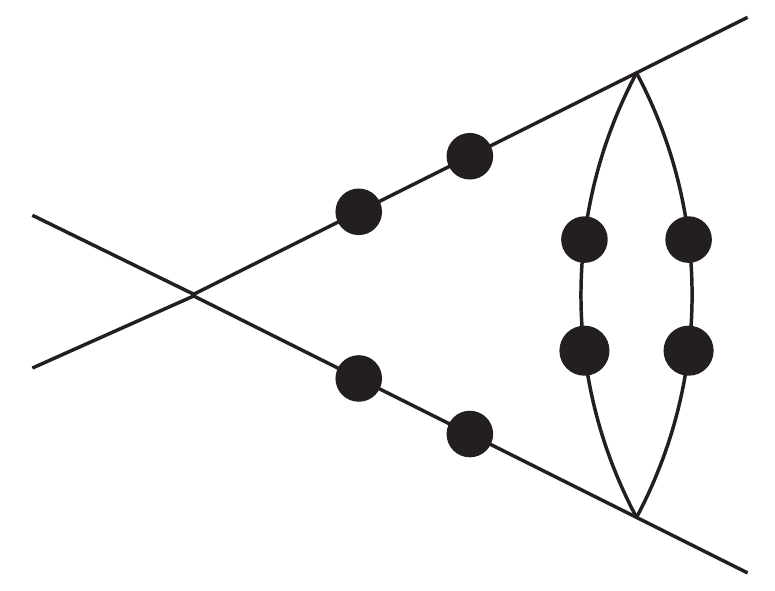}{10}(s)$      & $5~\mathrm{s}$     & $1.40 \times 10^{-3}$     &$\figgraph{.2}{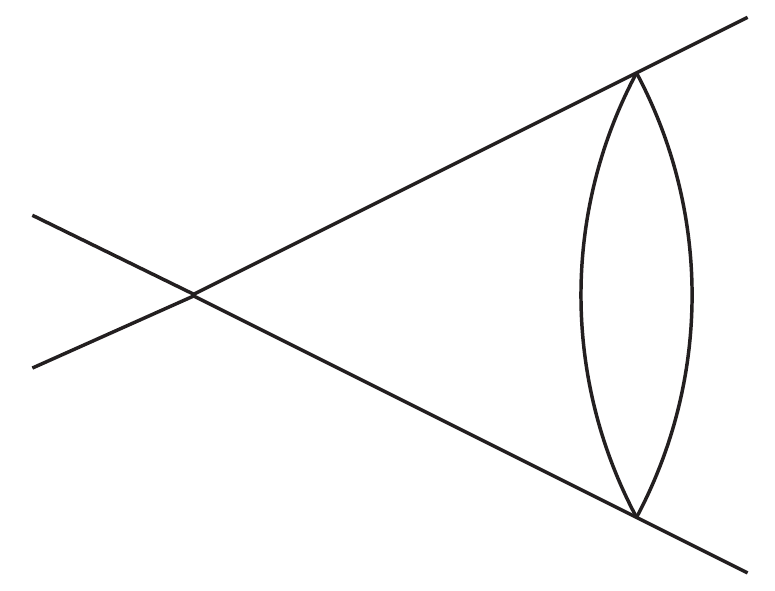}{4}(s)$      & $21~\mathrm{s}$     & $3.31 \times 10^{-3}$\\
\hline
$\figgraph{.2}{finf5andf6}{10}(t)$     & $4~\mathrm{s}$     & $7.20 \times 10^{-5}$     &$\figgraph{.2}{cornerf5andf6}{4}(t)$     & $5~\mathrm{s}$     & $1.11 \times 10^{-5}$\\
\hline
$\figgraph{.2}{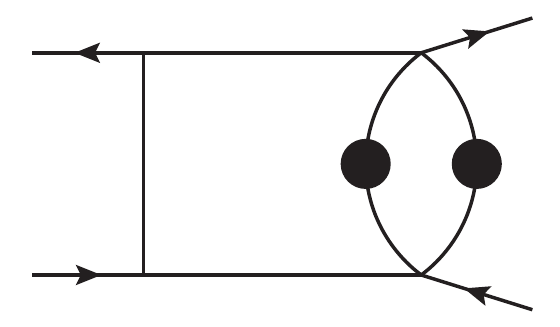}{6}(s,t)$    & $13~\mathrm{s}$     & $5.76 \times 10^{-3}$     &$\figgraph{.2}{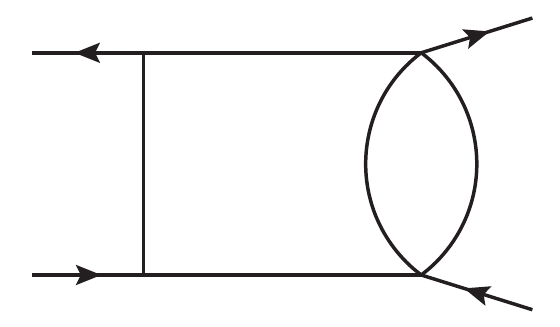}{4}(s,t)$     & $48~\mathrm{s}$     & $4.56 \times 10^{-3}$\\
\hline
$\figgraph{.2}{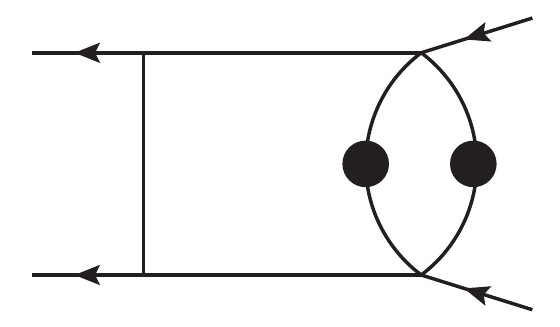}{6}(t,s)$      & $31~\mathrm{s}$     & $3.31 \times 10^{-3}$     &$\figgraph{.2}{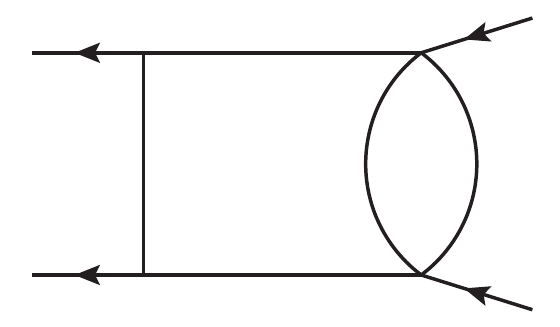}{4}(t,s)$      & $37~\mathrm{s}$     & $2.74 \times 10^{-3}$\\
\hline
$\figgraph{.2}{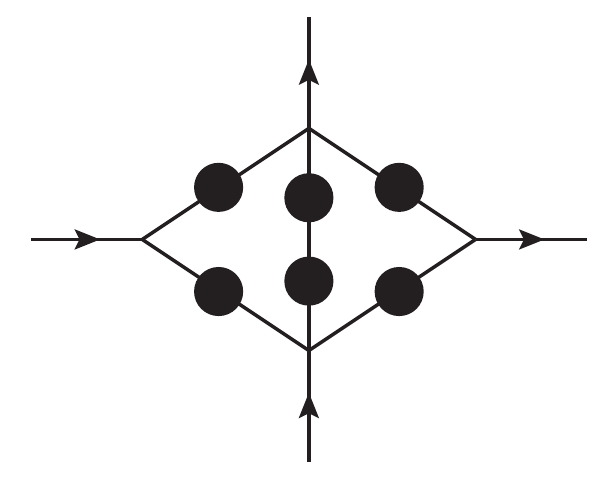}{10}(s,t)$    & $8~\mathrm{s}$     & $6.70 \times 10^{-4}$     &$\figgraph{.2}{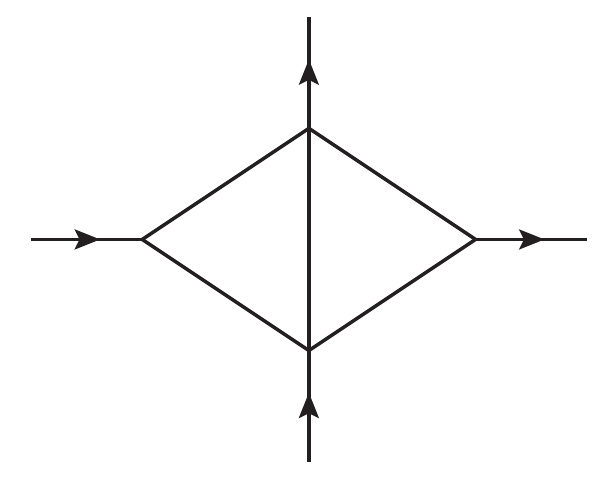}{4}(s,t)$     & $8~\mathrm{s}$     & $1.27 \times 10^{-3}$\\
\hline
$\figgraph{.2}{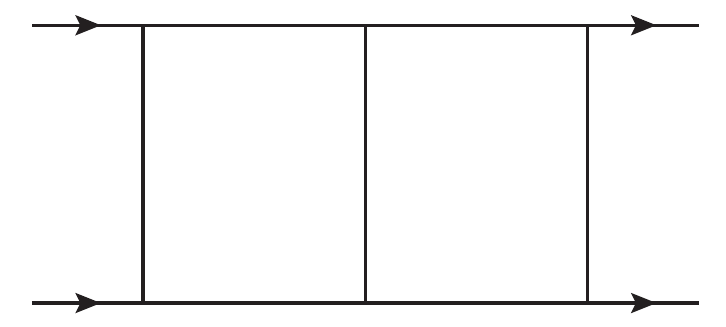}{6}(s,t)$      & $212~\mathrm{s}$     & $3.44 \times 10^{-3}$     &$\figgraph{.2}{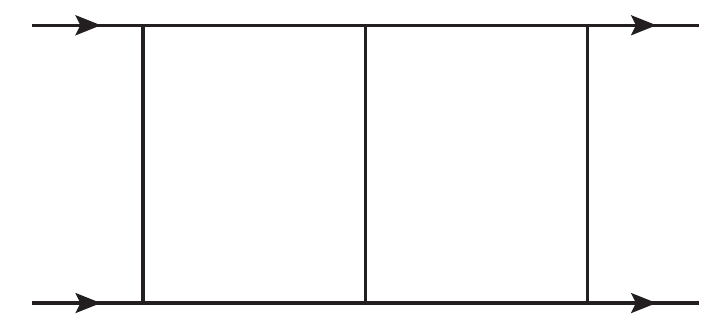}{4}(s,t)$      & $17503~\mathrm{s}$     & $3.59 \times 10^{-3}$\\
\hline
$\figgraph{.2}{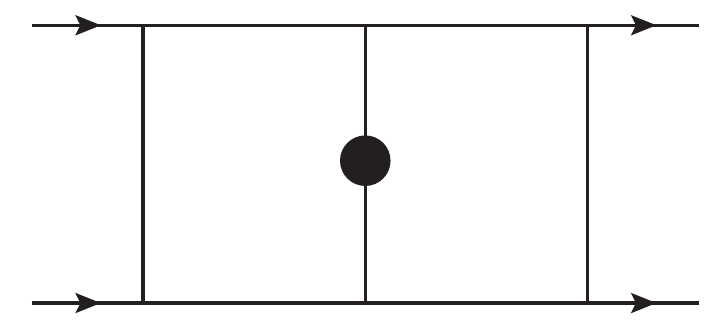}{6}(s,t)$      & $197~\mathrm{s}$     & $1.84 \times 10^{-4}$     &$\figgraph{.2}{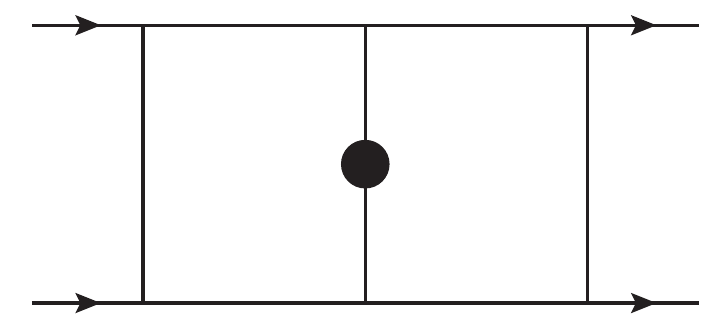}{4}(s,t)$      & $1337751~\mathrm{s}$     & $.263$\\
\hline
$\figgraph{.2}{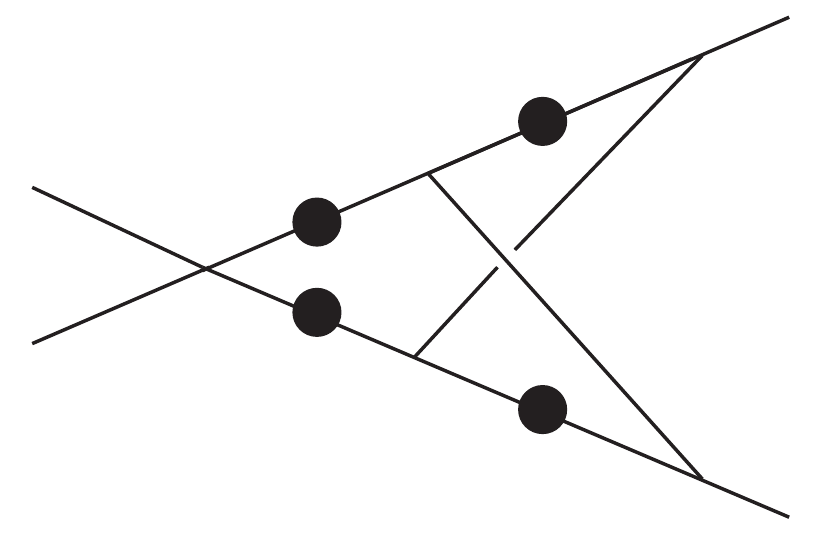}{8}(s)$     & $8~\mathrm{s}$     & $4.18 \times 10^{-3}$     &$\figgraph{.2}{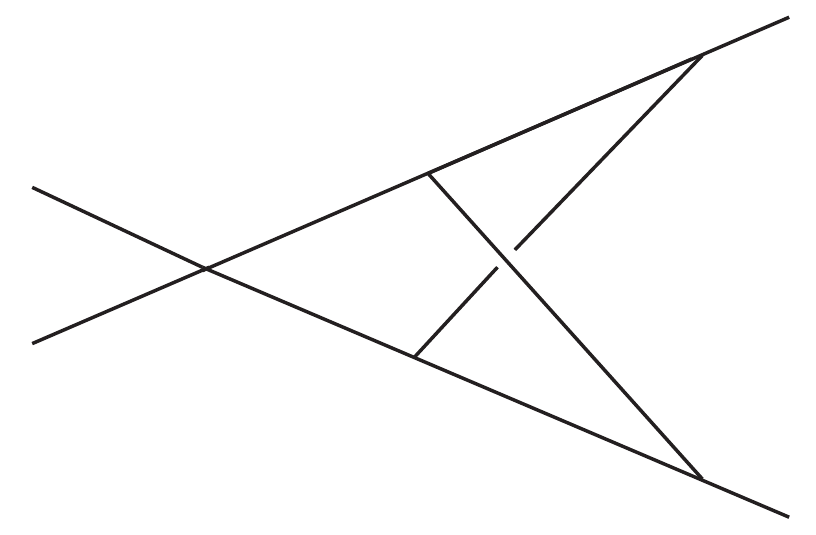}{4}(s)$     & $1261~\mathrm{s}$     & $1.63 \times 10^{-4}$\\
\hline
$\figgraph{.2}{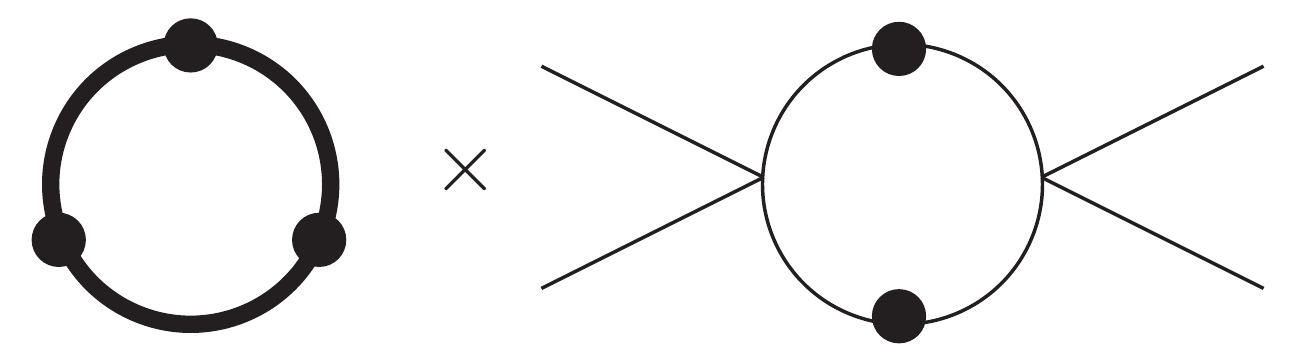}{6}(s)$     & $6~\mathrm{s}$     & $7.71 \times 10^{-4}$     &$\figgraph{.2}{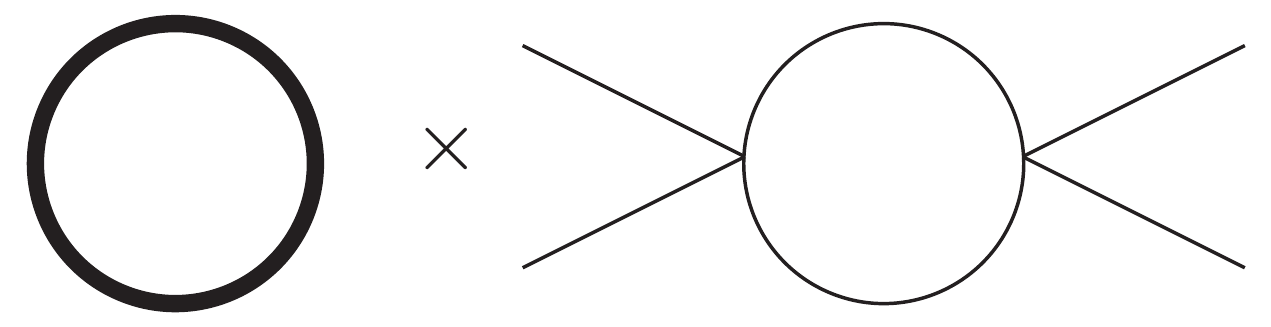}{4}(s)$     & $19~\mathrm{s}$     & $6.84 \times 10^{-3}$\\
\hline
$\figgraph{.2}{finf13andf14}{6}(t)$     & $3~\mathrm{s}$     & $1.03 \times 10^{-5}$     &$\figgraph{.2}{cornerf13andf14}{4}(t)$     & $5~\mathrm{s}$     & $4.60 \times 10^{-5}$\\
\hline
$\figgraph{.2}{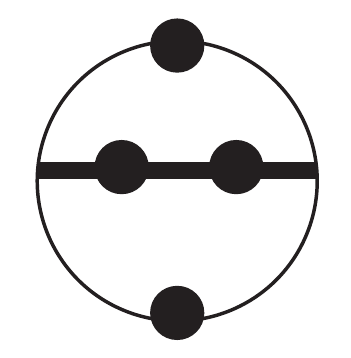}{6}$      & $4~\mathrm{s}$     & $4.60 \times 10^{-6}$     &$\figgraph{.2}{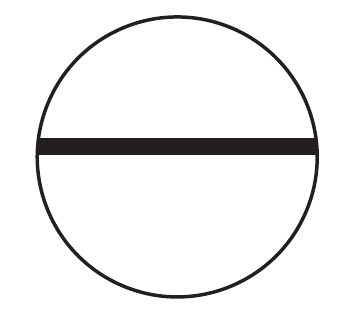}{4}$      & $5~\mathrm{s}$     & $5.39 \times 10^{-5}$\\
\hline
$\figgraph{.2}{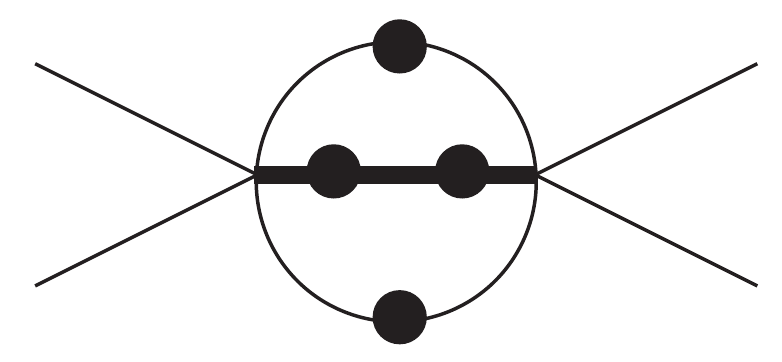}{6}(s)$     & $7~\mathrm{s}$     & $5.06 \times 10^{-4}$     &$\figgraph{.2}{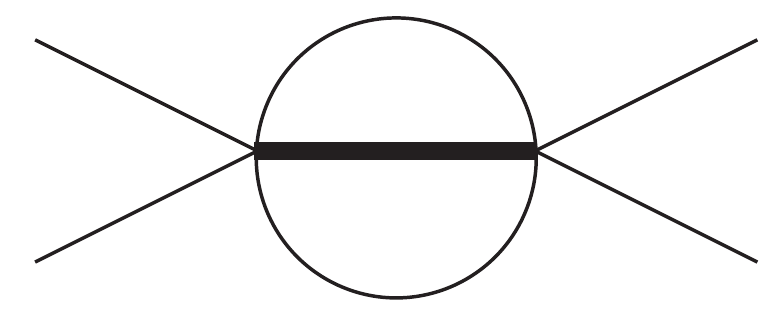}{4}(s)$     & $370~\mathrm{s}$     & $2.20 \times 10^{-4}$\\
\hline
$\figgraph{.2}{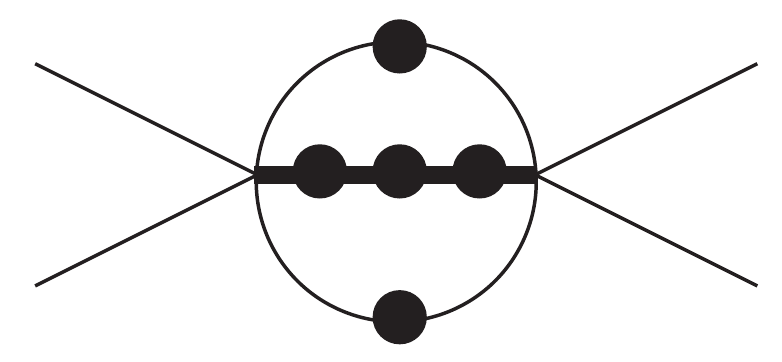}{6}(s)$     & $9~\mathrm{s}$     & $1.89 \times 10^{-4}$     &$\figgraph{.2}{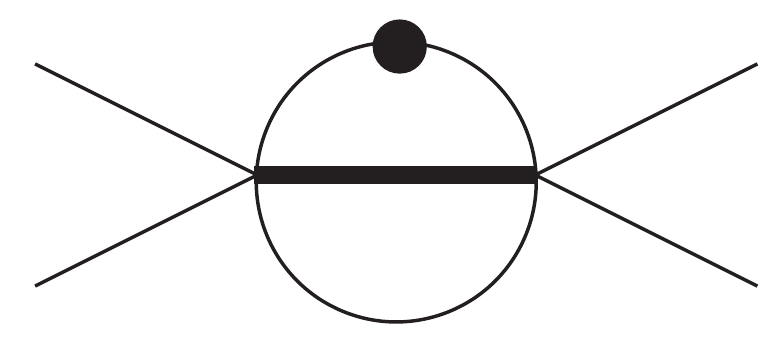}{4}(s)$     & $143~\mathrm{s}$     & $9.62 \times 10^{-3}$\\
\hline
$\figgraph{.2}{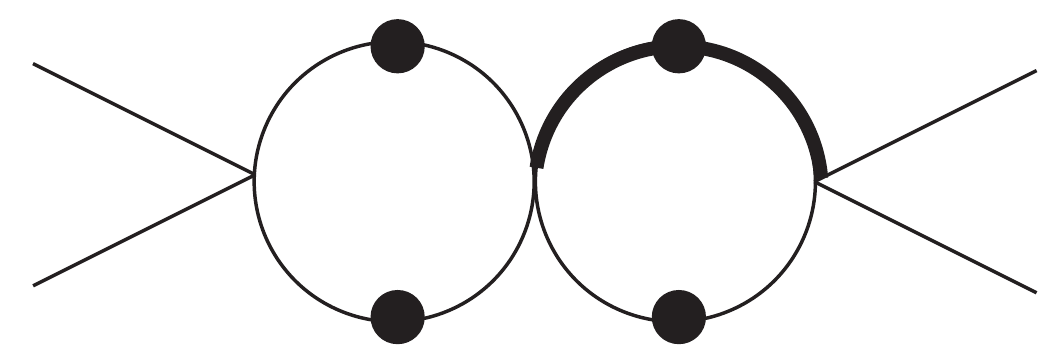}{6}(s)$     & $10~\mathrm{s}$     & $2.85 \times 10^{-4}$     &$\figgraph{.2}{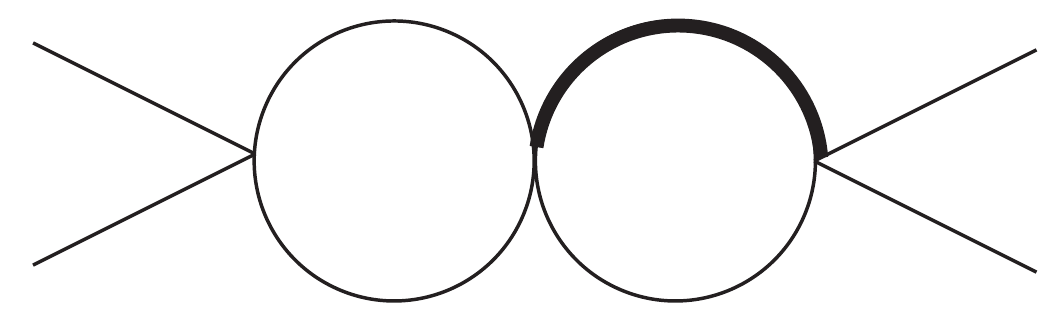}{4}(s)$     & $82~\mathrm{s}$     & $1.76 \times 10^{-3}$\\
\hline
$\figgraph{.2}{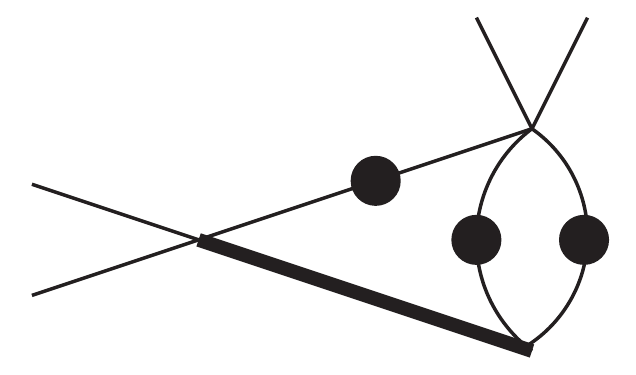}{6}(s)$     & $46~\mathrm{s}$     & $1.21 \times 10^{-4}$     &$\figgraph{.2}{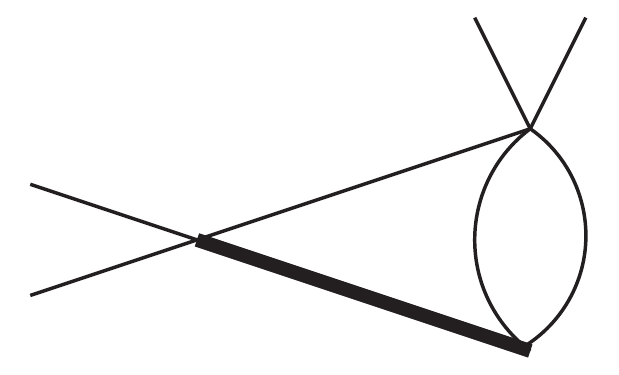}{4}(s)$     & $101~\mathrm{s}$     & $4.29 \times 10^{-3}$\\
\hline
$\figgraph{.2}{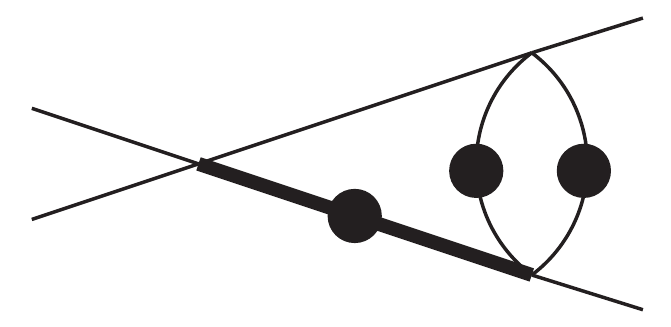}{6}(s)$     & $72~\mathrm{s}$     & $4.90 \times 10^{-4}$     &$\figgraph{.2}{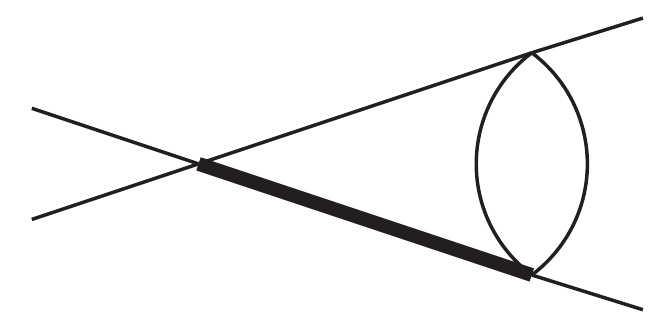}{4}(s)$     & $332~\mathrm{s}$     & $1.01 \times 10^{-3}$\\
\hline
$\figgraph{.2}{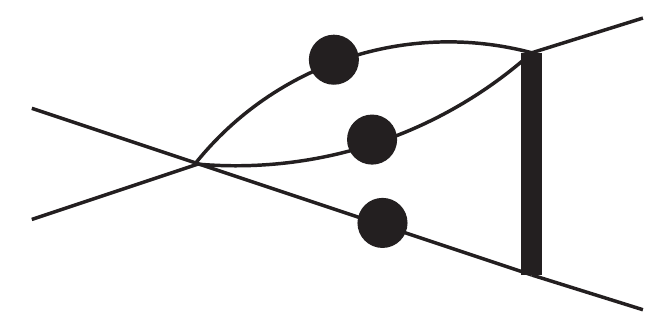}{6}(t)$     & $6~\mathrm{s}$     & $9.44 \times 10^{-6}$     &$\figgraph{.2}{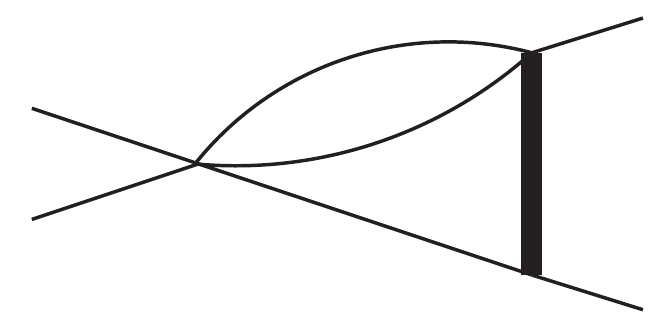}{4}(t)$     & $21~\mathrm{s}$     & $5.18\times 10^{-6}$\\
\hline
$\figgraph{.2}{finf21andf22}{6}(s)$     & $8~\mathrm{s}$     & $2.25 \times 10^{-4}$     &$\figgraph{.2}{cornerf21andf22}{4}(s)$     & $1004~\mathrm{s}$     & $4.82 \times 10^{-4}$\\
\hline
$\figgraph{.2}{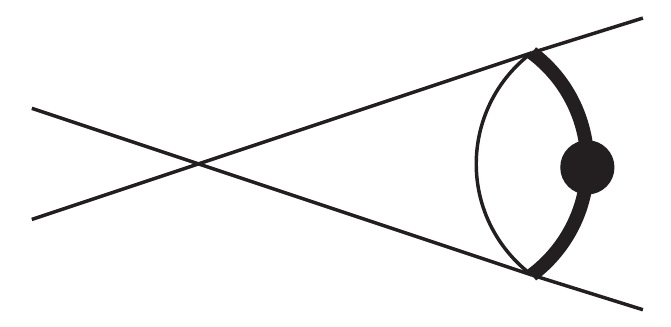}{4}(t)$     & $4~\mathrm{s}$     & $1.55 \times 10^{-5}$     &$\figgraph{.2}{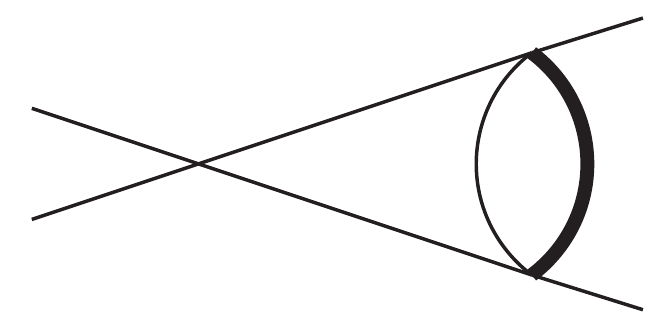}{4}(t)$      & $5~\mathrm{s}$     & $2.29 \times 10^{-5}$\\
\hline
$\figgraph{.2}{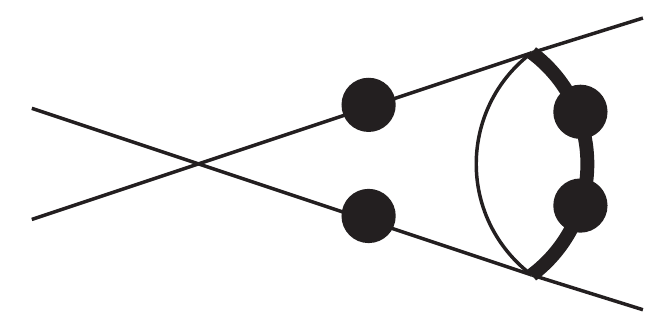}{6}(t)$      & $4~\mathrm{s}$     & $2.28 \times 10^{-5}$     &$\figgraph{.2}{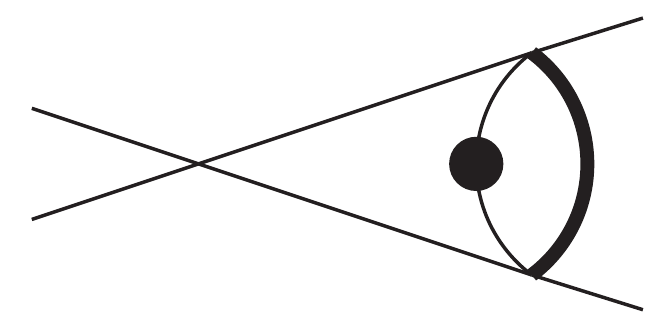}{4}(t)$      & $4~\mathrm{s}$     & $3.86 \times 10^{-6}$\\
\hline
$\figgraph{.2}{finf23andf25}{4}(s)$      & $9~\mathrm{s}$     & $2.88 \times 10^{-4}$     &$\figgraph{.2}{cornerf23andf25}{4}(s)$      & $441~\mathrm{s}$     & $7.73 \times 10^{-4}$\\
\hline
$\figgraph{.2}{finf24andf26}{6}(s)$     & $11~\mathrm{s}$     & $3.66 \times 10^{-4}$     &$\figgraph{.2}{cornerf24andf26}{4}(s)$      & $10~\mathrm{s}$     & $4.12 \times 10^{-4}$\\
\hline
$\figgraph{.2}{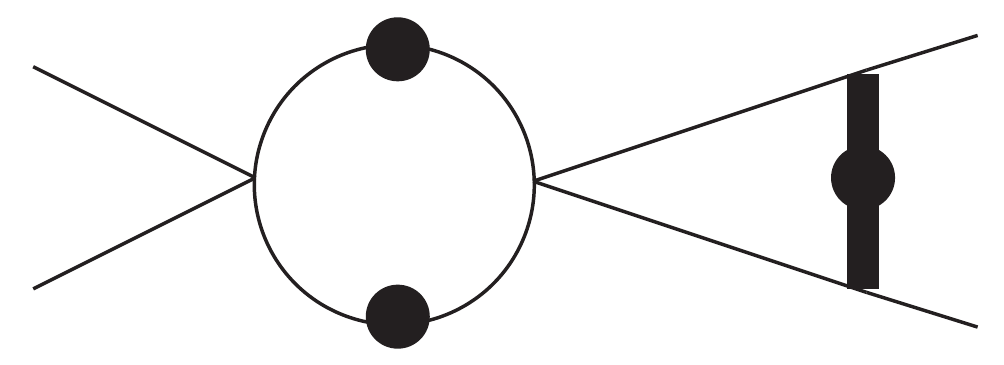}{6}(t)$      & $4~\mathrm{s}$     & $4.60 \times 10^{-5}$     &$\figgraph{.23}{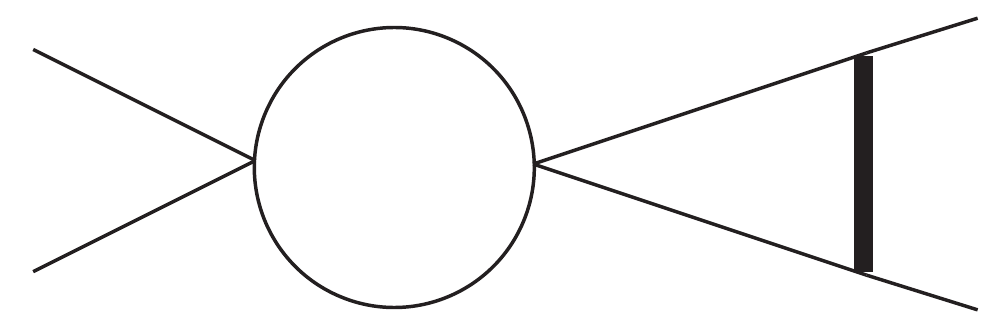}{4}(t)$      & $7~\mathrm{s}$     & $3.45 \times 10^{-5}$\\
\hline
$\figgraph{.2}{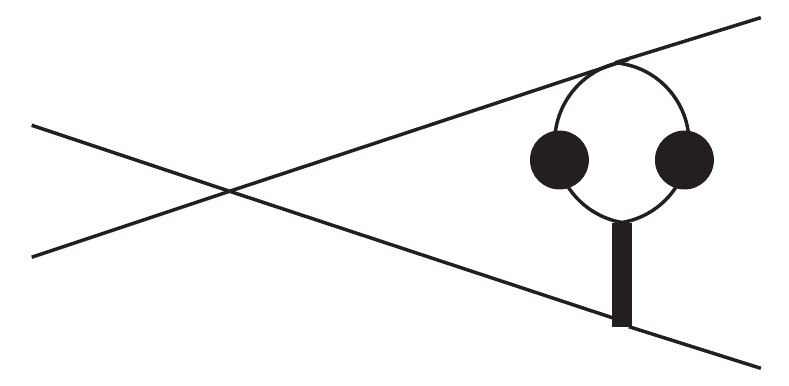}{6}(t)$     & $4~\mathrm{s}$     & $1.38 \times 10^{-5}$     &$\figgraph{.2}{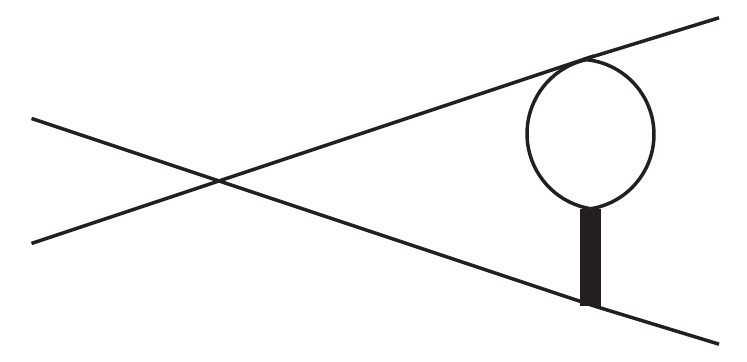}{4}(t)$     & $6~\mathrm{s}$     & $8.49 \times 10^{-6}$\\
\hline
$\figgraph{.2}{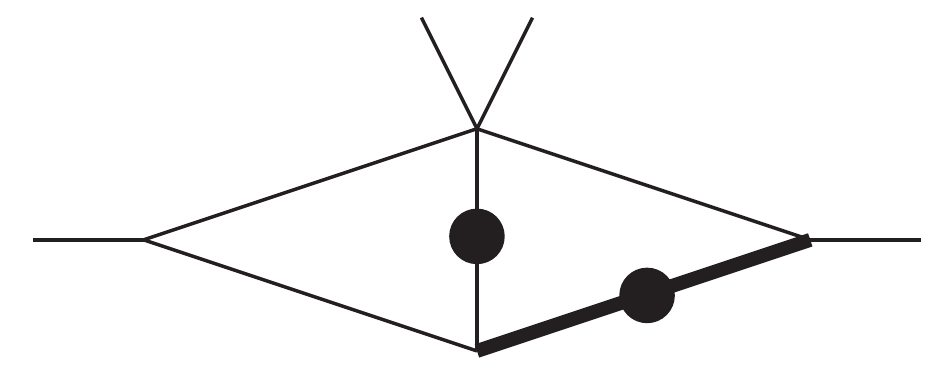}{6}(t)$    & $9~\mathrm{s}$     & $2.76 \times 10^{-5}$     &$\figgraph{.2}{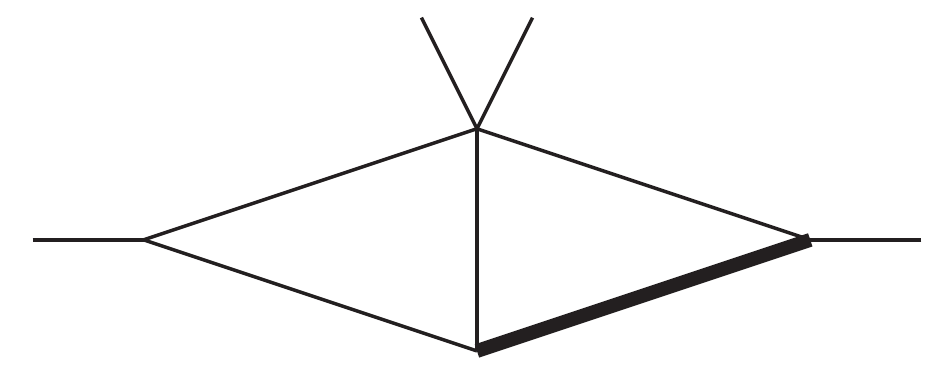}{4}(t)$     & $16~\mathrm{s}$     & $2.63 \times 10^{-5}$\\
\hline
$\figgraph{.2}{finf29andf30}{6}(s)$      & $18~\mathrm{s}$     & $1.04 \times 10^{-3}$     &$\figgraph{.2}{cornerf29andf30}{4}(s)$      & $41~\mathrm{s}$     & $6.47 \times 10^{-4}$\\
\hline
$\figgraph{.2}{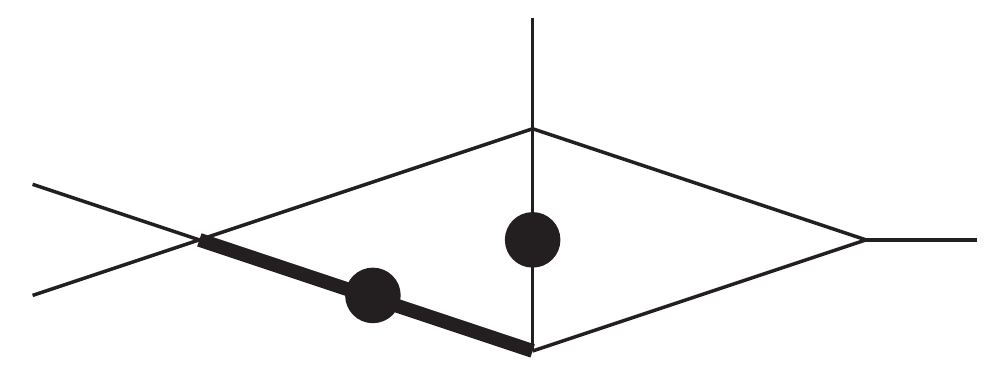}{6}(s)$    & $18~\mathrm{s}$     & $1.23 \times 10^{-4}$     &$\figgraph{.2}{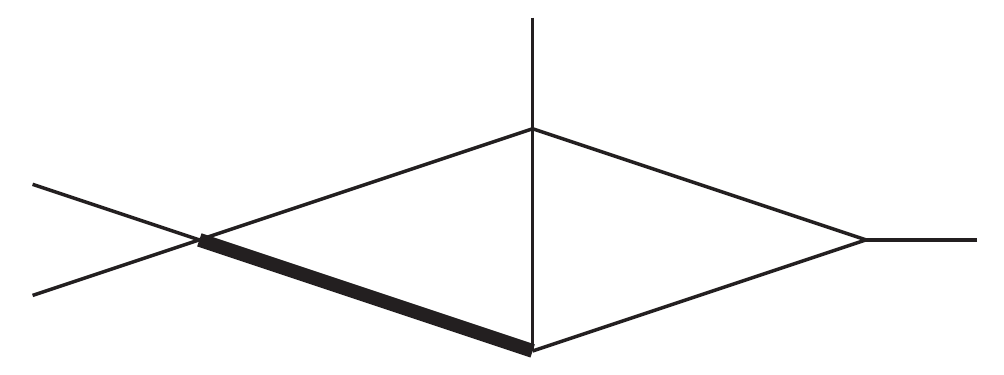}{4}(s)$     & $12~\mathrm{s}$     & $4.70 \times 10^{-4}$\\
\hline
$\figgraph{.2}{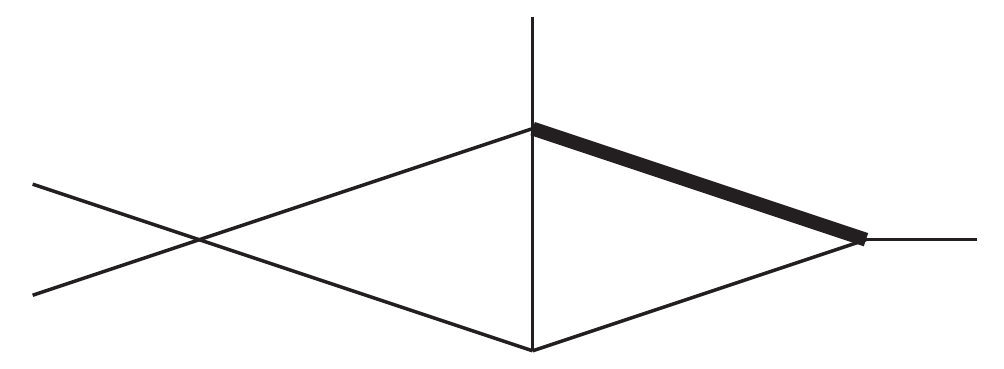}{4}(t)$      & $9~\mathrm{s}$     & $2.13 \times 10^{-5}$     &$\figgraph{.2}{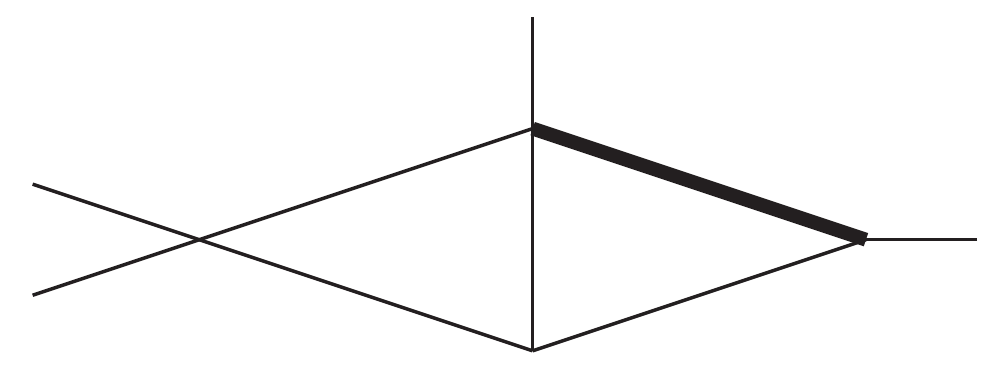}{4}(t)$      & $3~\mathrm{s}$     & $2.13 \times 10^{-5}$\\
\hline
$\figgraph{.2}{finf32andf33}{4}(s)$      & $31~\mathrm{s}$     & $7.51 \times 10^{-6}$     &$\figgraph{.2}{cornerf32andf33}{4}(s)$      & $17~\mathrm{s}$     & $7.51 \times 10^{-6}$\\
\hline
$\figgraph{.2}{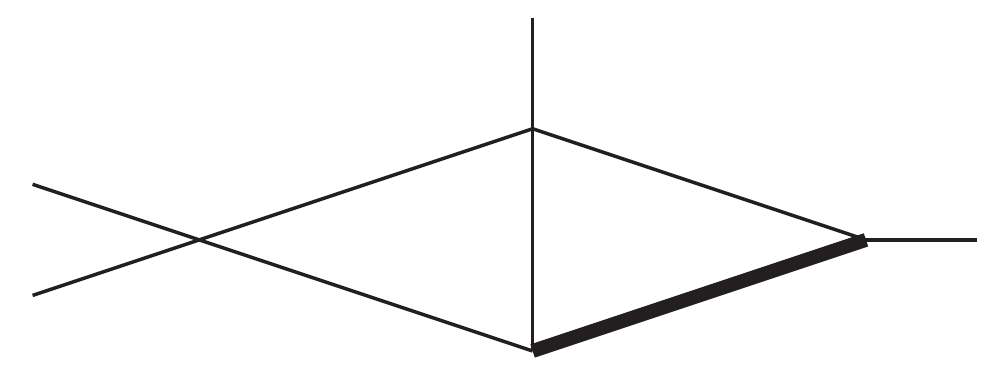}{4}(s)$     & $35~\mathrm{s}$     & $3.28 \times 10^{-5}$     &$\figgraph{.2}{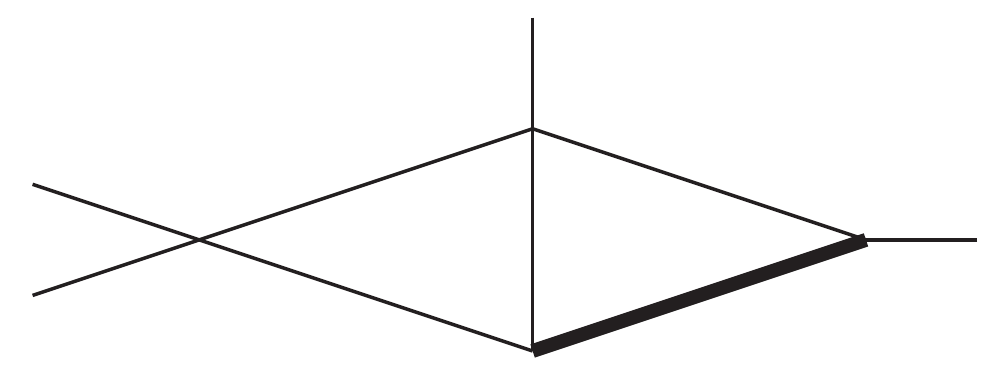}{4}(s)$     & $19~\mathrm{s}$     & $3.28 \times 10^{-5}$\\
\hline
$\figgraph{.2}{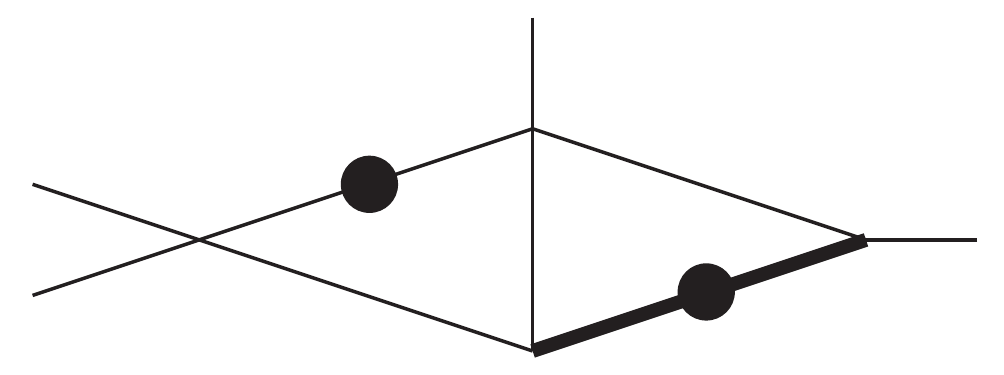}{6}(s)$     & $36~\mathrm{s}$     & $3.94 \times 10^{-4}$     &$\figgraph{.2}{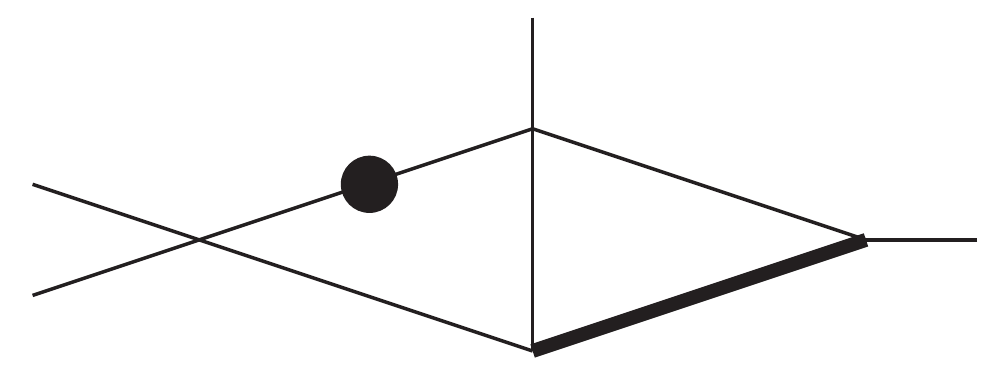}{4}(s)$     & $113~\mathrm{s}$     & $7.78 \times 10^{-4}$\\
\hline
$\figgraph{.2}{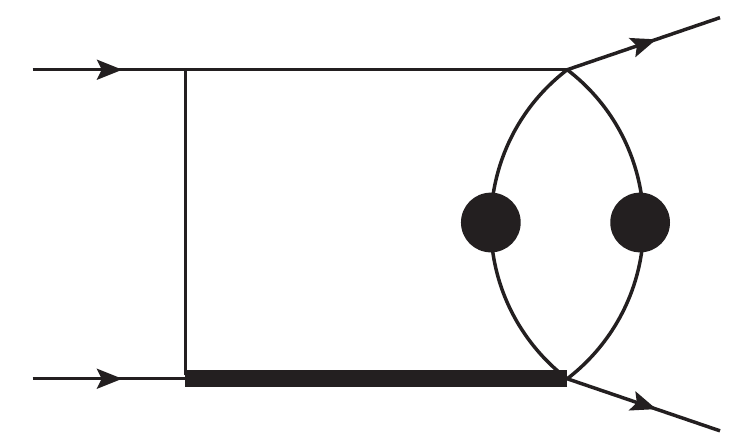}{6}(s,t)$     & $28~\mathrm{s}$     & $8.37 \times 10^{-4}$     &$\figgraph{.2}{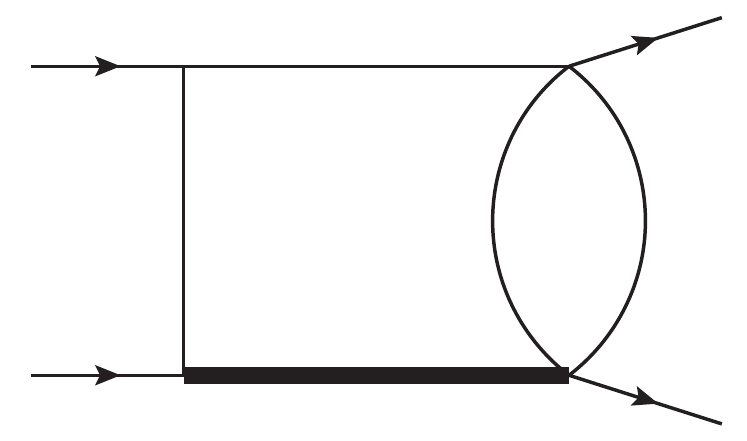}{4}(s,t)$     & $101~\mathrm{s}$     & $2.13 \times 10^{-3}$\\
\hline
$\figgraph{.2}{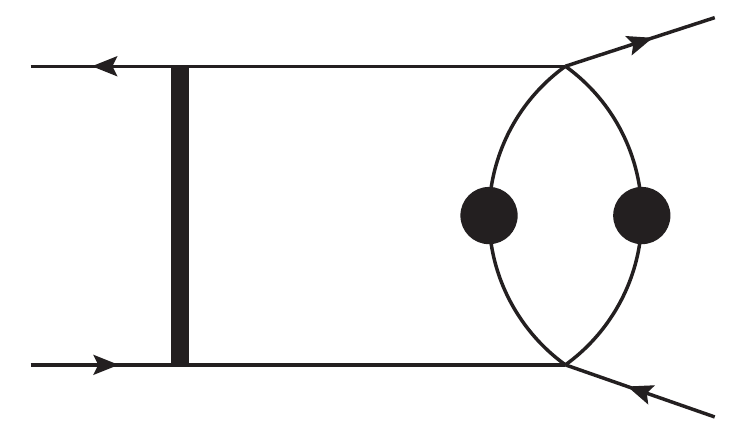}{6}(s,t)$      & $37~\mathrm{s}$     & $5.84 \times 10^{-3}$     &$\figgraph{.2}{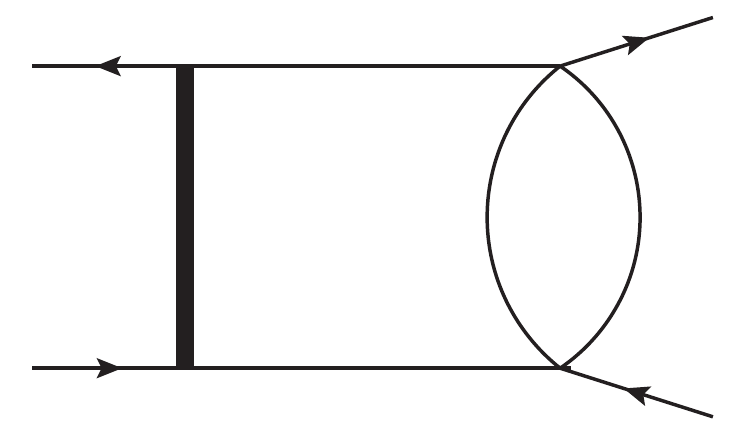}{4}(s,t)$      & $26~\mathrm{s}$     & $5.35 \times 10^{-4}$\\
\hline
$\figgraph{.2}{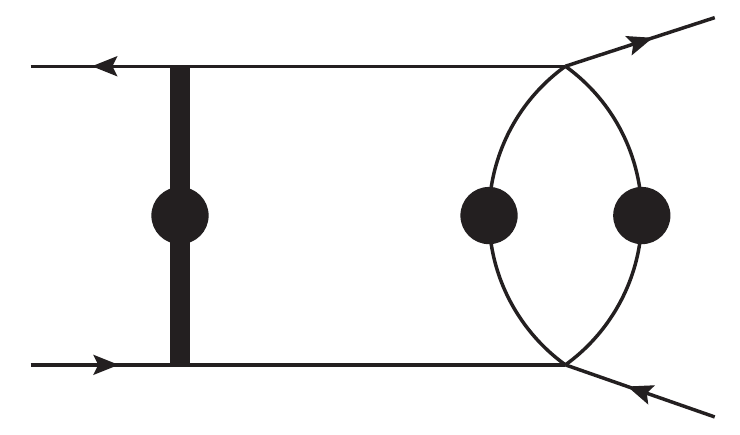}{6}(s,t)$     & $18~\mathrm{s}$     & $2.92 \times 10^{-3}$     &$\figgraph{.2}{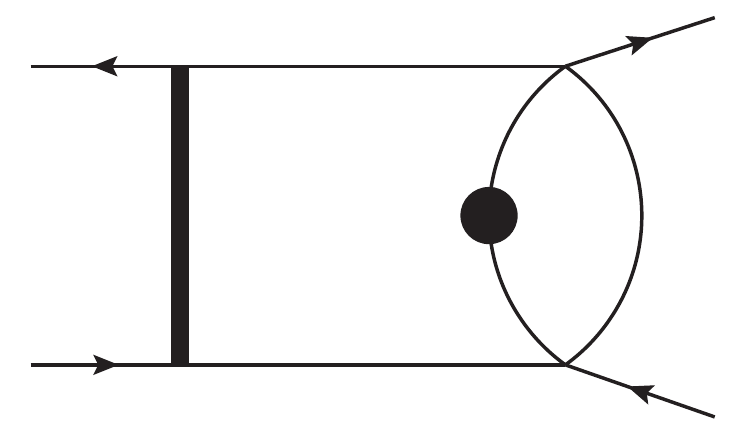}{4}(s,t)$     & $319~\mathrm{s}$     & $5.92 \times 10^{-2}$\\
\hline
$\figgraph{.2}{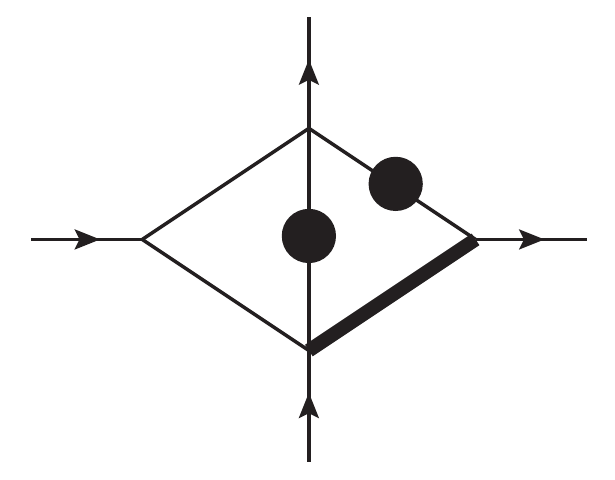}{6}(s,t)$     & $50~\mathrm{s}$     & $6.76 \times 10^{-4}$     &$\figgraph{.2}{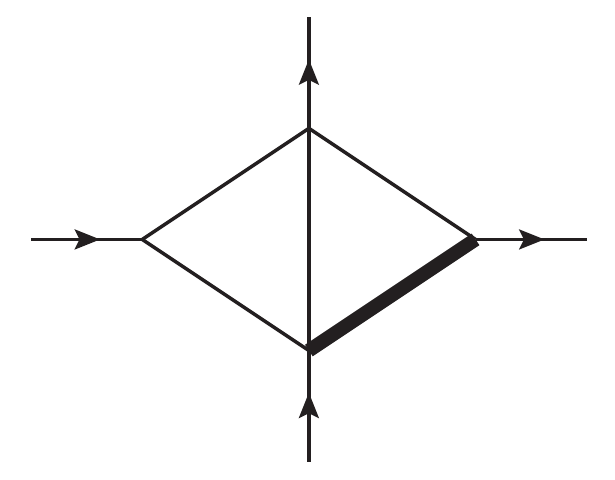}{4}(s,t)$     & $13~\mathrm{s}$     & $9.49 \times 10^{-4}$\\
\hline
$\figgraph{.2}{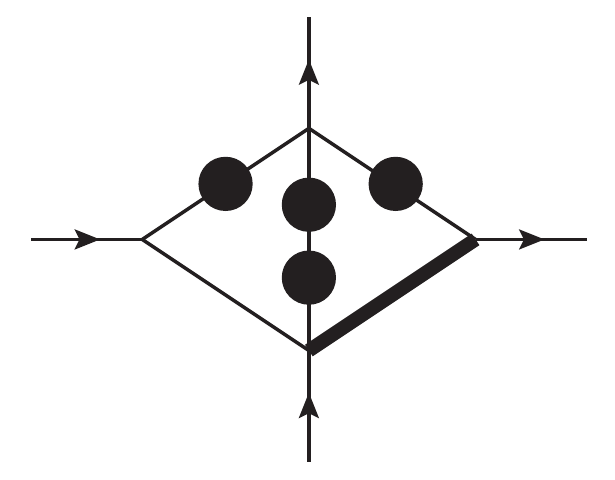}{8}(s,t)$     & $35~\mathrm{s}$     & $7.64 \times 10^{-4}$     &$\figgraph{.2}{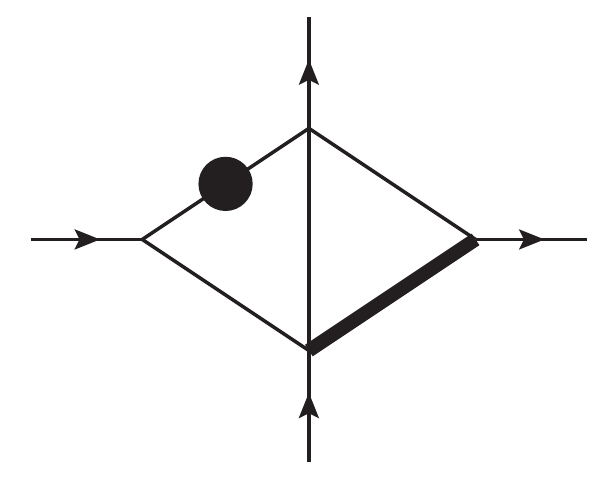}{4}(s,t)$     & $20605~\mathrm{s}$     & $9.87 \times 10^{-4}$\\
\hline
$\figgraph{.2}{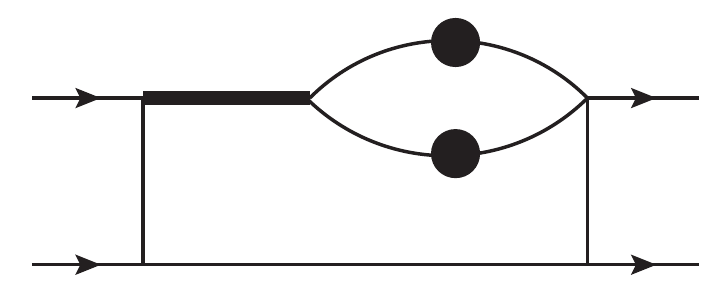}{6}(s,t)$     & $1609~\mathrm{s}$     & $4.39 \times 10^{-4}$     &$\figgraph{.2}{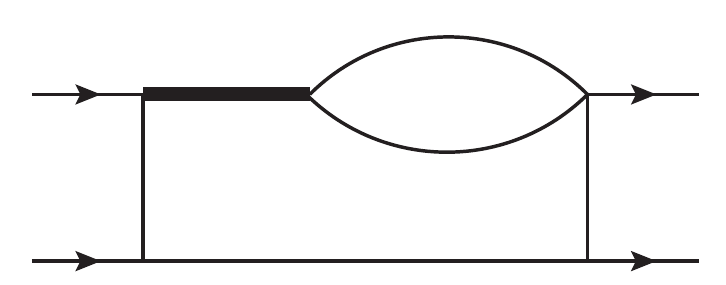}{4}(s,t)$     & $564~\mathrm{s}$     & $2.04 \times 10^{-2}$\\
\hline
$\figgraph{.2}{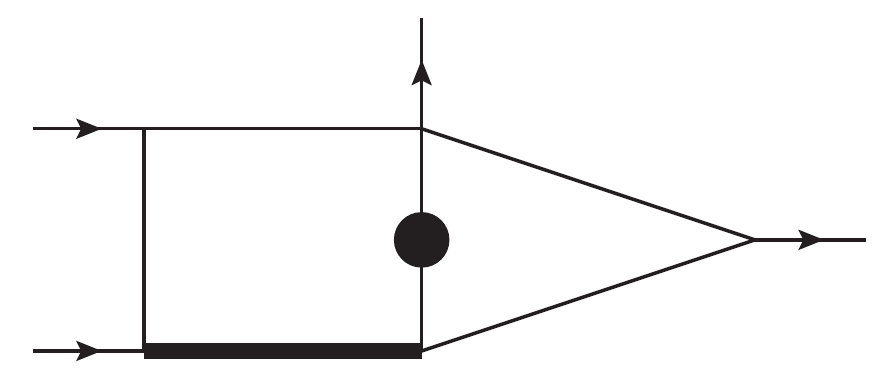}{6}(s,t)$     & $202~\mathrm{s}$     & $7.31 \times 10^{-4}$     &$\figgraph{.2}{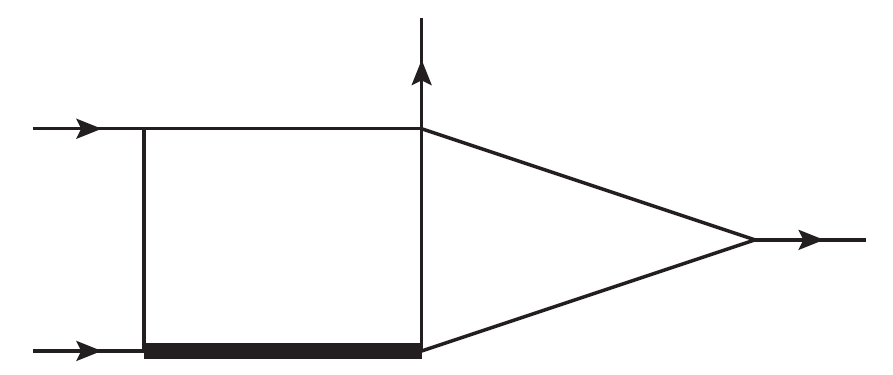}{4}(s,t)$     & $96~\mathrm{s}$     & $2.35 \times 10^{-3}$\\
\hline
$\figgraph{.2}{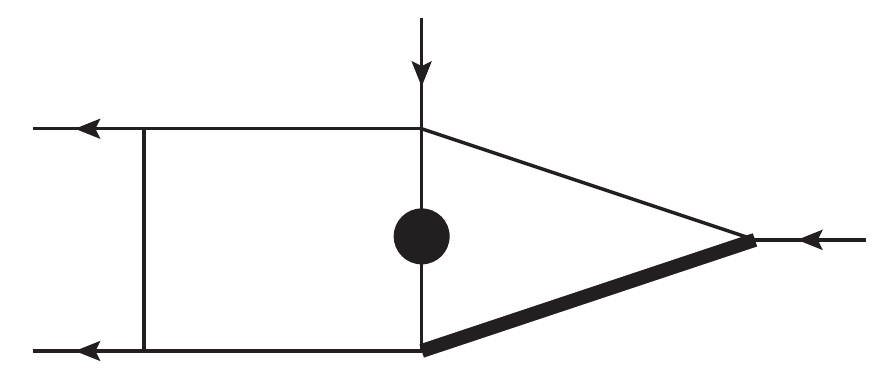}{6}(s,t)$     & $201~\mathrm{s}$     & $2.34 \times 10^{-4}$     &$\figgraph{.2}{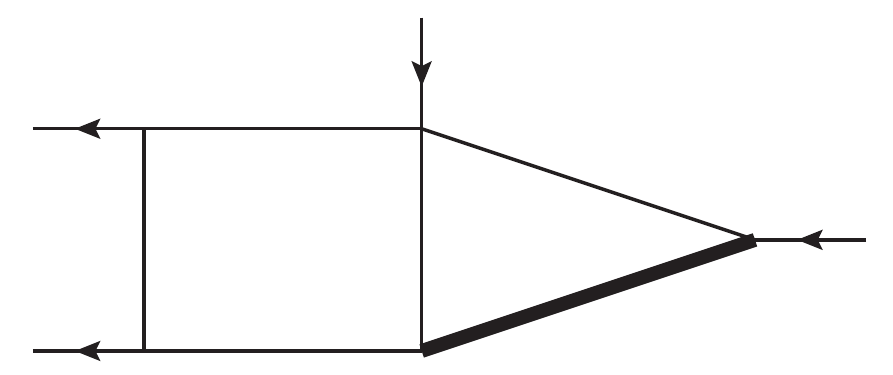}{4}(s,t)$     & $384~\mathrm{s}$     & $8.12 \times 10^{-4}$\\
\hline
$\figgraph{.2}{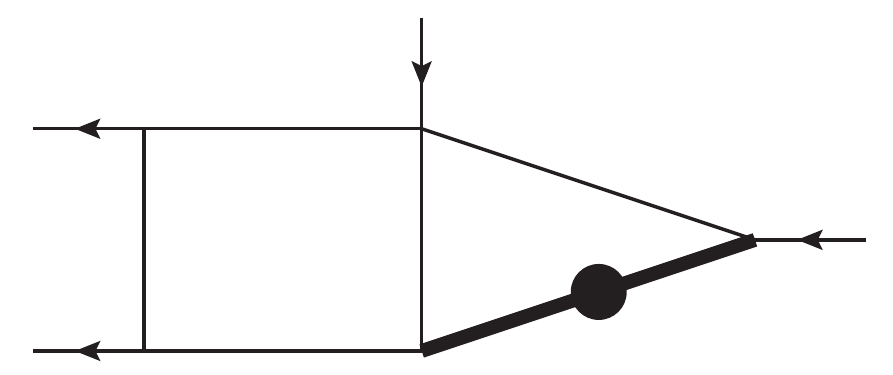}{6}(s,t)$     & $150~\mathrm{s}$     & $4.83 \times 10^{-4}$     &$\figgraph{.2}{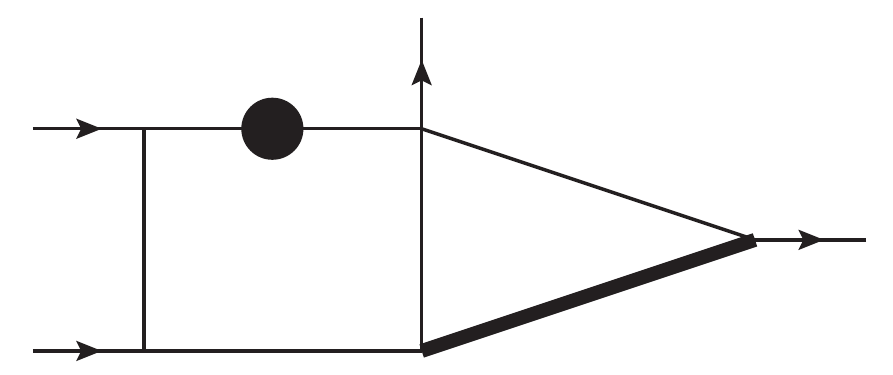}{4}(s,t)$     & $56538~\mathrm{s}$     & $1.67 \times 10^{-2}$\\
\hline
$\figgraph{.2}{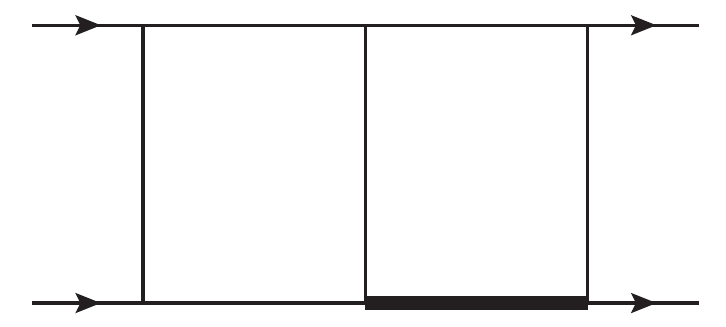}{6}(s,t)$     & $280~\mathrm{s}$     & $1.00 \times 10^{-3}$     &$\figgraph{.2}{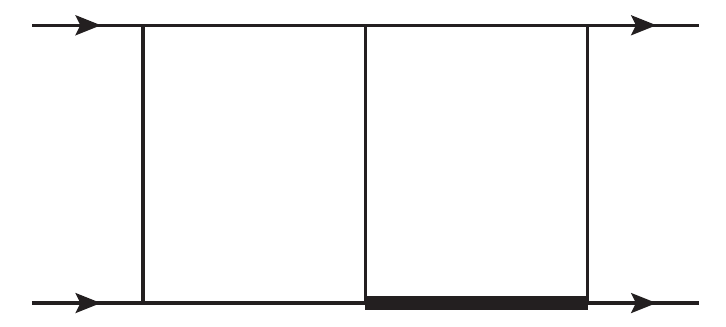}{4}(s,t)$     & $214135~\mathrm{s}$     & $8.29 \times 10^{-3}$\\
\hline
$\figgraph{.2}{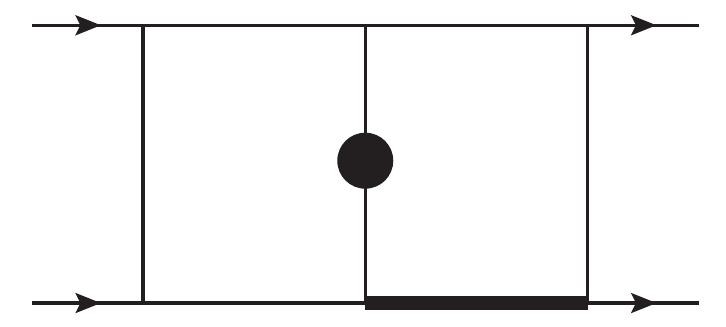}{6}(s,t)$     & $294~\mathrm{s}$     & $1.21 \times 10^{-3}$     &$\figgraph{.2}{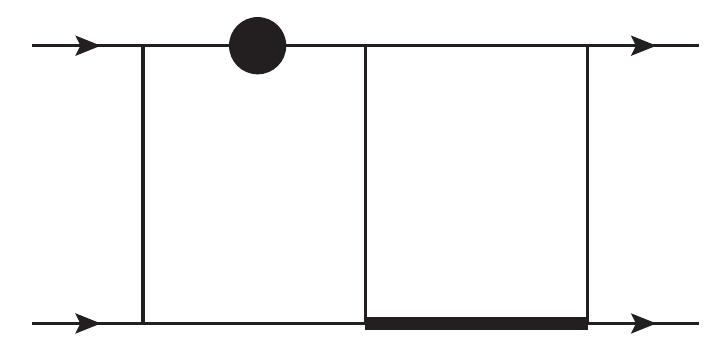}{4}(s,t)$     & $3484378~\mathrm{s}$     & $30.9$\\
\hline
$\figgraph{.2}{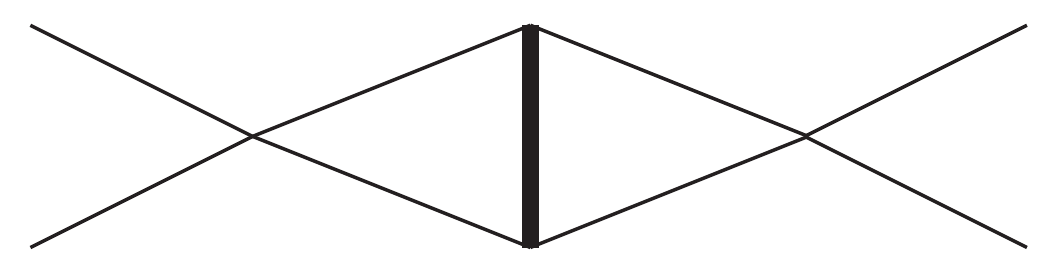}{4}(s)$     & $91~\mathrm{s}$     & $3.76 \times 10^{-4}$     &$\figgraph{.2}{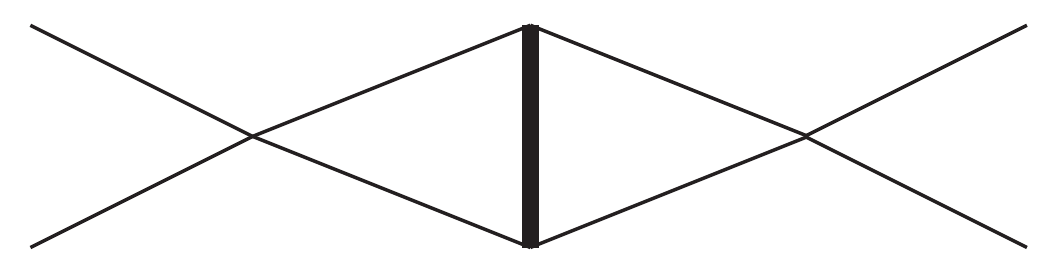}{4}(s)$     & $87~\mathrm{s}$     & $3.76 \times 10^{-4}$\\
\hline
$\figgraph{.2}{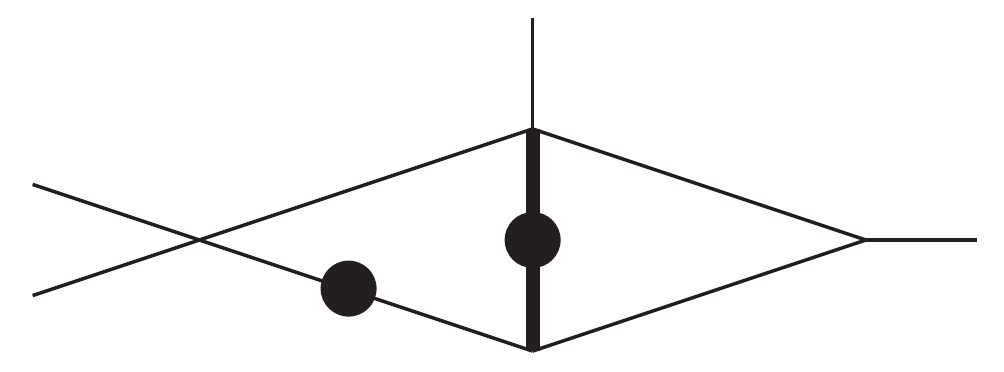}{6}(s)$     & $17~\mathrm{s}$     & $5.15 \times 10^{-4}$     &$\figgraph{.2}{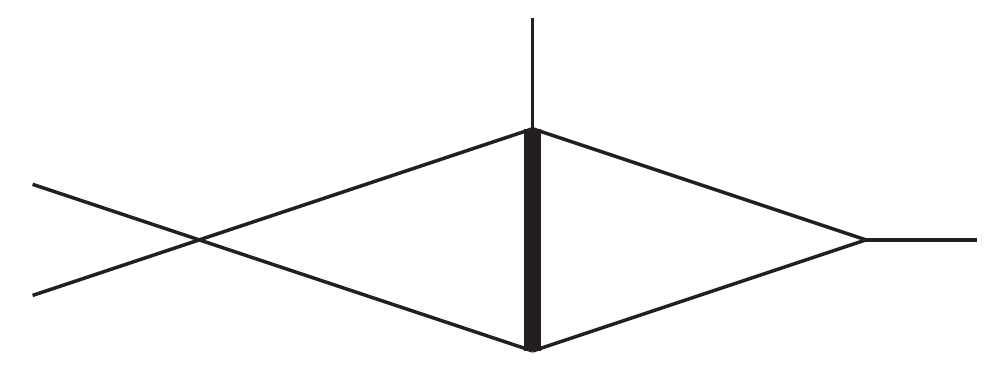}{4}(s)$     & $20~\mathrm{s}$     & $1.95 \times 10^{-4}$\\
\hline
$\figgraph{.2}{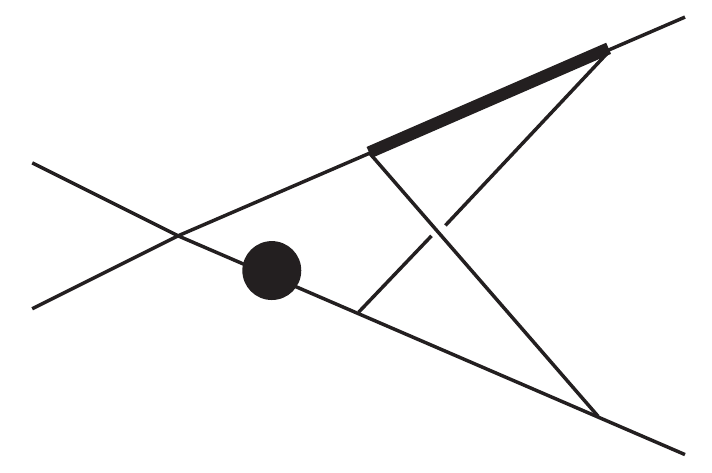}{6}(s)$     & $119~\mathrm{s}$     & $2.32 \times 10^{-3}$     &$\figgraph{.2}{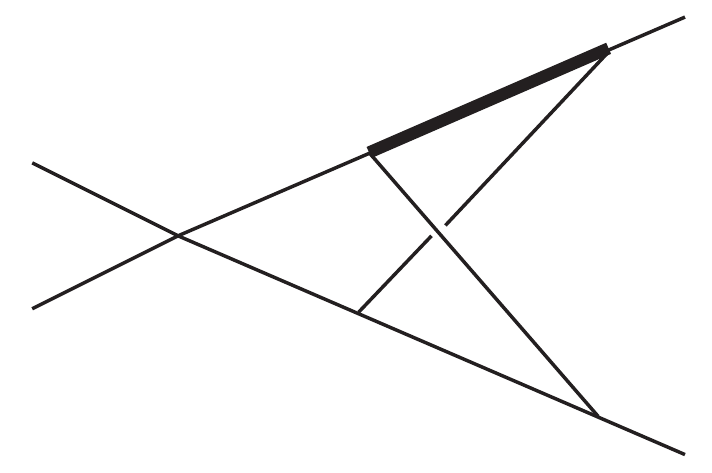}{4}(s)$     & $118~\mathrm{s}$     & $2.12 \times 10^{-3}$\\
\hline
Total/Max: & $3995~\mathrm{s}$     & $5.84 \times 10^{-3}$ & Total/Max: & $5136862~\mathrm{s}$     & $30.9$\\
\hline
\caption{\label{tab:mixedDYnums}
Numerical performance of finite and conventional integral bases
for two-loop mixed EW-QCD corrections to Drell-Yan lepton production
in the physical region using {\tt SecDec\;3}.
For each integral in the above, the run time is given in seconds and the fractional difference from the analytical solution is given for the expansion coefficient which first gives rise to weight four multiple polylogarithms.
In the final row of the table, total run times and worst-case relative accuracies are recorded for both integral bases.
}
\end{longtable}
\end{center}

We find that, in several other cases, significantly more mileage can be squeezed out
of publicly available sector decomposition programs simply by working with
a well-behaved, finite integral basis.
As a second example, we present the evaluation of three-loop form factors in
massless QCD in Appendix \ref{sec:3Lff}.
We employ the other publicly available program under active development, {\tt FIESTA\;4},
and observe dramatic gains in both speed  and numerical convergence when using
a basis of finite integrals.

\section{Conclusions and Outlook}
\label{sec:end}
In this paper, we considered the two-loop integrals relevant to the $\alpha\alpha_s$
corrections to Drell-Yan production with up to a single massive vector boson exchanged.
As a reference, we calculated their Laurent expansion through to weight five
in terms of multiple polylogarithms using the method of differential equations.
Our representation in terms of real-valued functions allows for fast and precise
numerical evaluations over the entire physical region of the phase space.
We found it challenging to even check our analytical solutions using available
sector decomposition programs.
Employing a basis of finite integrals systematically improved the situation and
rendered all integrals numerically accessible with {\tt SecDec\;3}, both in Euclidean and
physical kinematics.
Order of magnitude improvements both in program run time and integration error
were also found for massless three-loop form factor integrals using {\tt FIESTA\;4}
when using finite instead of conventional master integrals.
For the finite integrals, we allowed for shifts to higher numbers of spacetime
dimensions and additional powers of denominators;
the actual change of basis is performed with integration by parts reductions in a highly-automated way.
An implementation of the algorithm to construct a basis of finite integrals
is publicly available in the package {\tt Reduze\;2.1} on {\tt HepForge}.
Numerical evaluations along the lines discussed in this paper are not only
interesting for cross-checks of analytical solutions, but may actually serve as
the primary method for phenomenological applications in especially complicated
cases~\cite{Borowka:2016ehy,Borowka:2016ypz}.

\section*{Acknowledgments}

We would like to thank Gudrun Heinrich, Johannes Schlenk, and Alexander Smirnov for help with
{\tt SecDec} and {\tt FIESTA}. We would also like to thank Erik Panzer for his careful reading of our manuscript.
The work of RMS was supported in part by the European Research Council
through grants 291144 (EFT4LHC) and 647356 (CutLoops).
We are grateful to the Mainz Institute for Theoretical Physics (MITP) for its hospitality and support.
Our figures were generated using {\tt Jaxodraw} \cite{Binosi:2003yf}, based on {\tt AxoDraw} \cite{Vermaseren:1994je}.

\appendix
\section{Numerical Analysis for Massless Three-Loop Form Factors} 
\label{sec:3Lff}
In this appendix, we present the results of our {\tt FIESTA\;4}-based numerical study of the finite and conventional integral bases for the massless three-loop vertex functions ({\it i.e.} the bases employed in
references \cite{vonManteuffel:2015gxa} and \cite{Gehrmann:2010ue} respectively).
They show the advantages of using a basis of finite integrals in a setup independent
of what was employed in Section~\ref{sec:mixedDY}.
Due to the 
fact that the three-loop form factor master integrals have been calculated to weight eight \cite{Gehrmann:2005pd,Lee:2010ug,Lee:2010ik,Henn:2013nsa,vonManteuffel:2015gxa}, 
we can also study how well our finite basis performs for the evaluation of higher order coefficients
in the $\epsilon$ expansion.

By experimenting with the code and discussing some aspects of it with the developer, we settled on the {\tt FIESTA\;4} settings $\mathtt{ComplexMode = False}$, $\mathtt{STRATEGY = STRATEGY\_X}$, 
$\mathtt{BucketSize = 28}$, $\mathtt{NegativeTermsHandling = None}$, and $\mathtt{CurrentIntegratorSettings =}$ \\
$\mathtt{\{\{}$``$\mathtt{mineval}$''$\mathtt{, }$``$\mathtt{1000}$''$\mathtt{\},\{}$``$\mathtt{maxeval}$''$\mathtt{, }$``$\mathtt{500000}$''$\mathtt{\},\{}$``$\mathtt{nstart}$'',``$\mathtt{50000}$''$\mathtt{\}\}}$ 
for the trivial phase space point $q^2 = -1$.
Especially crucial is the setting $\mathtt{NegativeTermsHandling = None}$, which seems to substantially improve the performance 
for Euclidean Feynman integrals in general.
We also configured the code to run on all four cores of a i7-4940MX processor. 
Once again, we used the {\tt VEGAS} Monte Carlo integration routine provided by {\tt CUBA}, but, this time, we found that {\tt Mathematica 8.0.4} is a better choice than {\tt Mathematica 9.0.1} for the front end.

Let us begin by discussing Table \ref{tab:ff3L}, where all twenty-two irreducible integral topologies in both the finite and conventional integral bases are evaluated through to the terms of weight six. Reasonable results are obtained in both integral bases using the program settings described above. However, in going from the conventional basis to our chosen finite basis, the program run time drops by more than a factor of fifty:
from just shy of twenty-two hours to about half an hour. There is, however, another crucial difference between the two sets of integrals which plays an even more important role when one goes from three to four loops: for the fixed program settings
given above, one finds that the numerical convergence of the {\tt VEGAS} algorithm is far superior for the finite basis integrals.\footnote{Let us point out that, in recent
work on massless form factors~\cite{vonManteuffel:2015gxa}, 
we were able to numerically evaluate a finite non-planar twelve-line master integral through to weight eight to four decimal digit accuracy using {\tt FIESTA\;4} in about one day on a desktop computer.}

In fact, the average relative error of the weight six contributions with respect to the exact results is $2 \times 10^{-5}$
for the finite integral basis compared to $2 \times 10^{-4}$ for the conventional integral basis, an improvement of an order of magnitude. Finally, we see from Table \ref{tab:higherff3L} that, 
in going from weight six to weight eight with finite integrals, the run time increases by only a factor of three (with a slight loss of precision for fixed program settings).\footnote{This shows that,
to obtain acceptably fast and stable numerical evaluations of master integrals at higher orders in $\epsilon$,
our integral basis is an effective alternative to the basis suggested in \cite{Chetyrkin:2006dh}, where master integrals were systematically chosen to have the property that their coefficients remain finite in the $\epsilon \rightarrow 0$ limit.}
Given the long run times with conventional integrals already at weight six,
we refrained from trying to perform a similar weight eight evaluation of them.

\begin{center}
\begin{longtable}[h!]{|c|c|c||c|c|c|}
\hline
$\figgraph{.2}{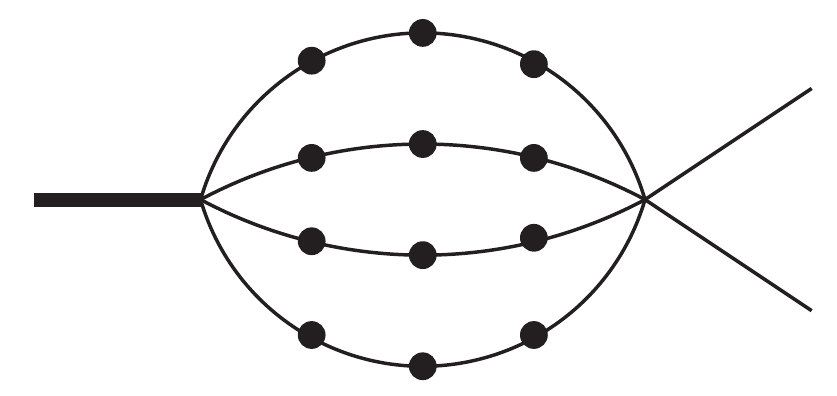}{10}$     & $6~\mathrm{s}$     & $2.46 \times 10^{-5}$     &$\figgraph{.2}{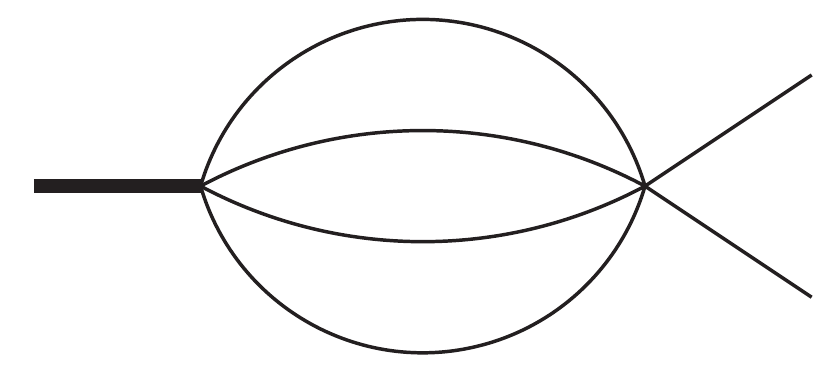}{4}$      & $7~\mathrm{s}$     & $3.84 \times 10^{-6}$\\
\hline
$\figgraph{.2}{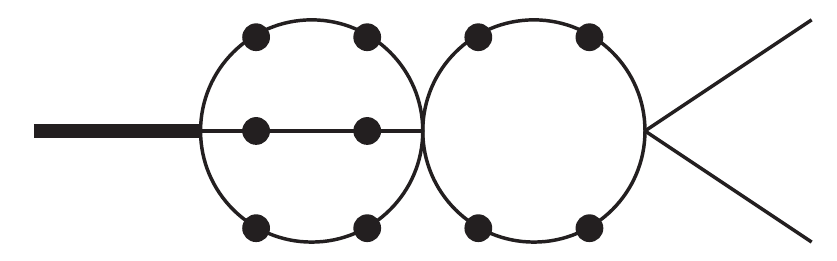}{8}$      & $49~\mathrm{s}$     & $8.88\times 10^{-7}$     &$\figgraph{.2}{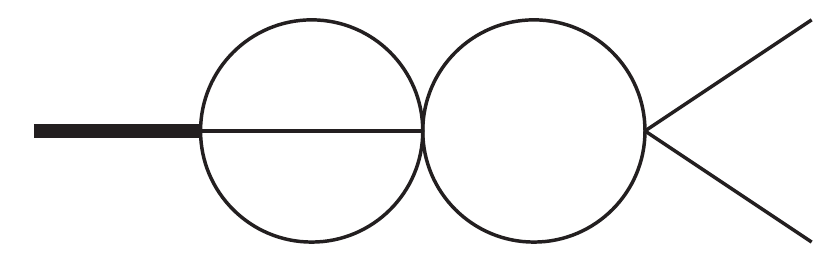}{4}$      & $57~\mathrm{s}$     & $1.36 \times 10^{-5}$\\
\hline
$\figgraph{.2}{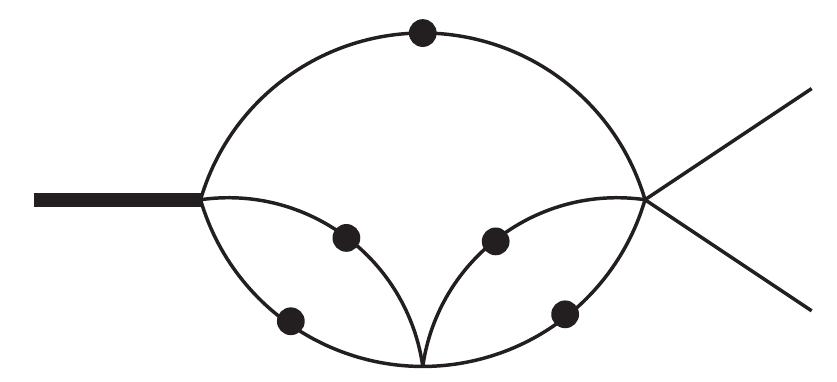}{6}$      & $35~\mathrm{s}$     & $4.31 \times 10^{-5}$     &$\figgraph{.2}{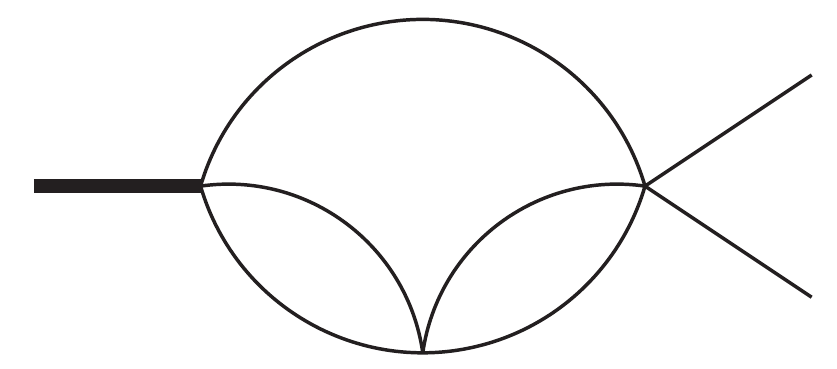}{4}$      & $33~\mathrm{s}$     & $2.68 \times 10^{-6}$\\
\hline
$\figgraph{.2}{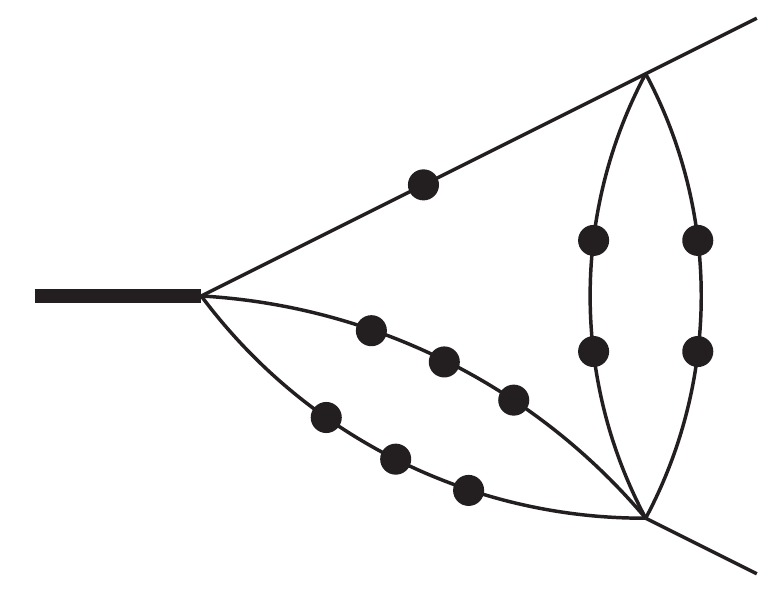}{10}$     & $17~\mathrm{s}$     & $4.09 \times 10^{-5}$     &$\figgraph{.2}{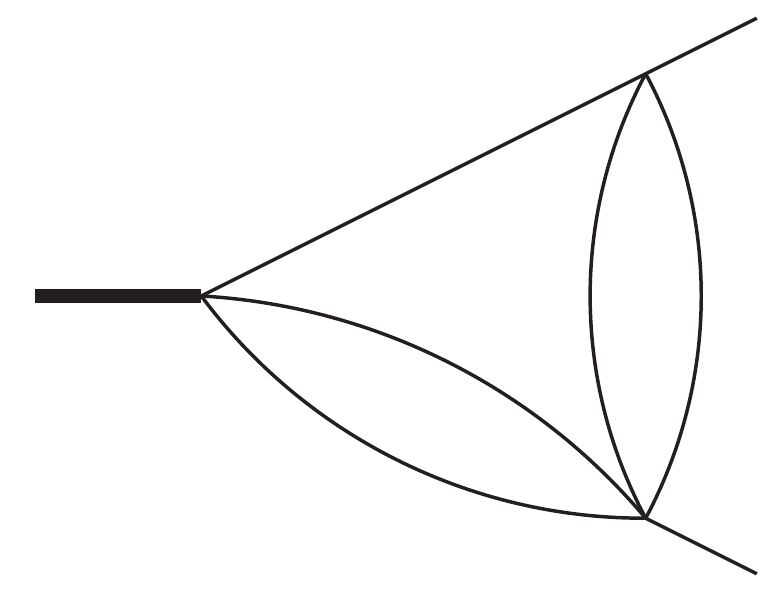}{4}$      & $15~\mathrm{s}$     & $6.97 \times 10^{-6}$\\
\hline
$\figgraph{.2}{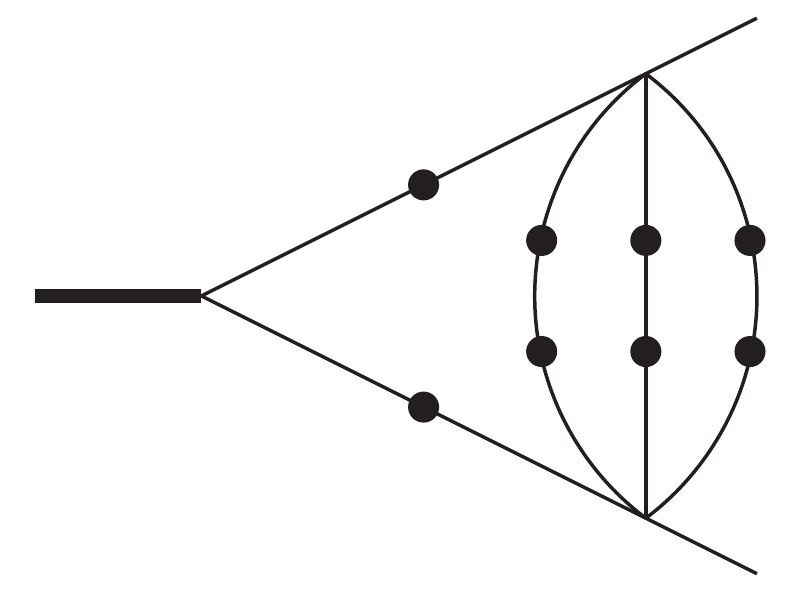}{8}$      & $16~\mathrm{s}$     & $1.28 \times 10^{-4}$     &$\figgraph{.2}{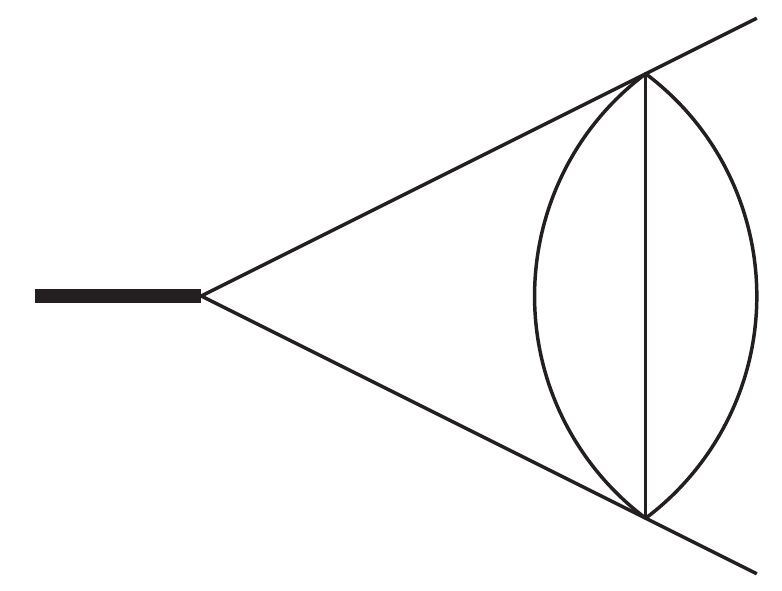}{4}$      & $22~\mathrm{s}$     & $1.45 \times 10^{-5}$\\
\hline
$\figgraph{.2}{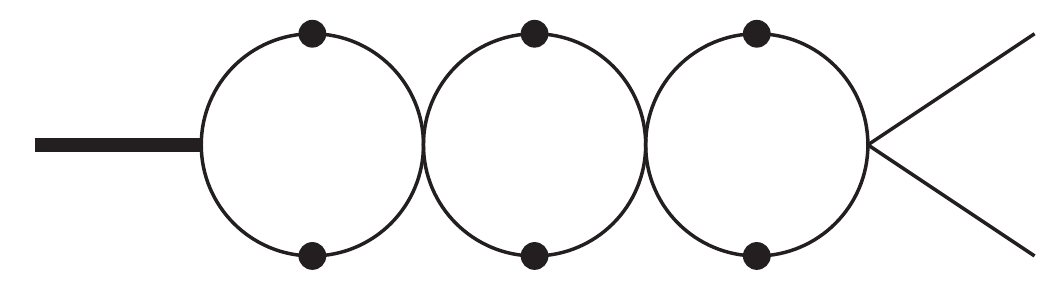}{6}$     & $121~\mathrm{s}$     & $5.42 \times 10^{-6}$     &$\figgraph{.2}{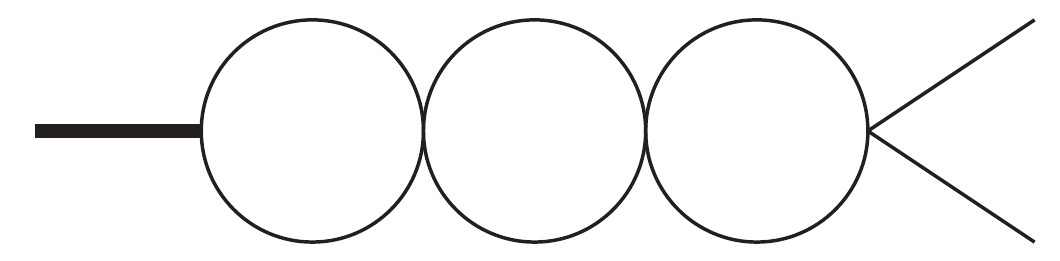}{4}$     & $92~\mathrm{s}$     & $5.33 \times 10^{-6}$\\
\hline
$\figgraph{.2}{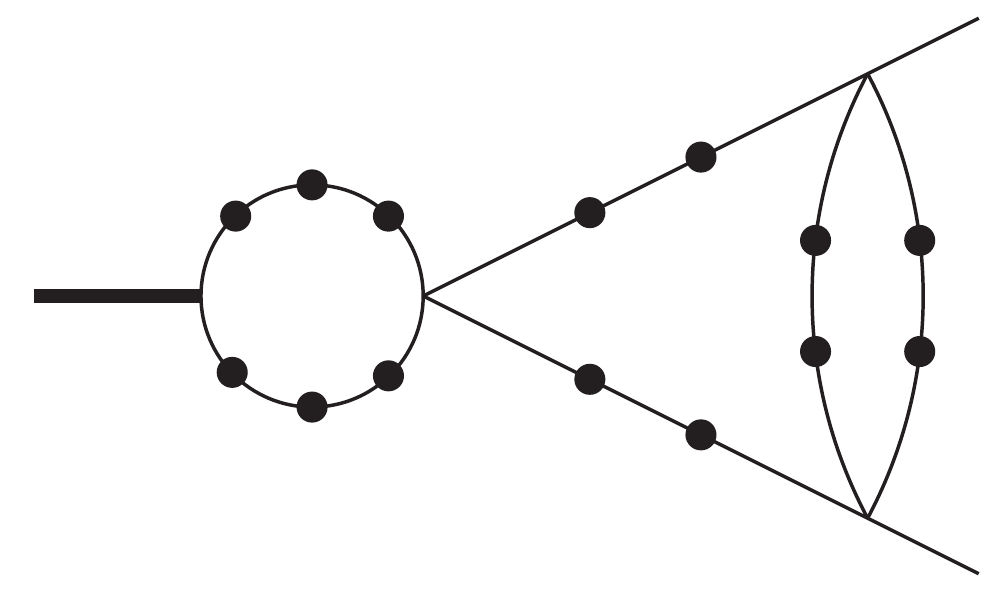}{10}$    & $93~\mathrm{s}$     & $1.37 \times 10^{-5}$     &$\figgraph{.2}{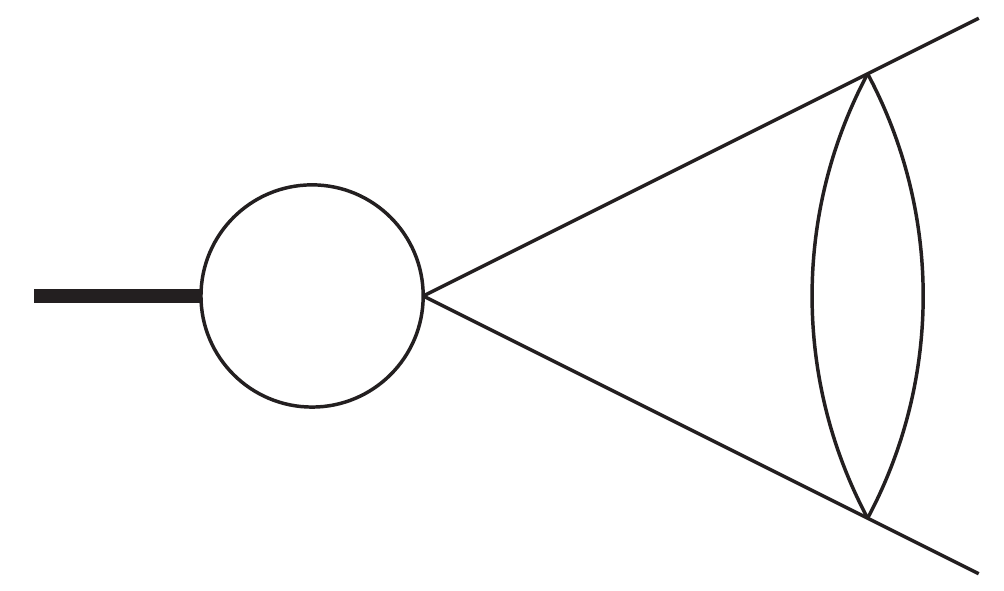}{4}$     & $62~\mathrm{s}$     & $1.96 \times 10^{-5}$\\
\hline
$\figgraph{.2}{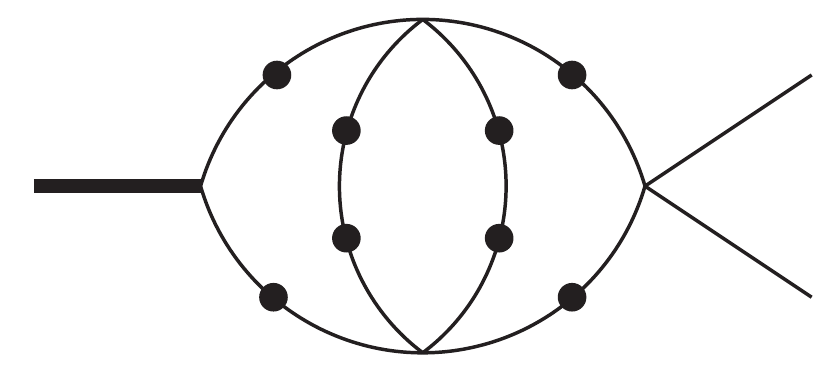}{8}$      & $39~\mathrm{s}$     & $2.25 \times 10^{-6}$     &$\figgraph{.2}{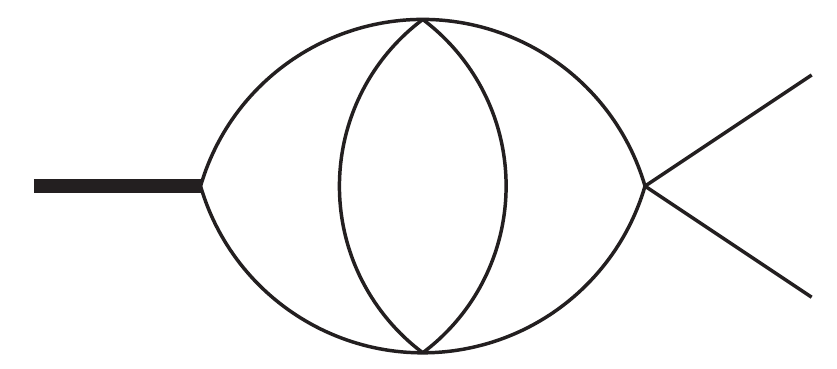}{4}$      & $87~\mathrm{s}$     & $1.82 \times 10^{-5}$\\
\hline
$\figgraph{.2}{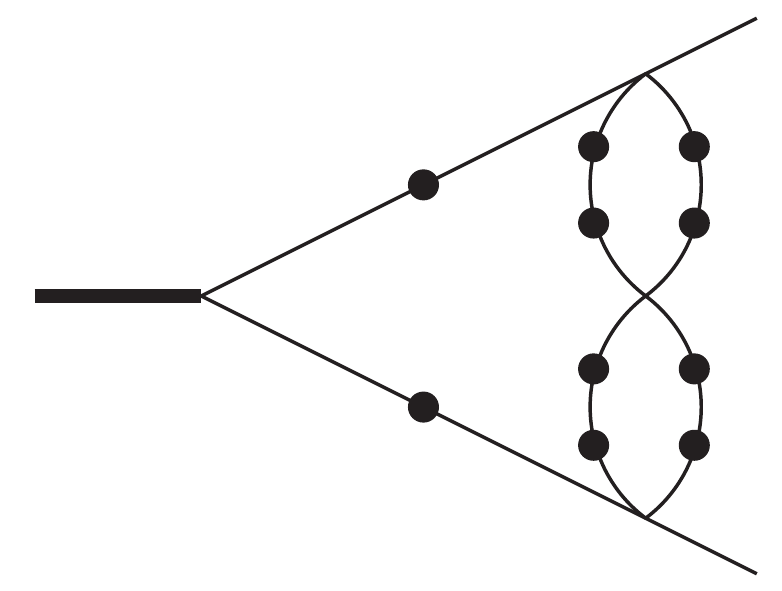}{10}$    & $34~\mathrm{s}$     & $2.90 \times 10^{-5}$     &$\figgraph{.2}{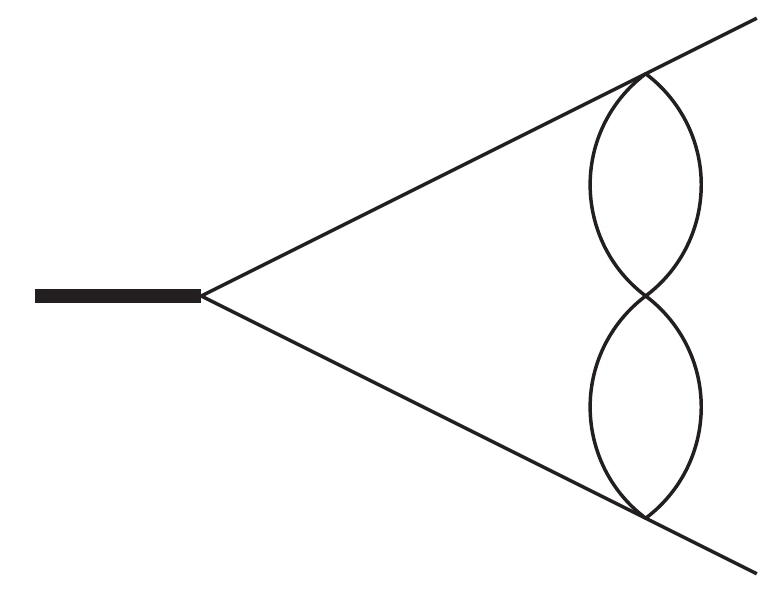}{4}$     & $23~\mathrm{s}$     & $4.55 \times 10^{-5}$\\
\hline
$\figgraph{.2}{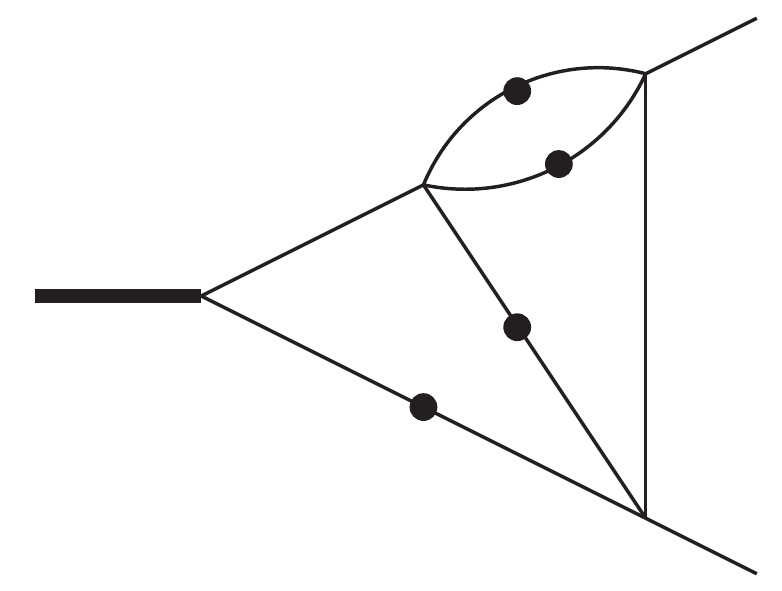}{6}$      & $30~\mathrm{s}$     & $2.02 \times 10^{-5}$     &$\figgraph{.2}{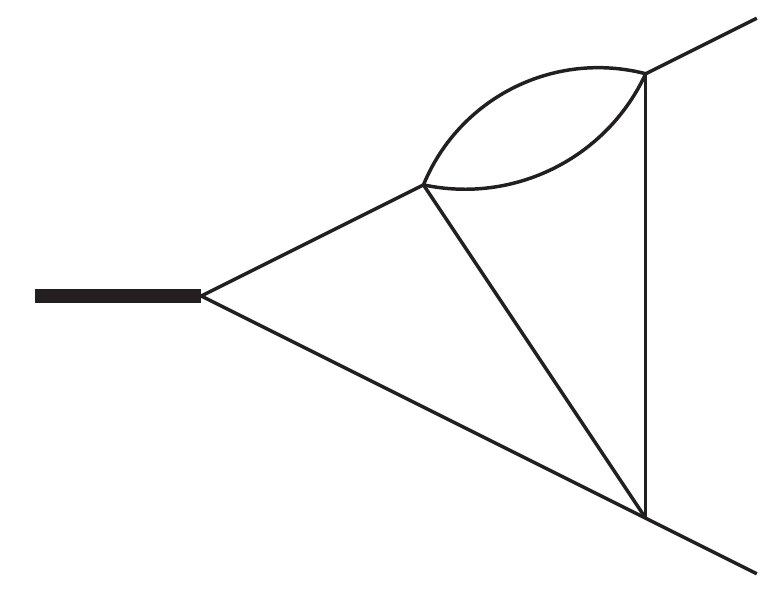}{4}$      & $43~\mathrm{s}$     & $5.05 \times 10^{-5}$\\
\hline
$\figgraph{.2}{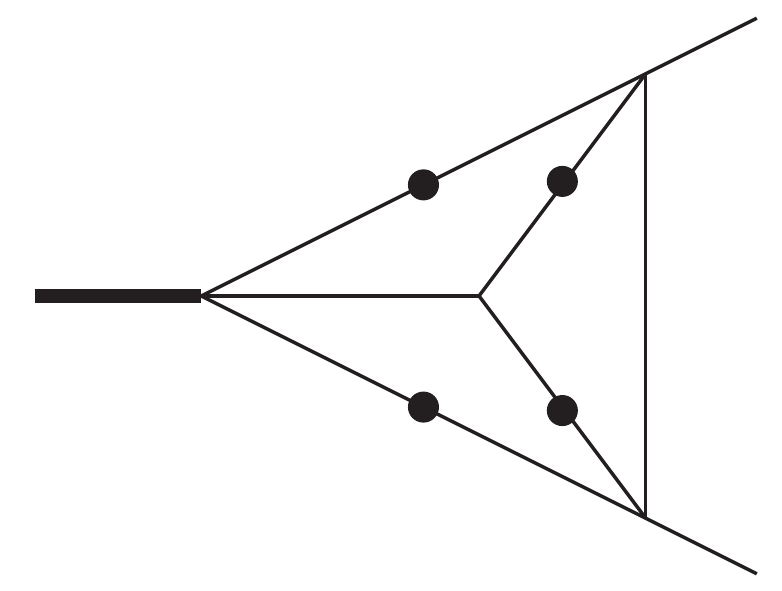}{6}$      & $19~\mathrm{s}$     & $1.86 \times 10^{-5}$     &$\figgraph{.2}{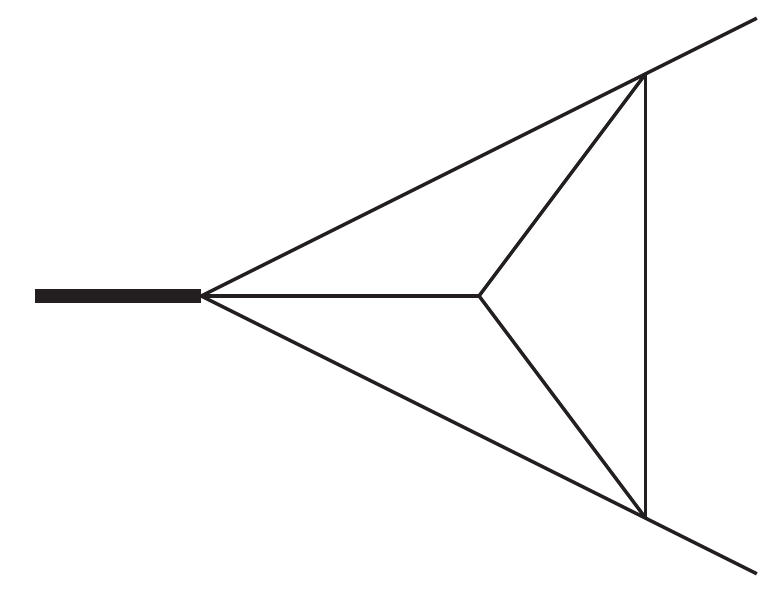}{4}$      & $16~\mathrm{s}$     & $1.42 \times 10^{-5}$\\
\hline
$\figgraph{.2}{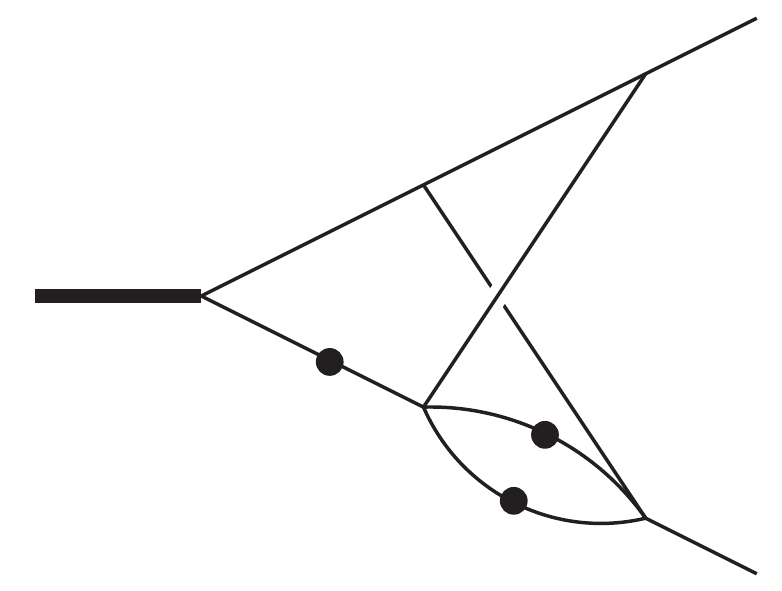}{6}$     & $69~\mathrm{s}$     & $8.83 \times 10^{-6}$     &$\figgraph{.2}{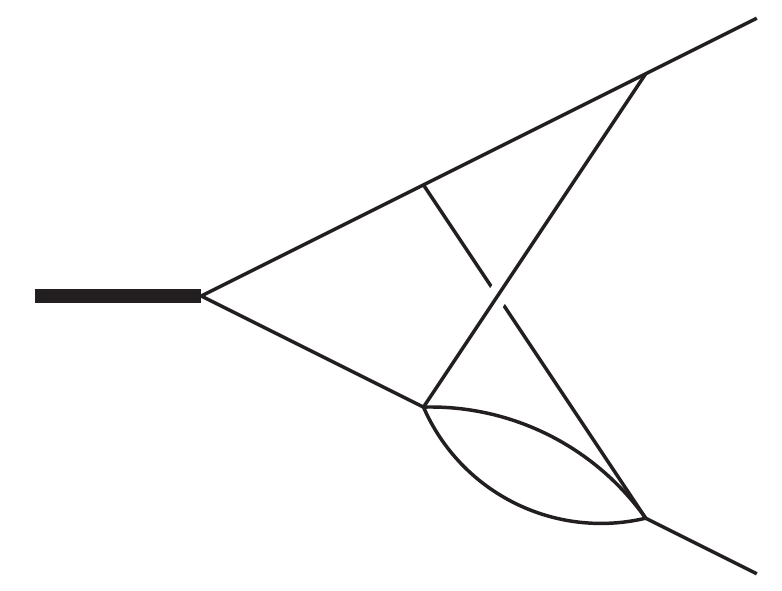}{4}$     & $46~\mathrm{s}$     & $1.57 \times 10^{-3}$\\
\hline
$\figgraph{.2}{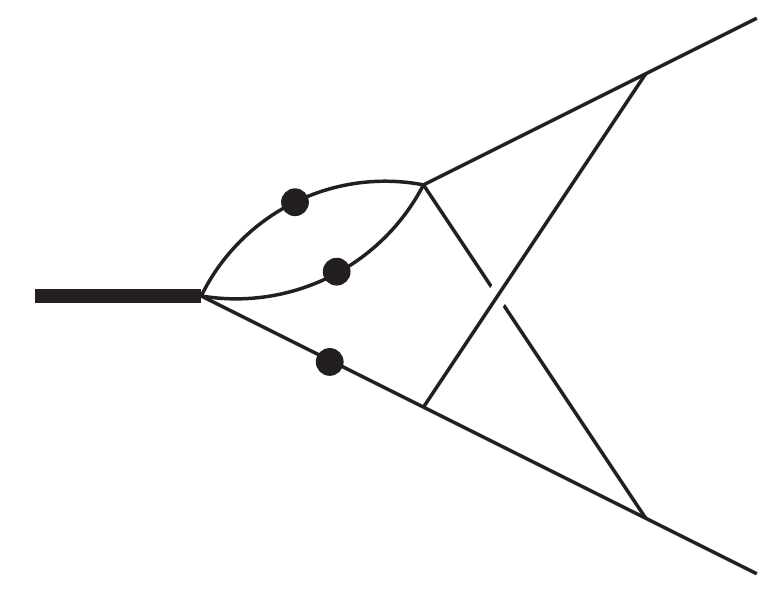}{6}$     & $88~\mathrm{s}$     & $1.93 \times 10^{-6}$     &$\figgraph{.2}{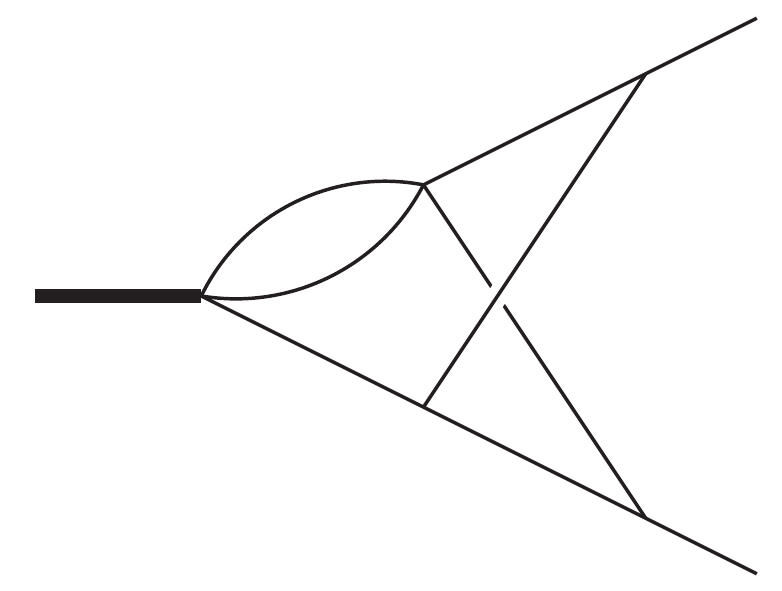}{4}$     & $117~\mathrm{s}$     & $1.35 \times 10^{-3}$\\
\hline
$\figgraph{.2}{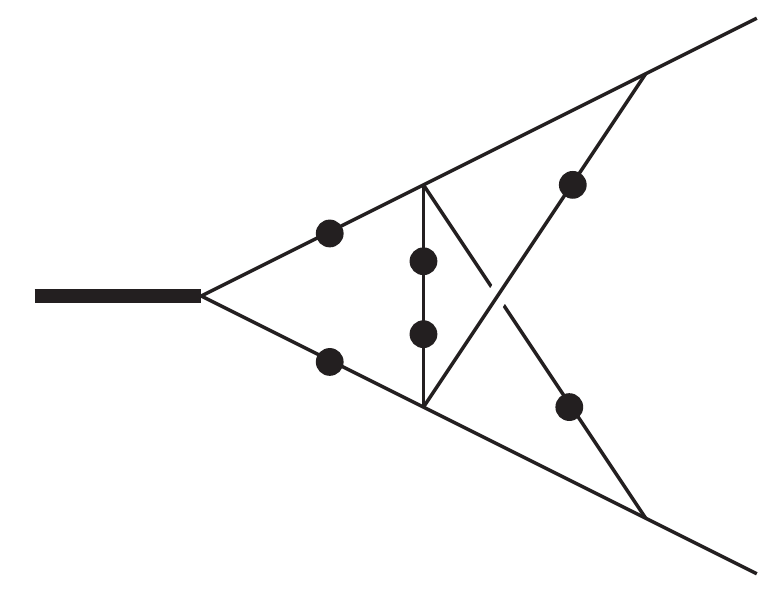}{8}$     & $74~\mathrm{s}$     & $6.62 \times 10^{-6}$     &$\figgraph{.2}{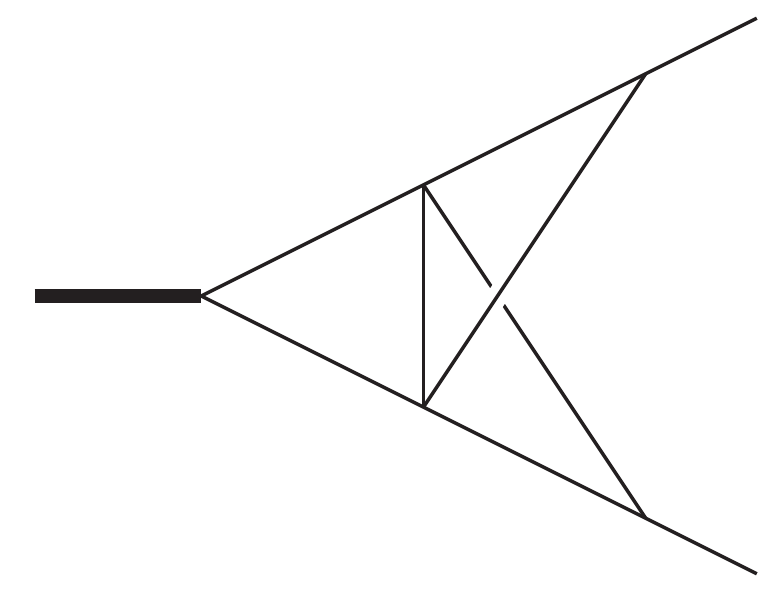}{4}$     & $46~\mathrm{s}$     & $1.40 \times 10^{-5}$\\
\hline
$\figgraph{.2}{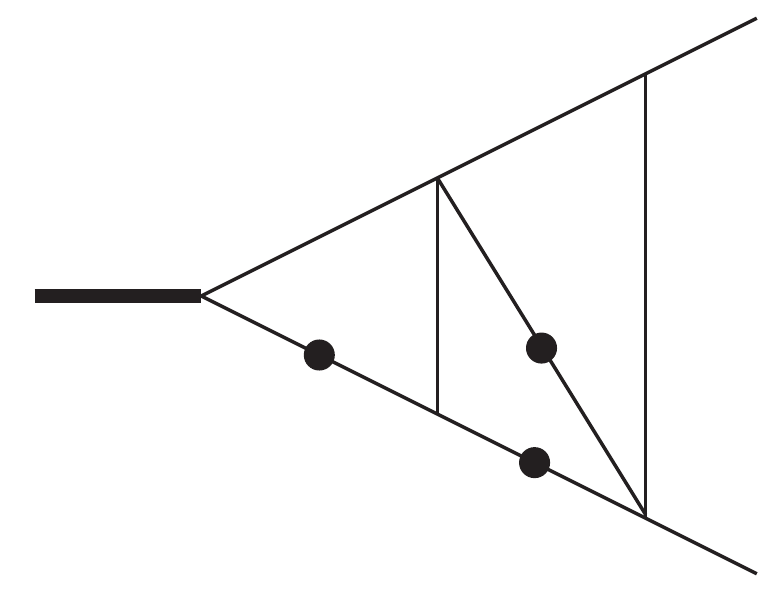}{6}$      & $26~\mathrm{s}$     & $4.29 \times 10^{-7}$     &$\figgraph{.2}{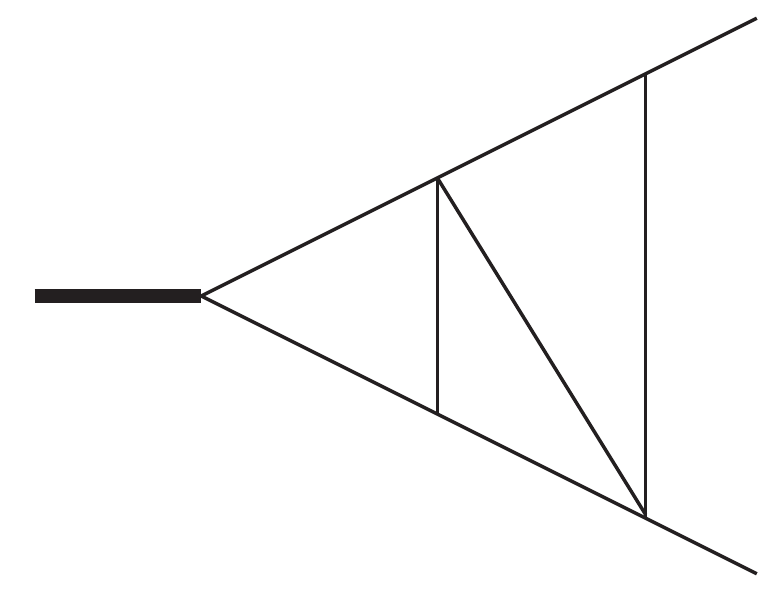}{4}$      & $17~\mathrm{s}$     & $4.89 \times 10^{-6}$\\
\hline
$\figgraph{.2}{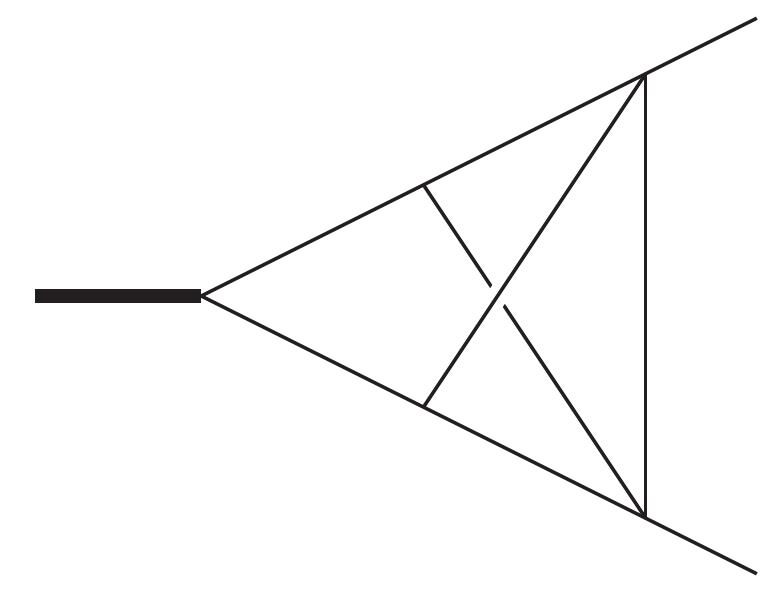}{4}$     & $20~\mathrm{s}$     & $6.27 \times 10^{-6}$     &$\figgraph{.2}{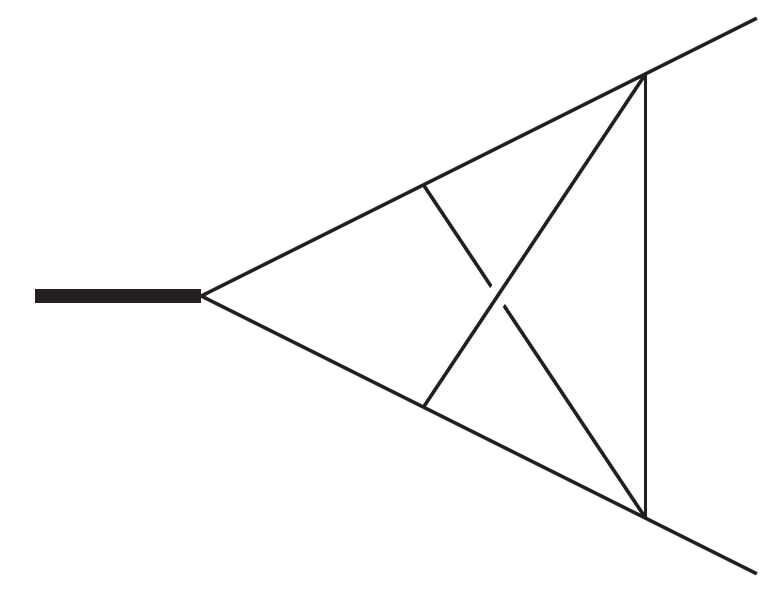}{4}$     & $20~\mathrm{s}$     & $6.27 \times 10^{-6}$\\
\hline
$\figgraph{.2}{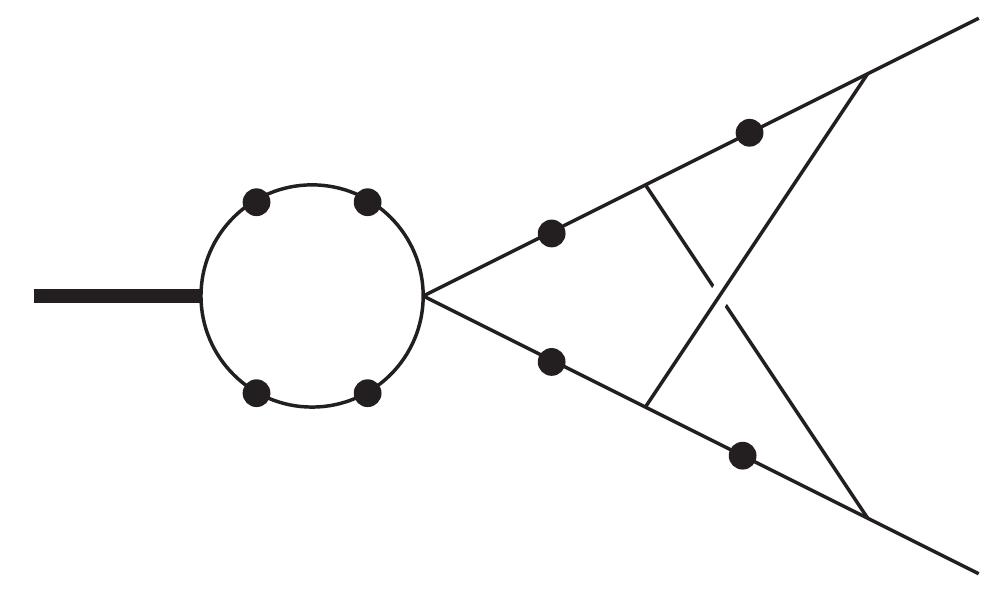}{8}$     & $212~\mathrm{s}$     & $1.50 \times 10^{-5}$     &$\figgraph{.2}{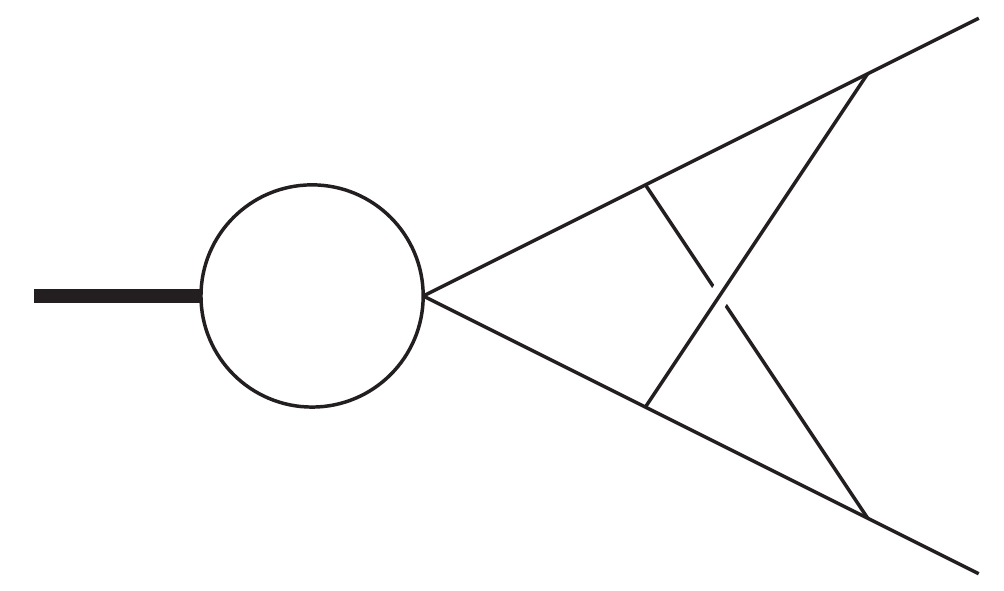}{4}$     & $407~\mathrm{s}$     & $5.66 \times 10^{-4}$\\
\hline
$\figgraph{.2}{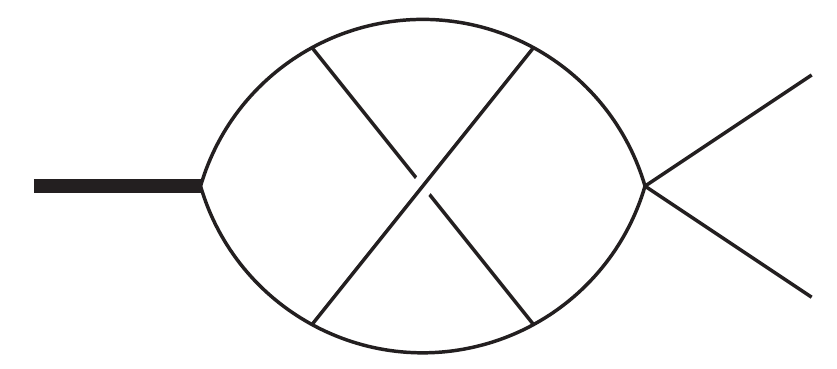}{4}$     & $72~\mathrm{s}$     & $3.11 \times 10^{-6}$     &$\figgraph{.2}{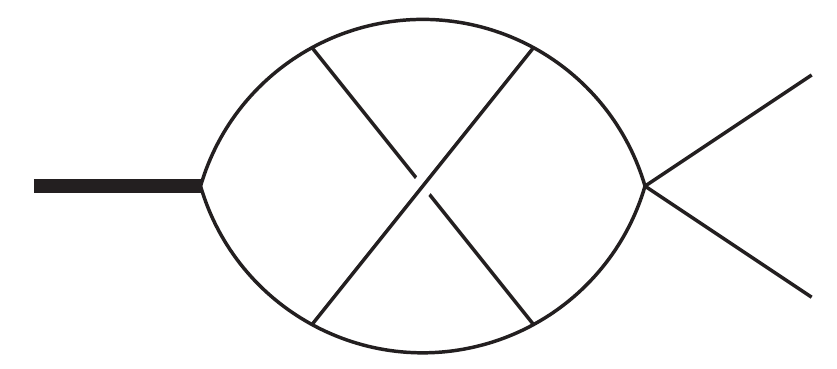}{4}$     & $75~\mathrm{s}$     & $3.11 \times 10^{-6}$\\
\hline
$\figgraph{.2}{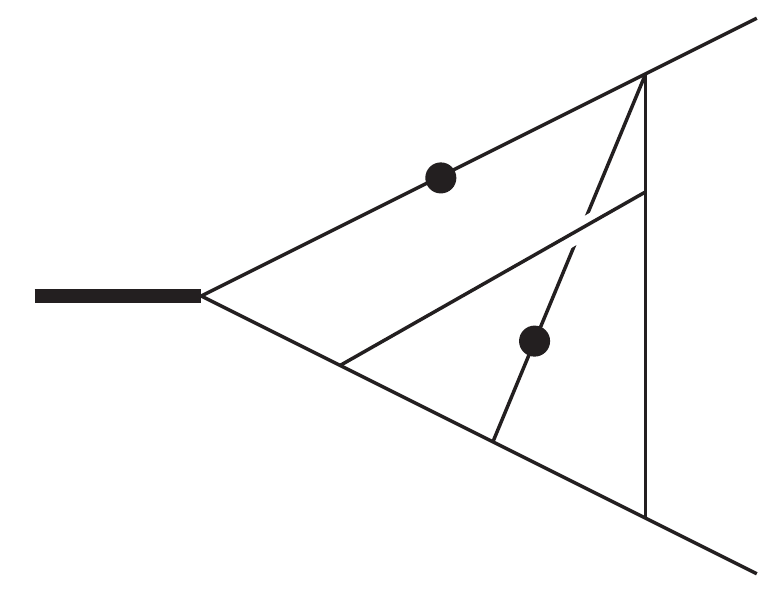}{6}$     & $74~\mathrm{s}$     & $9.04 \times 10^{-6}$     &$\figgraph{.2}{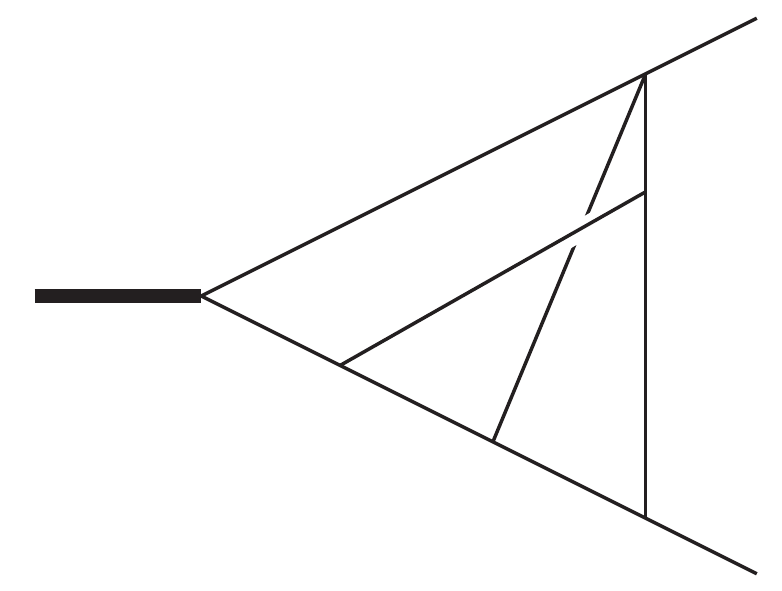}{4}$     & $126~\mathrm{s}$     & $2.18 \times 10^{-4}$\\
\hline
$\figgraph{.2}{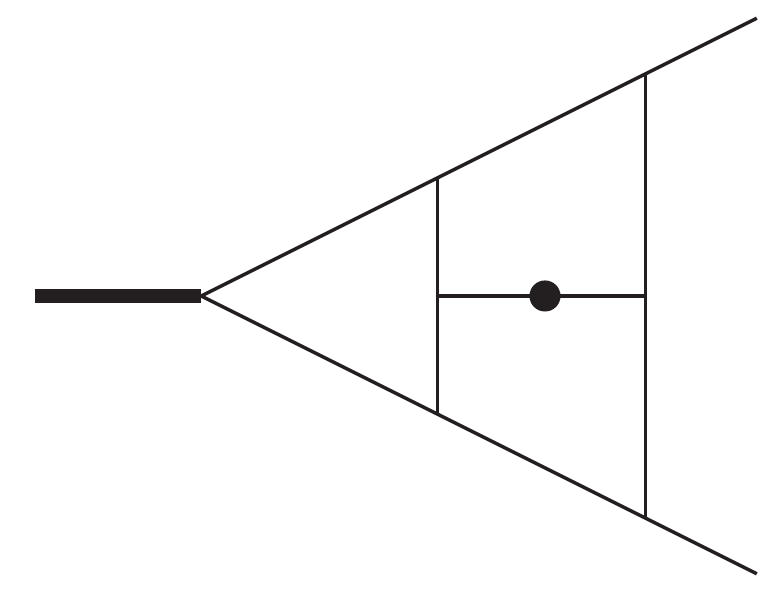}{6}$     & $128~\mathrm{s}$     & $5.12 \times 10^{-6}$     &$\figgraph{.2}{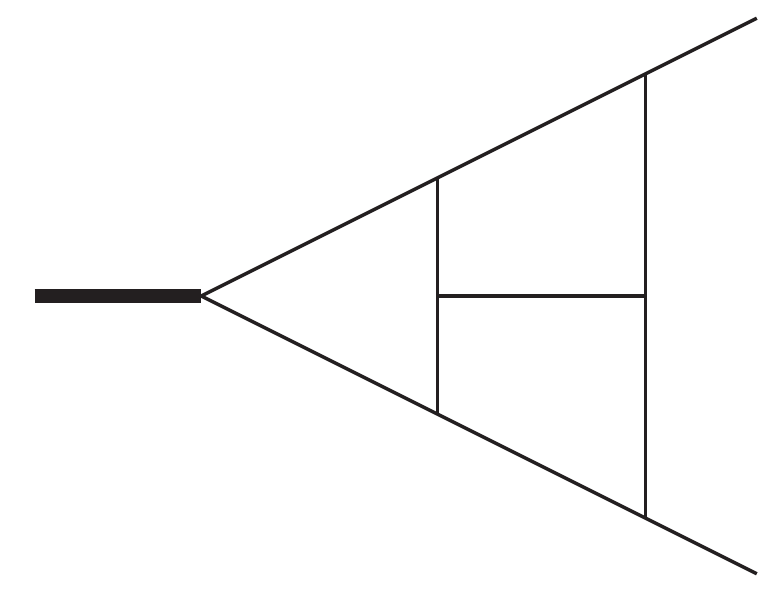}{4}$     & $39094~\mathrm{s}$     & $9.91 \times 10^{-4}$\\
\hline
$\figgraph{.2}{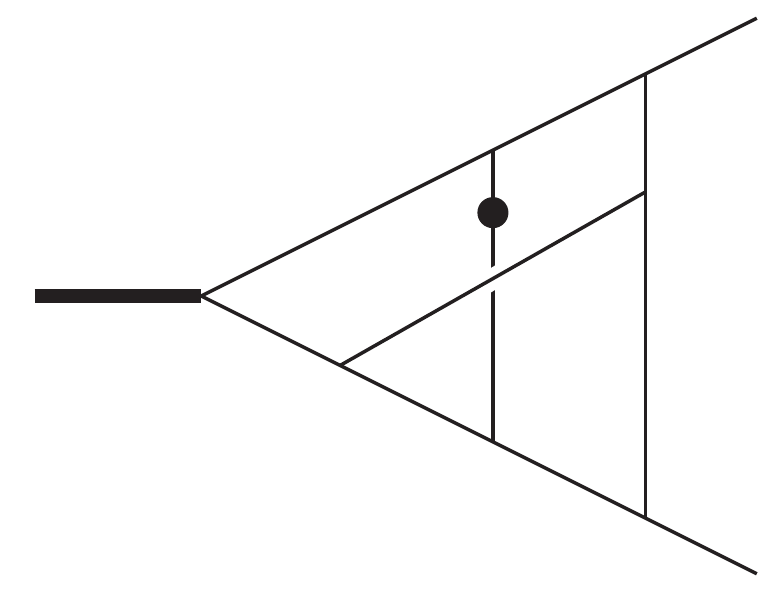}{6}$     & $192~\mathrm{s}$     & $2.68 \times 10^{-6}$     &$\figgraph{.2}{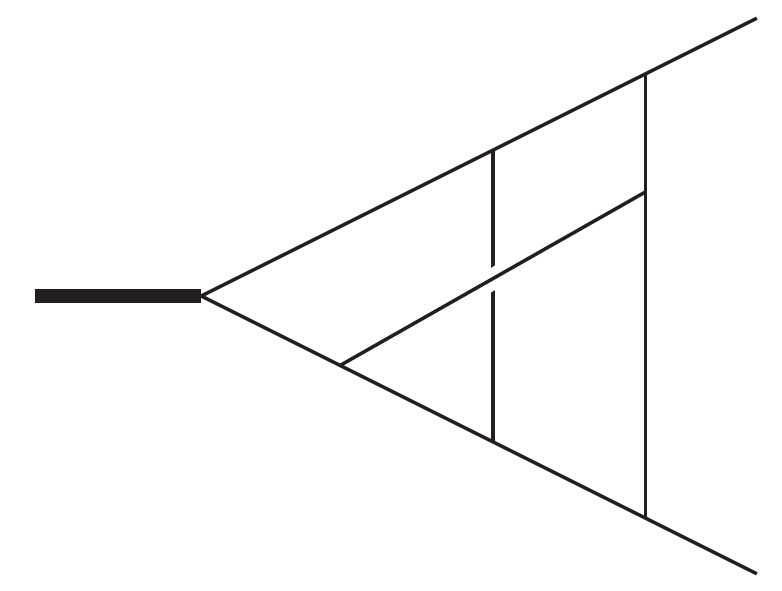}{4}$     & $19025~\mathrm{s}$     & $9.38 \times 10^{-5}$\\
\hline
$\figgraph{.2}{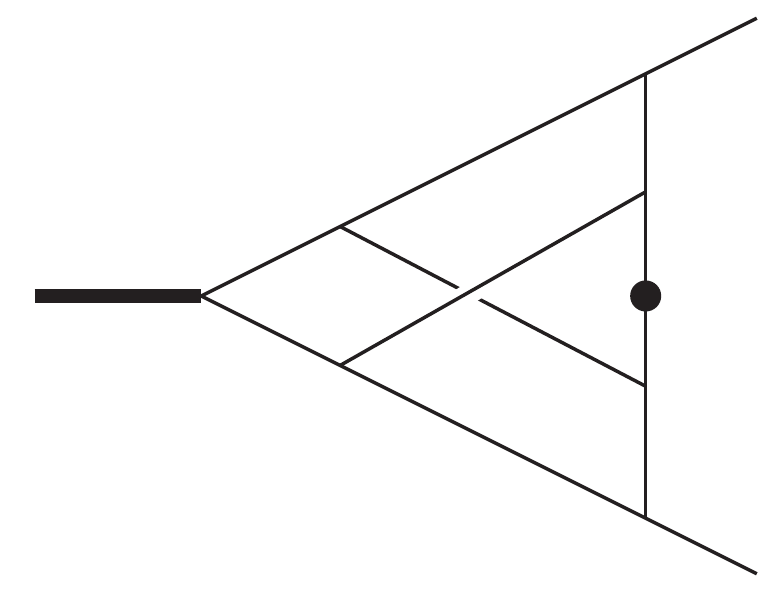}{6}$     & $127~\mathrm{s}$     & $2.26 \times 10^{-6}$     &$\figgraph{.2}{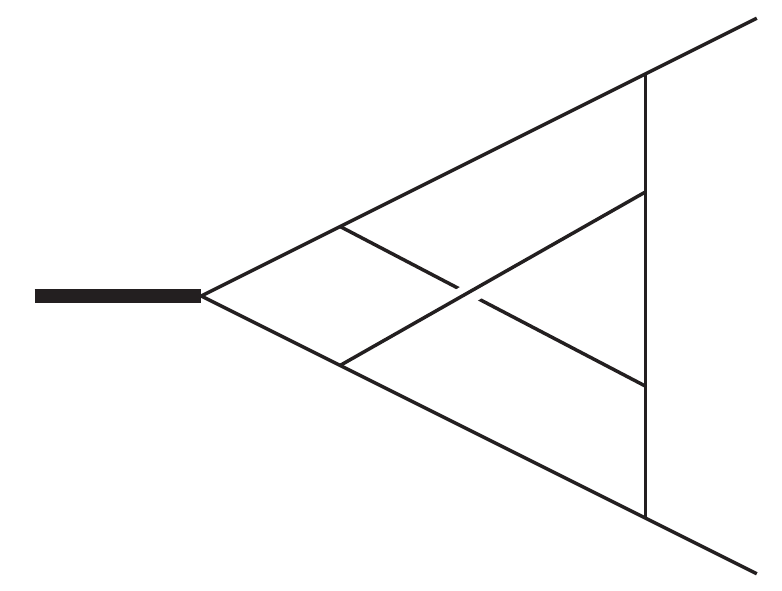}{4}$     & $19586~\mathrm{s}$     & $1.07 \times 10^{-4}$\\
\hline
Total/Average: & $1541~\mathrm{s}$     & $1.76 \times 10^{-5}$ & Total/Average: & $79016~\mathrm{s}$     & $2.33 \times 10^{-4}$\\
\hline
\caption{Numerical performance of finite and conventional integral bases for massless three-loop form factors with {\tt FIESTA\;4}.
In addition to the run times, the fractional difference from the exact value is shown.
In the final row of the table, total run times and average relative accuracies are recorded.}
\label{tab:ff3L}
\end{longtable}
\end{center}

\begin{center}
\begin{longtable}[h!]{|c|c|c||c|c|}
\hline
$\figgraph{.2}{ff3a_4_172}{10}$     & $6~\mathrm{s}$     & $2.46 \times 10^{-5}$     & $9~\mathrm{s}$     & $6.01 \times 10^{-5}$\\
\hline
$\figgraph{.2}{ff3a_5_662}{8}$      & $49~\mathrm{s}$     & $8.88 \times 10^{-7}$     & $123~\mathrm{s}$     & $7.72 \times 10^{-5}$\\
\hline
$\figgraph{.2}{ff3a_5_158}{6}$      & $35~\mathrm{s}$     & $4.31 \times 10^{-5}$     & $82~\mathrm{s}$     & $9.55 \times 10^{-5}$\\
\hline
$\figgraph{.2}{ff3a_5_412}{10}$     & $17~\mathrm{s}$     & $4.09 \times 10^{-5}$     & $46~\mathrm{s}$     & $1.37 \times 10^{-4}$\\
\hline
$\figgraph{.2}{ff3a_5_433}{8}$      & $16~\mathrm{s}$     & $1.28 \times 10^{-4}$     & $38~\mathrm{s}$     & $3.03 \times 10^{-4}$\\
\hline
$\figgraph{.2}{ff3a_6_2695}{6}$     & $121~\mathrm{s}$     & $5.42 \times 10^{-6}$     & $313~\mathrm{s}$     & $4.17 \times 10^{-5}$\\
\hline
$\figgraph{.2}{ff3a_6_1683}{10}$    & $93~\mathrm{s}$     & $1.37 \times 10^{-5}$     & $263~\mathrm{s}$     & $4.47 \times 10^{-5}$\\
\hline
$\figgraph{.2}{ff3a_6_691}{8}$      & $39~\mathrm{s}$     & $2.25 \times 10^{-6}$     & $103~\mathrm{s}$     & $3.96 \times 10^{-6}$\\
\hline
$\figgraph{.2}{ff3a_6_1433}{10}$    & $34~\mathrm{s}$     & $2.90 \times 10^{-5}$     & $90~\mathrm{s}$     & $9.54 \times 10^{-5}$\\
\hline
$\figgraph{.2}{ff3a_6_429}{6}$      & $30~\mathrm{s}$     & $2.02 \times 10^{-5}$     & $80~\mathrm{s}$     & $3.82 \times 10^{-5}$\\
\hline
$\figgraph{.2}{ff3a_6_444}{6}$      & $19~\mathrm{s}$     & $1.86 \times 10^{-5}$     & $45~\mathrm{s}$     & $4.43 \times 10^{-5}$\\
\hline
$\figgraph{.2}{ff3b_7_1770}{6}$     & $69~\mathrm{s}$     & $8.83 \times 10^{-6}$     & $202~\mathrm{s}$     & $1.29 \times 10^{-5}$\\
\hline
$\figgraph{.2}{ff3b_7_1780}{6}$     & $88~\mathrm{s}$     & $1.93 \times 10^{-6}$     & $274~\mathrm{s}$     & $9.25 \times 10^{-6}$\\
\hline
$\figgraph{.2}{ff3b_7_1766}{8}$     & $74~\mathrm{s}$     & $6.62 \times 10^{-6}$     & $220~\mathrm{s}$     & $4.79 \times 10^{-6}$\\
\hline
$\figgraph{.2}{ff3a_7_758}{6}$      & $26~\mathrm{s}$     & $4.29 \times 10^{-7}$     & $82~\mathrm{s}$     & $4.91 \times 10^{-6}$\\
\hline
$\figgraph{.2}{ff3b_7_1722}{4}$     & $20~\mathrm{s}$     & $6.27 \times 10^{-6}$     & $70~\mathrm{s}$     & $1.72 \times 10^{-5}$\\
\hline
$\figgraph{.2}{ff3c_8_2959}{8}$     & $212~\mathrm{s}$     & $1.50 \times 10^{-5}$     & $655~\mathrm{s}$     & $2.36 \times 10^{-5}$\\
\hline
$\figgraph{.2}{ff3b_8_2750}{4}$     & $72~\mathrm{s}$     & $3.11 \times 10^{-6}$     & $213~\mathrm{s}$     & $1.57 \times 10^{-5}$\\
\hline
$\figgraph{.2}{ff3b_8_1662}{6}$     & $74~\mathrm{s}$     & $9.04 \times 10^{-6}$     & $234~\mathrm{s}$     & $1.29 \times 10^{-5}$\\
\hline
$\figgraph{.2}{ff3a_9_1790}{6}$     & $128~\mathrm{s}$     & $5.12 \times 10^{-6}$     & $491~\mathrm{s}$     & $2.22 \times 10^{-5}$\\
\hline
$\figgraph{.2}{ff3b_9_1790}{6}$     & $192~\mathrm{s}$     & $2.68 \times 10^{-6}$     & $761~\mathrm{s}$     & $5.84 \times 10^{-6}$\\
\hline
$\figgraph{.2}{ff3c_9_1015}{6}$     & $127~\mathrm{s}$     & $2.26 \times 10^{-6}$    & $485~\mathrm{s}$     & $8.45 \times 10^{-6}$\\
\hline
Totals/Averages: & $1541~\mathrm{s}$     & $1.76 \times 10^{-5}$  & $4879~\mathrm{s}$     & $4.90 \times 10^{-5}$\\
\hline
\caption{Numerical performance of finite master integrals with {\tt FIESTA\;4}
for higher orders in the $\epsilon$ expansion.
On the left, the weight six results of Table \ref{tab:ff3L} are reproduced.
On the right, results for weight eight are shown.
}
\label{tab:higherff3L}
\end{longtable}
\end{center}
\bibliographystyle{JHEP}
\bibliography{dymasters1m}

\end{document}